%% file: main.tex
\documentclass[sigconf, nonacm]{acmart}

\input{extras/macros}
\input{extras/metadata}

\hypersetup{draft}

\begin{document}


\input{extras/title_and_authors}
\input{extras/abstract}
\maketitle
\input{extras/copyrightblock}


\section{Introduction}
\label{sec:intro}

\input{sections/introduction}

\input{sections/example-queries}

\paragraph*{Further Related Work}
\label{sec:related-work}
\input{sections/related-work}

\section{CEQL syntax and semantics}
\label{sec:ceql-examples}
\label{sec:ceql-semantics}

\input{sections/ceql-examples}
\input{sections/ceql-semantics}

\section{Query compilation}
\label{sec:cea}
\input{sections/cea}

\section{Evaluation algorithm} \label{sec:evaluation}

\input{sections/algorithm}

\section{Experiments}
\label{sec:experiments}

\input{sections/experiments}

\section{Conclusions and future work}
\label{sec:conclusions}
\input{sections/conclusions}

\begin{acks}
	The authors are grateful to Mart\'in Ugarte for stimulating discussions that lie at the basis of this work.
	A. Grez, A. Quintana, and C. Riveros were supported by ANID - Millennium Science Initiative Program - Code \verb|ICN17_002|.  S. Vansummeren was supported by the Bijzonder Onderzoeksfonds (BOF) of Hasselt University (Belgium) under Grant No. \verb|BOF20ZAP02|.
\end{acks}

\newpage

\bibliographystyle{ACM-Reference-Format}
{\bibliography{extras/references}}

 \inFullVersion{%
 \newpage


 \onecolumn
 \appendix


 \section{Proofs of Section~\ref{sec:ceql-semantics}}

 \input{appendix/ceql-semantics}

 \section{Proofs of Section~\ref{sec:evaluation}}

 \input{appendix/algorithm}

 \section{Other baseline systems}
 \label{sec:app-other-competitors}
 \input{appendix/app-other-competitors}

 \section{Stock market dataset}
 \label{sec:app-seq-stocks}
 \input{appendix/app-seq-stocks}

 \section{Smart homes dataset}
 \label{sec:app-seq-homes}
 \input{appendix/app-seq-homes}

 \section{Taxi trips dataset}
 \label{sec:app-seq-taxis}
 \input{appendix/app-seq-taxis}

 \section{Stock market queries with other operators}

\input{appendix/app-stock-queries}

 \section{Experiments with selection strategies}
 \label{sec:app-seq-strategies}
 \input{appendix/app-sel-strategies}

%

%
%
 }

\end{document}


\onecolumn

\title{CORE: a COmplex event Recognition Engine \\ Supplementary Material}

\author{Marco Bucchi}
\affiliation{%
	\institution{PUC Chile}
}
\email{mabucchi@uc.cl}

\author{Alejandro Grez}
\affiliation{%
	\institution{PUC Chile}
}
\email{ajgrez@uc.cl}

\author{Andrés Quintana}
\affiliation{%
	\institution{PUC Chile}
}
\email{afquintana@uc.cl}

\author{Cristian Riveros}
\affiliation{%
	\institution{PUC Chile}
}
\email{cristian.riveros@uc.cl}


\author{Stijn Vansummeren}
\orcid{0000-0001-7793-9049}
\affiliation{%
  \institution{UHasselt -- Hasselt University}
}
\email{stijn.vansummeren@uhasselt.be}

\maketitle

The following supplementary material is available at \url{https://github.com/CORE-cer}:

\begin{itemize}
\item A full, extended version of the submitted paper whose appendix contains proofs of the formal statements, as well as the full syntax of the stock market queries considered in the experiments.
  
\item The full source code of the CORE engine.
\item The scripts to replicate the experiments considered in the paper.
\end{itemize}

%% file: extras/macros.tex
\usepackage[utf8]{inputenc}
\usepackage{wrapfig}
\usepackage{stmaryrd}
\usepackage{thmtools, thm-restate}
\usepackage{epsfig}
\usepackage{bbold}
\usepackage{multicol}
\usepackage{multirow}
\usepackage{varwidth}
\usepackage{tikz}
\usepackage{pgfplots}
\usepackage{xspace}

\pgfplotsset{compat=1.12}
\usetikzlibrary{arrows,decorations.pathmorphing,backgrounds,positioning,fit,calc,automata}
\usetikzlibrary{trees}
\usetikzlibrary{shapes}
\usetikzlibrary{chains}
\usetikzlibrary{patterns}
\usepackage{calc}
\usepackage[textwidth=2cm,textsize=small]{todonotes}
\usepackage{algorithm}
\usepackage[noend]{algpseudocode}
\usepackage{paralist}
\usepackage{enumitem}
\usepackage{balance}
\usepackage{footnote}
\makesavenoteenv{tabular}
\usepackage{subfig}

\setitemize{noitemsep, topsep=6pt, parsep=1pt, partopsep=6pt, leftmargin=12pt, itemindent=0pt}

\usepackage{xpatch}
\usepackage{textcase}
\makeatletter
\xpatchcmd{\@sect}{\uppercase}{\MakeTextUppercase}{}{}
\xpatchcmd{\@sect}{\uppercase}{\MakeTextUppercase}{}{}
\makeatother

\usepackage[explicit]{titlesec}

\titleformat{\paragraph}[runin]
    {\normalfont\bfseries}
    {}
    {0pt}
    {#1}

\titlespacing{\paragraph}
    {0pt} 
    {1ex plus .1ex minus .2ex}
    {1ex} 

\newcommand{\inFullVersion}[1]{#1}

\newcommand{\inConfVersion}[1]{}

\newcommand{\altVersion}[2]{\inFullVersion{#1}\inConfVersion{#2}}

\newcommand{\newtext}[1]{#1}

\newcommand{\systemname}{CORE}
\newcommand{\systemql}{CEQL}
\newcommand{\cer}{CER\xspace}
\newcommand{\cel}{CEL\xspace}

\newcommand{\lset}{\mathbf{L}}
\newcommand{\pset}{\mathbf{P}}

\newcommand{\xset}{\mathbf{X}}

\newcommand{\cA}{\mathcal{A}}

\newcommand{\bbN}{\mathbb{N}}

\newcommand{\sem}[1]{{\llbracket{}{#1}\rrbracket}}
\newcommand{\seq}[1]{\ensuremath{\bar{#1}}}

\newcommand{\dset}{\mathbf{D}}
\newcommand{\aset}{\mathbf{A}}
\newcommand{\tset}{\mathbf{T}}

\newcommand{\type}{\operatorname{type}}
\newcommand{\nullvalue}{\texttt{NULL}}

\newcommand{\ctime}{\operatorname{time}}
\newcommand{\cstart}{\operatorname{start}}
\newcommand{\cend}{\operatorname{end}}
\newcommand{\cdata}{\operatorname{data}}
\newcommand{\cnormal}[1]{C_{#1}}

\newcommand{\FILTER}{~\mathtt{FILTER}~}

\newcommand{\cor}{~\mathtt{OR}~}

\newcommand{\tFILTER}{\mathtt{FILTER}}

\newcommand{\tcor}{\mathtt{OR}}

\newcommand{\sq}{\,;\,}

\newcommand{\tsq}{;}
\newcommand{\ks}{+}
\newcommand{\IN}{~\operatorname{AS}~}
\newcommand{\sIN}{\operatorname{AS}}
\newcommand{\PROJ}[1]{~\pi_{#1}~}

\newcommand{\within}{~\mathtt{WITHIN}~}

\newcommand{\semcel}[1]{{\llbracket{}{#1}\rrbracket}}
\newcommand{\semaux}[1]{{\llceil{#1}\rrfloor}}

\newcommand{\CEA}{CEA}
\newcommand{\amark}{\bullet}
\newcommand{\umark}{\circ}
\newcommand{\trans}[2][]{\raisebox{-1pt}[10pt][0pt]{$\overset{#2}{\underset{^{#1}}{\raisebox{0pt}[3pt][0pt]{$\relbar\mspace{-8mu}\longrightarrow$}}}$}}

\newcommand{\ecs}{\mathcal{E}}
\newcommand{\nnodes}{N}
\newcommand{\bnodes}{N_B}
\newcommand{\onodes}{N_O}
\newcommand{\unodes}{N_U}
\newcommand{\mnode}{\texttt{m}}
\newcommand{\nnode}{\texttt{n}}
\newcommand{\bnode}{\texttt{b}}
\newcommand{\onode}{\texttt{o}}

\newcommand{\unode}{\texttt{u}}
\newcommand{\uleft}{\textsf{left}}
\newcommand{\uright}{\textsf{right}}
\newcommand{\uleftcomp}{\textsf{l}}
\newcommand{\urightcomp}{\textsf{r}}
\newcommand{\opos}{\textsf{pos}}
\newcommand{\onext}{\textsf{next}}
\newcommand{\outce}[1]{\sem{#1}_{\ecs}}
\newcommand{\maxstart}{\text{max}}
\newcommand{\dbottom}{\textsf{new-bottom}}
\newcommand{\dextend}{\textsf{extend}}
\newcommand{\dunion}{\textsf{union}}
\newcommand{\ulist}{\texttt{ul}}

\newcommand{\linit}{\textsf{new-ulist}}
\newcommand{\linsert}{\textsf{insert}}
\newcommand{\lmerge}{\textsf{merge}}

\newcommand{\stack}{\texttt{st}}
\newcommand{\sinit}{\textsf{new-stack}}
\newcommand{\pop}{\textsf{pop}}
\newcommand{\push}{\textsf{push}}

\newcommand{\ctable}{\operatorname{T}}
\newcommand{\ckeys}{\textsf{keys}}
\newcommand{\cordkeys}{\textsf{ordered-keys}}

\newcommand{\cO}{\mathcal{O}}

\newcommand{\ssd}[3]{\arraycolsep=0pt
	\left[
	\renewcommand{\arraystretch}{1}
	\begin{array}{c}
          \textnormal{\textsf{#1}} \\ \textnormal{\textsf{#2}} \\ #3
	\end{array}
	\right]}

\newcommand{\TRUE}{\mathtt{TRUE}}

\newcommand{\cdotlor}{\mathrel{\ooalign{$\lor$\cr\hidewidth\raise.225ex\hbox{$\cdot\mkern3.3mu$}\cr}}}

\newcommand{\strans}[2][]{\raisebox{-1pt}[10pt][0pt]{$\overset{#2}{\underset{^{#1}}{\raisebox{0pt}[3pt][0pt]{$\relbar\mspace{-8mu}\rightarrow$}}}$}}

\newcommand{\yield}{\operatorname{yield}}
\newcommand{\odepth}{\operatorname{odepth}}

\newtheorem{lemma}{Lemma}

\newtheorem{example}{Example}

\tikzset{
	defaultstyle/.style={
		>=stealth, 
		semithick, 
		auto,
		initial text= {},
		initial distance= {3mm},
		accepting distance= {3mm}}}



%% file: extras/metadata.tex
\newcommand\vldbdoi{XX.XX/XXX.XX}
\newcommand\vldbpages{XXX-XXX}
\newcommand\vldbvolume{15}
\newcommand\vldbissue{9}
\newcommand\vldbyear{2022}
\newcommand\vldbauthors{\authors}
\newcommand\vldbtitle{\shorttitle} 
\newcommand\vldbavailabilityurl{https://github.com/CORE-cer}
\newcommand\vldbpagestyle{empty} 

%% file: extras/title_and_authors.tex
\title{CORE: a Complex Event Recognition Engine}

\author{Marco Bucchi}
\affiliation{%
	\institution{PUC Chile}
}
\email{mabucchi@uc.cl}

\author{Alejandro Grez}
\affiliation{%
	\institution{PUC Chile and IMFD Chile}
}
\email{ajgrez@uc.cl}

\author{Andrés Quintana}
\affiliation{%
	\institution{PUC Chile and IMFD Chile}
}
\email{afquintana@uc.cl}

\author{Cristian Riveros}
\affiliation{%
	\institution{PUC Chile and IMFD Chile}
}
\email{cristian.riveros@uc.cl}


\author{Stijn Vansummeren}
\orcid{0000-0001-7793-9049}
\affiliation{%
  \institution{UHasselt, Data Science Institute}
}
\email{stijn.vansummeren@uhasselt.be}


%% file: extras/abstract.tex

\begin{abstract}
Complex Event Recognition (CER) systems are a prominent technology for finding user-defined query patterns over large data streams in real time. CER query evaluation is known to be computationally challenging, since it requires maintaining a set of partial matches, and this set quickly grows super-linearly in the number of processed events. We present CORE, a novel COmplex event Recognition Engine that focuses on the efficient evaluation of a large class of complex event queries, including time windows as well as the partition-by event correlation operator. This engine uses a novel automaton-based evaluation algorithm that circumvents the super-linear partial match problem: under data complexity, it takes constant time per input event to maintain a data structure that compactly represents the set of partial matches and, once a match is found, the query results may be enumerated from the data structure with output-linear delay. We experimentally compare CORE against state-of-the-art CER systems on real-world data. We show that (1) CORE's performance is stable with respect to both query and time window size, and (2) CORE outperforms the other systems by up to five orders of magnitude on different workloads.
\end{abstract}


%% file: extras/copyrightblock.tex
\pagestyle{\vldbpagestyle}
\begingroup\small\noindent\raggedright\textbf{PVLDB Reference Format:}\\
\vldbauthors. \vldbtitle. PVLDB, \vldbvolume(\vldbissue): \vldbpages, \vldbyear.\\
\href{https://doi.org/\vldbdoi}{doi:\vldbdoi}
\endgroup
\begingroup
\renewcommand\thefootnote{}\footnote{\noindent
This work is licensed under the Creative Commons BY-NC-ND 4.0 International License. Visit \url{https://creativecommons.org/licenses/by-nc-nd/4.0/} to view a copy of this license. For any use beyond those covered by this license, obtain permission by emailing \href{mailto:info@vldb.org}{info@vldb.org}. Copyright is held by the owner/author(s). Publication rights licensed to the VLDB Endowment. \\
\raggedright Proceedings of the VLDB Endowment, Vol. \vldbvolume, No. \vldbissue\ %
ISSN 2150-8097. \\
\href{https://doi.org/\vldbdoi}{doi:\vldbdoi} \\
}\addtocounter{footnote}{-1}\endgroup

\ifdefempty{\vldbavailabilityurl}{}{
\vspace{.3cm}
\begingroup\small\noindent\raggedright\textbf{PVLDB Artifact Availability:}\\
The source code, data, and/or other artifacts have been made available at \url{\vldbavailabilityurl}.
\endgroup
}


%% file: sections/introduction.tex
\emph{Complex Event Recognition (CER for short)}, also called Complex Event
Processing, has emerged as a prominent technology for supporting streaming
applications like maritime monitoring~\cite{10.1145/3328905.3329762}, network
intrusion detection~\cite{mukherjee1994}, industrial control
systems~\cite{groover2007} and real-time analytics~\cite{sahay2008}. CER systems
operate on high-velocity streams of primitive events and evaluate expressive
event queries to detect \emph{complex events}: collections of primitive events
that satisfy some pattern. In particular, CER queries match incoming events on
the basis of their content; where they occur in the input stream; and how this
order relates to other events in the stream~\cite{cugola2012,
  DBLP:journals/vldb/GiatrakosAADG20, DBLP:journals/csur/AlevizosSAP17}. 

CER systems hence aim to detect situations of interest, in the form of complex
events, in order to give timely insights for implementing reactive responses to
them when necessary.  As such, they strive for low latency query evaluation. CER
query evaluation, however, is known to be computationally challenging
~\cite{SASEautomata,mei2009zstream,Cugola:2012,SASEcomplexity,DBLP:conf/sigmod/0019ANNW21,
  DBLP:conf/icde/ZhaoHW20}. Indeed, conceptually, evaluating a CER query
requires maintaining or recomputing a set of partial matches, so that when a new
event arrives all partial matches that---together with the newly arrived
event--now form a complete answer can be found. \newtext{Unfortunately, even for simple
CER patterns, the set of partial matches quickly becomes polynomial in the
number $N$ of previously processed events (or, when time windows are used, the
number of events in the current window). Even worse, under the so-called
skip-till-any-match selection strategy~\cite{SASEautomata}, queries that include
the iteration operator may have sets of partial matches that grow
\emph{exponentially} in $N$~\cite{SASEautomata}. As a result, the arrival of
each new event requires a computation that is super-linear in $N$, which is
incompatible with the small latency requirement.}

In recognition of the computational challenge of CER query evaluation, a
plethora of research has proposed innovative evaluation
methods~\cite{DBLP:journals/vldb/GiatrakosAADG20,cugola2012,DEBStutorial}. These
methods range from proposing diverse execution
models~\cite{SASE,cayuga,mei2009zstream,DBLP:journals/ker/ArtikisSPP12},
including cost-based database-style query optimizations to trade-off between
materialization and lazy computation~\cite{mei2009zstream,DBLP:conf/debs/KolchinskySS15,DBLP:journals/pvldb/KolchinskyS18a}; to focusing on
specific query fragments (e.g., event selection policies~\cite{SASEautomata})
that somewhat limit the super-linear partial match explosion; to using load
shedding~\cite{DBLP:conf/icde/ZhaoHW20} to obtain low latency at the expense of
potentially missing matches; and to employing distributed
computation~\cite{Cugola:2012,DBLP:conf/debs/MayerTR17}.  All of these still
suffer, however, from a processing overhead per event that is super-linear
in~$N$. As such, their scalability is limited to CER queries over a short time
window, as we show in Section~\ref{sec:experiments}.  Unfortunately, for
applications such as maritime monitoring~\cite{10.1145/3328905.3329762}, network
intrusion~\cite{mukherjee1994} and fraud
detection~\cite{DBLP:conf/debs/ArtikisKCBMSFP17}, long time windows are
necessary and a solution based on new principles hence seems desirable.

\newtext{In recent work~\cite{CELJOURNAL,ICDT2019}, a subset of the authors have proposed
a theoretical algorithm for evaluating CER queries that circumvents the
super-linear partial match problem in theory: under data complexity the
algorithm takes \emph{constant} time per input event to maintain a data
structure that compactly represents the set of partial and full matches in a
size that is at most linear in $N$. Once a match is found, complex event(s) may
be enumerated from the data structure with \emph{output-linear delay}, meaning
that the time required to output recognized complex event $C$ is linear in the
size of $C$. This complexity is asymptotically optimal since any evaluation
algorithm needs to at least inspect every input event and list the query
answers.}

\newtext{Briefly, the evaluation algorithm consists of translating a CER query into a
particular kind of automaton model, called a Complex Event Automaton
(CEA). Given a CEA and an input event stream, the algorithm simulates the CEA on
the stream, and records all CEA runs by means of a graph data
structure. Interestingly, this graph succinctly encodes all the partial and
complete matches. When a new event arrives, the graph can be updated in constant
time to represent the new (partial) runs. Once a final state of the automaton is
reached, a traversal over the graph allows to enumerate the complex events with
output-linear delay.}

\newtext{To date, this approach to CER query evaluation has only been the subject of
theoretical investigation.  Because it is the only known algorithm that provably circumvents the super-linear partial match problem, however, the question is whether it can serve as the basis of a practical CER
system. In this paper, we answer this question affirmatively by presenting
\systemname, a novel COmplex event Recognition Engine based on the principles
of~\cite{CELJOURNAL,ICDT2019, DBLP:conf/icdt/GrezRUV20}.}

\newtext{A key limitation of \cite{CELJOURNAL,ICDT2019} that we need to resolve in
designing \systemname\xspace is that time windows are not supported in \cite{CELJOURNAL,ICDT2019},  neither semantically at the query  language level, nor at the evaluation algorithm level.  
Indeed, the algorithm records \emph{all} runs, no matter how long
ago the run occurred in the stream and no matter whether the run still occurs in
the current time window. As such, it also does not prune the graph to clear
space for further processing, which quickly becomes a memory bottleneck. Another
limitation is that the algorithm cannot deal with simple equijoins, such as
requiring all matching events to have identical values in a particular
attribute. Such equijoins are typically expressed by means of the so-called
\emph{partition-by} operator~\cite{DBLP:journals/vldb/GiatrakosAADG20}. We lift
both limitations in \systemname.}

\newtext{Technically, to ensure that CER queries with
time windows can still be processed in constant update per event and
output-linear delay enumeration we need to be able to represent CEA runs in a
\emph{ranked order fashion}: during enumeration, runs that start later in the
stream must be traversed before runs that start earlier, so that once a run does
not satisfy a time window restriction all runs that succeed it in the ranked
order will not satisfy the time window either, and can therefore be pruned. The
challenge is to encode this ranking in the graph representation of runs while
ensuring that we continue to be able to update it in constant time per new
event. This requires a completely new kind of graph data structure for encoding
runs, and a correspondingly new CEA evaluation algorithm. This new algorithm is
compatible with the partition-by operator, as we will see. We stress that, by
this new algorithm, the runtime of \systemname's complexity per event is
independent of the number of partial matches, as well as the length of the time
window being used. It therefore supports long time windows by design.}

\paragraph*{Contributions.} Our contributions  are as follows.

(1) \newtext{To be precise about the query language features supported by
  \systemname's evaluation algorithm, we formally introduce \systemql, a functional query
  language that focuses on \emph{recognizing} complex events. It supports common
  event recognition operators including sequencing, disjunction, filtering,
  iteration, and projection~\cite{ICDT2019,DBLP:conf/icdt/GrezRUV20,CELJOURNAL},
  extended with partition-by, and time-windows. Other features considered in the
  literature that focus on \emph{processing} of complex events, such as
  aggregation~\cite{PoppeLRM17,DBLP:conf/sigmod/PoppeLR019}, integration of
  non-event data sources~\cite{DBLP:conf/sigmod/0019ANNW21}, and parallel or
  distributed~\cite{nextCEP,distCED,Cugola:2012} execution are currently not
  supported by \systemql, and left for future work. }

(2) \newtext{We present an evaluation algorithm for \systemql\xspace that is based on entirely
new evaluation algorithm for CEA that deals with time windows and partitioning,
thereby lifting the limitations of~\cite{ICDT2019,DBLP:conf/icdt/GrezRUV20,
  CELJOURNAL}.}

(3) \newtext{We show that the algorithm is practical. We implement it inside of \systemname, and  experimentally compare \systemname\xspace against state-of-the-art CER engines on real-world data. Our experiments show that \systemname's performance is stable: the throughput is not affected by the size of the query or size of the time window. Furthermore,  \systemname\xspace outperforms existing systems by one to five orders of magnitude in throughput on different query workloads. }

The structure of the paper is as follows. 
We finish this section by discussing further related work not already mentioned above.
\newtext{We continue by introducing CEQL in Section~\ref{sec:ceql-examples}. We present the CEA computation model in Section~\ref{sec:cea}. The algorithm and its data structures are described in Section~\ref{sec:evaluation}, which also discusses implementation aspects of CORE. We dedicate Section \ref{sec:experiments} to experiments and conclude in
Section~\ref{sec:conclusions}.}

\inConfVersion{Because of space limitations, certain details, formal statements and proofs are omitted. A full version of this paper, whose Appendix contains those items, is available online~\cite{core-full-paper-version}.}
\inFullVersion{Because of space limitations, certain details, formal statements and proofs are deferred to the Appendix.}


%% file: sections/example-queries.tex
\begin{figure*}[tbp]
  \footnotesize
  \hspace{-1cm}
  \begin{tabular}{p{5.6cm}p{3.3cm}p{7cm}}
\begin{verbatim}
SELECT * FROM Stock
WHERE SELL as msft; SELL as intel; SELL as amzn
FILTER msft[name="MSFT"] AND msft[price > 100]
   AND intel[name="INTC"] 
   AND amzn[name="AMZN"] AND amzn[price < 2000]
\end{verbatim}
    &
\begin{verbatim}
SELECT b FROM Stock
WHERE (BUY or SELL) as s; 
      (BUY or SELL) as b
PARTITION BY [name], [volume]
WITHIN 1 minute
\end{verbatim}
&    
\begin{verbatim}
SELECT MAX * FROM Stock
WHERE (BUY OR SELL) as low; (BUY OR SELL)+ as mid; (BUY OR SELL as) high
FILTER low[price < 100] AND mid[price >= 100] 
   AND mid[price <= 2000] AND high[price > 2000] 
PARTITION BY [name]
\end{verbatim}
   \\
\multicolumn{1}{c}{$(Q_1)$} & \multicolumn{1}{c}{\newtext{$(Q_2)$}} & \multicolumn{1}{c}{\newtext{$(Q_3)$}}
  \end{tabular}
  \caption{CEQL queries on a Stock stream.}
  \label{fig:ceql-queries}
\end{figure*}


%% file: sections/related-work.tex
CER systems are usually divided into three approaches:
automata-based, tree-based, and logic-based, with some systems (e.g.,
\cite{EsperTech, TESLA,DBLP:conf/icde/ZhaoHW20}) being hybrids. We refer to recent surveys~\cite{cugola2012, DBLP:journals/vldb/GiatrakosAADG20,DBLP:journals/csur/AlevizosSAP17,DEBStutorial} of the field for in-depth discussion of these classes of systems.
\systemname\ falls within the class of automata-based systems~\cite{cayuga,cayuga2,SASEcomplexity,SASEautomata,SASE,nextCEP,distCED,TESLA,DBLP:conf/sigmod/0019ANNW21,DBLP:conf/sigmod/PoppeLR019,PoppeLAR17,PoppeLRM17,DBLP:conf/debs/KolchinskySS15}. These systems use automata as their underlying execution model. As already mentioned, these systems either materialize a super-linear number of partial matches, or recompute them on the fly. Conceptually, almost all of these works propose a method to reduce the materialization/recomputation cost by representing partial matches in a more compact manner. \systemname\ is the first system to propose a representation of partial matches with formal, proven, and optimal performance guarantees: linear in the number of seen events, with constant update cost, and output-linear enumeration delay. 

\newtext{
Tree- and logic-based systems~\cite{mei2009zstream,liu2011cube,EsperTech,anicic2010rule,artikis2015event,chesani2010logic,DBLP:journals/pvldb/KolchinskyS18a} typically evaluate queries by constructing and evaluating  a tree of CER operators, much like relational database systems evaluate relational algebra queries. Cost models can be used to identify efficient trees, and stream characteristics monitored  to re-optimize trees during processing when necessary~\cite{DBLP:journals/pvldb/KolchinskyS18a}.
These evaluation trees do no have the formal, optimal performance guarantees offered by \systemname.}

Query evaluation with bounded delay has been extensively studied for relational database queries~\cite{Segoufin13,Bagan:2007,DBLP:journals/vldb/IdrisUVVL20,DBLP:conf/pods/BerkholzKS17,DBLP:conf/pods/BerkholzKS17,DBLP:journals/tods/CarmeliK21}, where it forms an attractive evaluation method, especially when query output risks being much bigger than the size of the input. \systemname\ applies this methodology to the CER domain which differs from the relational setting in the choice of query operators, in particular the presence of operators like sequencing and Kleene star (iteration). In this respect, \systemname's evaluation algorithm is closer the work on query evaluation with bounded delay over words and trees~\cite{DBLP:journals/tods/FlorenzanoRUVV20,DBLP:journals/tods/AmarilliBMN21,AmarilliBJM17,AmarilliBMN19}. Those works, however, do not consider enumeration with time window constraints as we do here.


%% file: sections/ceql-examples.tex
\newtext{\systemname's query language is similar in spirit to existing languages for
expressing CER queries (see, e.g.,
\cite{cugola2012,DBLP:journals/vldb/GiatrakosAADG20,DBLP:journals/csur/AlevizosSAP17}
for a language survey). In particular, \systemname\ shares with these languages
a common set of operators for expressing CER patterns, including, for example
sequencing, disjunction, iteration (a.k.a. Kleene closure), and filtering, among
others~\cite{DBLP:journals/vldb/GiatrakosAADG20,DBLP:journals/csur/AlevizosSAP17}. While
hence conceptually similar, it is important to note that existing system
implementations and their evaluation algorithms differ in (1) the exact set of
operators supported, (2) how these operators can be composed, and (3) even in
the semantics ascribed to operators. 
Given these sometimes subtle (semantic)
differences, we wish to be unambiguous about the
class of queries supported by \systemname's evaluation algorithm. In this
section, we therefore formally define the syntax and semantics of \systemql, the query language that is fully implemented by 
\systemname.}

\newtext{\systemql\xspace is based on \emph{Complex Event Logic} (\cel for short)---a
formal logic that is built from the above-mentioned common operators without
restrictions on composability, and whose expressiveness and complexity have been
studied in \cite{ICDT2019,DBLP:conf/icdt/GrezRUV20, CELJOURNAL}.
\systemql\xspace extends \cel by adding support for time windows as well as the
partition-by event correlation operator. We first introduce
\systemql\xspace by means of  examples, and then proceed with the formal syntax and semantics.}

\subsection{\systemql\xspace by example}

Consider that we have a stream \texttt{Stock} that is emitting \texttt{BUY} and \texttt{SELL} events of particular stocks. The events carry the stock name, the volume bought or sold, the price, and a timestamp.  
Suppose
that we are interested in all triples of \textsf{SELL} events where the first is a
sale of Microsoft over 100 USD, the second is a sale of Intel (of any price), and the third is a sale of Amazon below 2000
USD. Query $Q_1$ in Figure~\ref{fig:ceql-queries} expresses this in
\systemql. In $Q_1$, the \textsf{FROM}
clause indicates the streams to read events from, while the \textsf{WHERE} clause indicates
the pattern of atomic events that need to be matched in the stream. This can be
any unary Complex Event Logic (CEL) expression~\cite{CELJOURNAL}. In $Q_1$, the CEL
expression
\begin{center}
	\texttt{SELL as msft; SELL as intel; SELL as amzn} 
\end{center}
indicates that we wish to see three 
\textsf{SELL} events and that we will
refer to the first, second and third events by means of the variables \textsf{msft}, \textsf{intel} and \textsf{amzn}, respectively. In particular, the semicolon
operator (\textsf{;}) indicates sequencing among events. Sequencing in CORE is non-contiguous. As such, the \textsf{msft} event needs
not be followed immediately by the \textsf{intel} event---there may be other
events in between, and similarly for \textsf{amzn}. The \textsf{FILTER} clause requires the \textsf{msft} event
to have \textsf{MSFT} in its \textsf{name} attribute, and a price above 100. It
makes similar requirements on the \textsf{intel} and \textsf{amzn} events.

The conditions in a \systemql\xspace \textsf{FILTER} clause can only express predicates on single events. Correlation among events, in the form of equi-joins,  is supported in CEQL by the \textsf{PARTITION BY} clause. \newtext{This feature is illustrated by query $Q_2$ in Figure~\ref{fig:ceql-queries}, which detects all pairs of \textsf{BUY} or \textsf{SELL} events of the same stock and the same volume.} In particular, there, the \textsf{PARTITION BY} clause requests that all matched events have the same values in the \textsf{name} and \textsf{volume} attributes. The \textsf{WITHIN} clause specifies that the matched pattern must be detected within 1 minute. In \systemname, each event is assigned the time at which it arrives to the system, so we do not assume that events include a special attribute representing time, as some other systems do. Finally, the \textsf{SELECT} clause ensures that, from the matched pair of events, only the event in variable \textsf{b} is returned.

In general, the pattern specified in the \textsf{WHERE} clause in a \systemql~query may include other operators such as disjunction (denoted $\cor$\!) and iteration (also known as Kleene closure, denoted \textsf{+}). These may be freely nested in the \textsf{WHERE} clause. Query $Q_2$ illustrates the use of disjunction. Query $Q_3$ illustrates the use of iteration. \newtext{In $Q_3$, \textsf{100} and \textsf{2000} are two values representing a lower and upper limit price, respectively. $Q_3$ looks for an upward trend: a sequence of \textsf{BUY} or \textsf{SELL} events pertaining to the same stock symbol where the sale price is initially below \textsf{100} (captured by the \textsf{low} variable), then between \textsf{100} and \textsf{2000} (captured by \textsf{mid}), then above \textsf{2000} (\textsf{high}). Importantly, because of the Kleene closure iteration  operator, variable \textsf{mid} captures all sales of the stock in the $[\textsf{100},\textsf{2000}]$ price range in such a trend. 
The \textsf{MAX} operator in the \textsf{SELECT} clause is an example of a selection strategy~\cite{SASEautomata,DBLP:journals/vldb/GiatrakosAADG20,CELJOURNAL}: it ensures that within a trend \textsf{mid} is bound to a maximal sequence of events in the $[\textsf{100},\textsf{2000}]$ price range. 
If this policy were not specified, \systemql\xspace would adopt the \textsf{skip-till-any-match} policy~\cite{SASEautomata,DBLP:journals/vldb/GiatrakosAADG20} by default, which also returns complex events with \textsf{mid}  containing only subsets of this maximal sequences.}


%% file: sections/ceql-semantics.tex

\subsection{\systemql\xspace syntax and semantics}

We start by defining \systemname's event model.

\paragraph*{Events, complex events, and valuations.} We assume given a set of \emph{event types} $\tset$ (consisting, e.g., of the event types \textsf{BUY} and \textsf{SELL} in our running example), a set of \emph{attribute names} $\aset$ (e.g., \textsf{name}, \textsf{price}, etc) and a set of \emph{data values}  $\dset$  (e.g. integers, strings, etc.). A \emph{data-tuple} $t$ is a partial mapping that maps attribute names from~$\aset$ to data values in~$\dset$.  Each data-tuple is associated to an event type. We denote by $t(a) \in \dset$ the value of the attribute $a \in \aset$ assigned by $t$, and by $t(\type) \in \tset$ the event type of $t$. If $t$ is not defined on attribute $a$, then we write $t(a) = \nullvalue$. 

A \emph{stream} is a possibly infinite sequence 
$S = t_0 t_1 t_2 \ldots$ of data-tuples.  Given a set $D \subseteq \bbN$, we define the set of data
tuples $S[D] = \{t_i \mid i \in D\}$.
%
A \emph{complex event}  is a pair $C = ([i,j], D)$ where $i \leq j \in \bbN$ and $D$ is a subset of $\{i,\ldots,j\}$.
Intuitively, given a stream $S = t_0 t_1 \ldots$ the interval $[i, j]$ of $C$ represents the subsequence $t_i t_{i+1} \ldots t_j$ of $S$ where the complex event $C$ happens and $S[D]$ represents the data-tuples from $S$ that are relevant for $C$. We write $C(\ctime)$ to denote the time-interval $[i,j]$, and $C(\cstart)$ and $C(\cend)$ for $i$ and $j$, respectively. Furthermore, we write $C(\cdata)$ to denote the set~$D$.

To define the semantics of \systemql, we will also need the following
notion. Let $\xset$ be a set of \emph{variables}, which includes all event
types, $\tset \subseteq \xset$. A \emph{valuation} is a pair $V = ([i,j], \mu)$ with $[i,j]$ a time
interval as above and $\mu$ a mapping that assigns subsets of $\{i,\ldots,j\}$
to variables in $\xset$. Similar to complex events, we write $V(\ctime)$,
$V(\cstart)$, and $V(\cend)$ for $[i, j]$, $i$, and $j$, respectively, and
$V(X)$ for the subset of $\{i,\ldots,j\}$ assigned to $X \in \xset$ by $\mu$. 

We write $\cnormal{V}$ for the complex
event that is obtained from valuation $V$ by forgetting the variables in $V$, and retaining only its positions: $\cnormal{V}(\ctime) = V(\ctime)$ and
$\cnormal{V}(\cdata) = \bigcup_{X \in \xset} V(X)$. The semantics of \systemql{} will be defined in terms of valuations, which are subsequently transformed into complex events in this manner.

\paragraph*{Predicates} A (unary) \emph{predicate} is a possibly infinite set $P$ of data-tuples. For example, $P$ could be the set of all tuples $t$ such that $t(\mathsf{price}) \geq 100$. A data-tuple $t$ \emph{satisfies} predicate $P$, denoted $t \models P$, if, and only if, $t \in P$. We generalize this definition from data-tuples to sets by taking a ``for all'' extension: a set of data-tuples $T$ satisfies $P$, denoted by $T \models P$, if, and only if, $t \models P$ for all $t \in T$. 

\paragraph*{\systemql.} Syntactically, a
\systemql{} query has the form:
\begin{center}
  \small
	\begin{tabular}{ll}
		\textsf{SELECT} & [$\text{selection-strategy}$] $<\text{list-of-variables}>$ \\
		\textsf{FROM} &  $<\text{list-of-streams}>$ \\
		\textsf{WHERE} & $<\text{CEL-formula}>$ \\
		\textsf{[PARTITION BY} & $<\text{list-of-attributes}>$\textsf{]} \\
		\textsf{[WITHIN} & $<\text{time-value}>$\textsf{]} \\
	\end{tabular}
      \end{center}
      Specifically, the \textsf{WHERE} clause consists of a formula in Complex Event Logic (\cel)~\cite{CELJOURNAL}, whose abstract syntax is given by the following grammar:
      \[
	\varphi  :=  R \mid \varphi \IN X  \mid  \varphi \FILTER X\texttt{[}P\texttt{]} \mid   \varphi \cor \varphi  \mid  \varphi \sq \varphi  \mid  \varphi \ks  \mid \PROJ{L}(\varphi).
      \]
      In this grammar, $R$ is a event type in $\tset$, $X$ is a variable in $\xset$, $P$ is a  predicate, and $L$ is a subset of variables in $\xset$.\footnote{Observe that \cel includes \textsf{FILTER}, there is hence no separate \textsf{FILTER} clause in \systemql. For convenience, in CEQL queries we use $\varphi \FILTER \theta_1 \textsf{AND } \theta_2$ in the \textsf{WHERE} clause as a shorthand for $(\varphi \FILTER \theta_1) \FILTER \theta_2$, and $\varphi \FILTER \theta_1 \cor \theta_2$ as shorthand for $(\varphi \FILTER \theta_1) \cor (\varphi \FILTER \theta_2)$.}

      The semantics of \systemql\xspace is now as follows. Conceptually, a \systemql{} query first evaluates its \textsf{FROM} clause, then its \textsf{PARTITION BY} clause, and subsequently its  \textsf{WHERE},  \textsf{SELECT}, and \textsf{WITHIN} clauses (in that order). The \textsf{FROM} clause merely specifies the list of streams registered to the system from which events should be inspected. All these streams are logically merged into a single stream $S$ that is processed by the subsequent clauses. The \textsf{PARTITION BY} clause, if present,  logically partitions this stream into multiple substreams $S_1, S_2, \dots$, and executes the \textsf{WHERE-SELECT-WITHIN} clauses on each substream separately. The union of the outputs generated for each substream constitute the final output. Concretely, every $S_i$ is a maximal subsequence of $S$ such that for every pair of tuples  $t$ and $t'$ occurring in $S_i$, and for every attribute $a$ mentioned in the \textsf{PARTITION BY} clause, it holds that $t(a) \not = \nullvalue$, $t'(a) \not = \nullvalue$, and $t(a) = t'(a)$. As such, all tuples in $S_i$ share the same value in every attribute of the \textsf{PARTITION BY} clause. 	 

The semantics of the \textsf{WHERE-SELECT-WITHIN} clauses is as follows. 
\systemql's \textsf{WHERE} clause is derived from the semantics of \cel\footnote{Note that the semantics used in this paper is an extension of the semantics of \cel in~\cite{CELJOURNAL} since we also consider the time interval as part of the complex event.}, which is inductively defined in Table~\ref{tab-semantics}. Concretely, given a stream $S = t_0t_1t_2\dots$ (or one of the substreams $S_i$ if the query has a \textsf{PARTITION BY} clause), a \cel formula $\varphi$ evaluates to a set of valuations, denoted $\semaux{\varphi}(S)$.
The base case is when $\varphi$ is an event type $R$. In that case $\semaux{\varphi}(S)$ contains all valuations whose time-interval is a single  position $i$, such that the data-tuple $t_i$ at position $i$ in $S$ is of type $R$. Furthermore, the valuation is such that variable $R$ (recall that $\tset \subseteq \xset$) stores only position $i$ and all other variables are empty. The $\sIN$ clause is a variable assignment that takes an existing valuation $V \in \semaux{\varphi}(S)$ and extends it by gathering all positions $\cup_{Y} V(Y)$ in variable $X$, keeping all other variables as in $V$. The filter clause $\tFILTER~X\texttt{[}P\texttt{]}$ retains only those valuations for which the content of variable $X$ satisfies predicate $P$, and the $\tcor$ clause takes the union of two sets of valuations. The sequencing operator uses the time-interval for capturing all pairs of valuations in which the first is chronologically followed by the second. Specifically, $\semaux{\varphi_1 \sq \varphi_2}(S)$ takes $V_1 \in \semaux{\varphi_1}(S)$ and $V_2 \in \semaux{\varphi_2}(S)$ such that $V_2$ is after $V_1$ (i.e., $V_1(\cend) < V_2(\cstart)$) and joins them into one valuation $V$, where the time interval is given by the start of $V_1$ and the end of $V_2$.
The semantics of iteration $\varphi\ks$ is defined as the application of sequencing ($\tsq$) one or more times over the same formula. The projection $\PROJ{L}$ modifies valuations by  setting all variables that are not in $L$ to empty.

\begin{figure}[t]
  \small
	\centering
	\begin{align*}
		\semaux{R}(S) & \ = \ \{ V \ \mid \!\!
		\begin{array}[t]{l}
			V(\ctime) = [i,i] \ \wedge \ t_i(\type) = R  \\
			\wedge \ \ V(R) = \{i\} \ \wedge \  \forall X \neq R. \ V(X) = \emptyset \ \} 
		\end{array}  \\ 
          \semaux{\varphi \IN X}(S) & \  = \  \{ V \ \mid \!\!
		\begin{array}[t]{l}
			\exists \ V' \in \semaux{\varphi}(S). \ \ V(\ctime) = V'(\ctime) \\ \wedge \ \ V(X) = \cup_{Y} V'(Y)   \\
			\wedge \ \ \forall Z \neq X. \  V(Z) = V'(Z) \ \} 
		\end{array} \\
		\semaux{\varphi \FILTER X\texttt{[}P\texttt{]}}(S) & \ = \ \{ V \ \mid \ V \in \semaux{\varphi}(S) \wedge V(X) \models P \ \} \\
		\semaux{\varphi_1 \cor \varphi_2}(S) & \ = \ \semaux{\varphi_1}(S) \ \cup \  \semaux{\varphi_2}(S) \\
		\semaux{\varphi_1 \sq \varphi_2}(S) & \ = \ \{ V \ \mid \!\!
		\begin{array}[t]{l}
			\exists \ V_1 \in \semaux{\varphi_1}(S), V_2 \in \semaux{\varphi_2}(S). \\
			V_1(\cend) < V_2(\cstart)  \\ 
			\wedge \ \ V(\ctime) = [V_1(\cstart), V_2(\cend)]  \\
			\wedge \ \ \forall X. \  V(X) = V_1(X) \cup V_2(X) \ \}
		\end{array}  \\
		\semaux{\varphi\ks}(S) & \ = \ \semaux{\varphi}(S) \ \cup \ \semaux{\varphi \sq \varphi \ks}(S) \\
\semaux{\PROJ{L}\!(\varphi)}(S) & \ = \ \{ V \ \mid \!\!
		\begin{array}[t]{l}
			\exists \ V' \in \semaux{\varphi}(S). \ V(\ctime) = V'(\ctime)\\ \wedge
\ \forall X \in L.\ V(X) = V'(X) \\
			\wedge \ \  \forall X \notin L. \ V(X) = \emptyset  \}
		\end{array}
	\end{align*}
	\vspace{-5mm}
	\caption{The semantics of a CEL formulas.
      }
      \label{tab-semantics}
      \vspace{-2mm}
      \end{figure}

The \textsf{WHERE} part of a \systemql{} query hence returns a set of
valuations when evaluated over a stream.  The \textsf{SELECT} clause, if
it does not mention a selection strategy, corresponds to a projection in
\cel, and hence operates on this set accordingly. If it does specify a
selection strategy, then a \cel projection is applied, followed by removing certain valuations from the set.
We refer the interested reader to \cite{CELJOURNAL} for a  definition and discussion of selection strategies. 
Finally, if $\epsilon$ is a time-interval, then  the \textsf{WITHIN} clause operate on the resulting set of valuations as follows:
\begin{align*}
\semaux{\varphi \within \epsilon}(S) & \ = \ \{ V  \in \semaux{\varphi}(S) \mid V(\cend) -V(\cstart)  \leq \epsilon \}. 
\end{align*}


\paragraph*{Complex Event Semantics.} The semantics defined above is one where \cel and \systemql\xspace queries return valuations. In \cer systems, it is customary, however, to return complex events instead. The complex event semantics of \cel and \systemql\xspace is obtained by first evaluating the query under the valuation semantics, and then removing variables altogether. That is, if $\varphi$ is a \cel formula or \systemql\xspace query, its complex event semantics $\semcel{\varphi}(S)$ is defined by
$
\semcel{\varphi}(S) := \{\cnormal{V} \mid V \in \semaux{\varphi}(S) \}.
$
For the rest of this paper, we will be interested in efficiently computing the complex event semantics $\semcel{\varphi}(S)$. We stress, however, that our techniques can be extended to also efficiently compute the valuation semantics 
instead. 

%


%% file: sections/cea.tex

At the heart of evaluating a \systemql{} query lies the problem of evaluating the query's
\textsf{SELECT-WHERE-WITHIN} clauses on either a single stream, or multiple
different substreams thereof (for a query with \textsf{PARTITION BY}). In this section and the next, we discuss how to do so efficiently, focusing on evaluation over a single stream. To this end, let $Q$ be a \systemql\ query without \textsf{PARTITION BY} and with only
one stream mentioned in the \textsf{FROM} clause.


In \systemname{}  we first compile the
\textsf{SELECT-WHERE} part of $Q$ into a \emph{Complex Event
  Automaton (\CEA{} for short)}~\cite{ICDT2019,CELJOURNAL}, which is a form of
finite state automaton that produces complex events. \systemname{}'s evaluation algorithm is then defined in terms of \CEA{}: it takes as input a \CEA{} $\cA$, the (optional) time window $\epsilon$ specified in the \textsf{WITHIN} clause of $Q$, and a stream $S$, and uses this to compute $\sem{Q}(S)$. This evaluation algorithm is
described in Section~\ref{sec:evaluation}. Here, we introduce \CEA.

Roughly speaking, a \CEA{} is similar to  a standard finite state automaton. 
The difference is that  a standard finite state automaton 
processes finite strings and also has transitions
of the form $p \trans{\sigma} q$ with $q,p$ states and $\sigma$ a symbol from
some finite alphabet, whereas a \CEA{} processes possibly unbounded streams of data-tuples and has
transitions of the form $p \trans{P / m} q$ with $p$ and $q$ states, $P$ a
predicate and $m$ an action, which can be \emph{marking} ($\amark$) or
\emph{unmarking}~($\umark$). The semantics of such a transition
$p \trans{P/m} q$ is that, when a new tuple $t$ arrives in the stream and the
\CEA{} is in state $p$, if $t$ satisfies $P$ then the \CEA{} moves to state $q$
and applies the action $m$: if $m$ is a marking action then the event $t$ will
be part of the output complex event once a final state is reached, otherwise it
will not.

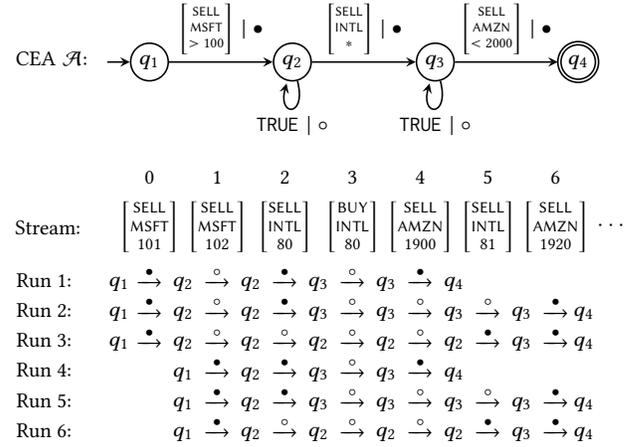
\begin{figure}[tbp]
  \begin{center}
    \small
	\begin{tikzpicture}[->,>=stealth, semithick, auto, initial text= {}, initial distance= {3mm}, accepting distance= {4mm}, node distance=1.9cm, semithick]
	\tikzstyle{every state}=[draw=black,text=black,inner sep=0pt, minimum size=5mm]
        
        \begin{scope}
	\node[initial,state]	(1)			{$q_1$};
	\node[state]		(2) [right of=1]	{$q_2$};
	\node[state]		(3) [right of=2]	{$q_3$};
	\node[accepting,state]	(4) [right of=3]	{$q_4$};

	\path
	(1) edge node {${\tiny \ssd{SELL}{MSFT}{>100}} \mid \amark$} (2)
	(2) edge node {${\tiny \ssd{SELL}{INTL}{*}} \mid \amark $} (3)
   		    edge [loop below] node {$\TRUE \mid \umark$} (2)
        (3) edge node {${\tiny \ssd{SELL}{AMZN}{<2000}} \mid \amark$} (4)     	 edge [loop below] node {$\TRUE \mid \umark$} (3);

	\node at ($(1) + (-1.3,0.05)$) {CEA $\cA$:};

      \end{scope}

      \begin{scope}[yshift=-2.2cm, node distance=0.9cm]
            	\node[label=0] (1)             {$\scriptsize \ssd{SELL}{MSFT}{101}$};
	\node[label=1,right of=1] (2) {$\scriptsize \ssd{SELL}{MSFT}{102}$};
	\node[label=2,right of=2] (3) {$\scriptsize \ssd{SELL}{INTL}{80}$};
	\node[label=3,right of=3] (4) {$\scriptsize \ssd{BUY}{INTL}{80}$};
	\node[label=4,right of=4] (5) {$\scriptsize \ssd{SELL}{AMZN}{1900}$};
	\node[label=5,right of=5] (6) {$\scriptsize \ssd{SELL}{INTL}{81}$};
	\node[label=6,right of=6] (7) {$\scriptsize \ssd{SELL}{AMZN}{1920}$};
        \node[right of=7, node distance=0.75cm] (8) {$\scriptsize \cdots$};
	\node at ($(1) + (-1.35, 0)$) {Stream:};

	\node[below of=1, node distance=0.6cm] (t1) {$\strans{\amark}$};
	\node[below of=2, node distance=0.6cm] (t2) {$\strans{\umark}$};
	\node[below of=3, node distance=0.6cm] (t3) {$\strans{\amark}$};
	\node[below of=4, node distance=0.6cm] (t4) {$\strans{\umark}$};
	\node[below of=5, node distance=0.6cm] (t5) {$\strans{\amark}$};
	\node[below of=6, node distance=0.6cm,white] (t6) {$\strans{\amark}$};
	\node[below of=7, node distance=0.6cm,white] (t7) {$\strans{\umark}$};
	\node[below of=8, node distance=0.6cm,white] (t8) {$\strans{\amark}$};

	\node at ($(t1) + (0,-0.15) + (-0.4,0)$) (q0) {$q_1$};
	\node at ($0.5*(t2) + 0.5*(t1) + (0,-0.15)$) (q1) {$q_2$};
	\node at ($0.5*(t3) + 0.5*(t2) + (0,-0.15)$) (q2) {$q_2$};
	\node at ($0.5*(t4) + 0.5*(t3) + (0,-0.15)$) (q3) {$q_3$};
	\node at ($0.5*(t5) + 0.5*(t4) + (0,-0.15)$) (q4) {$q_3$};
	\node at ($0.5*(t6) + 0.5*(t5) + (0,-0.15)$) (q5) {$q_4$};
	\node[white] at ($0.5*(t7) + 0.5*(t6) + (0,-0.15)$) (q6) {$q_2$};
	\node[white] at ($0.5*(t8) + 0.5*(t7) + (0,-0.15)$) (q7) {$q_3$};
	\node at ($(q0) + (-1,0.05)$) {Run 1:};

	\node[below of=t1, node distance=0.4cm] (t1) {$\strans{\amark}$};
	\node[below of=t2, node distance=0.4cm] (t2) {$\strans{\umark}$};
	\node[below of=t3, node distance=0.4cm] (t3) {$\strans{\amark}$};
	\node[below of=t4, node distance=0.4cm] (t4) {$\strans{\umark}$};
	\node[below of=t5, node distance=0.4cm] (t5) {$\strans{\umark}$};
	\node[below of=t6, node distance=0.4cm] (t6) {$\strans{\umark}$};
	\node[below of=t7, node distance=0.4cm] (t7) {$\strans{\amark}$};
	\node[below of=t8, node distance=0.4cm,white] (t8) {$\strans{\amark}$};

	\node at ($(t1) + (0,-0.15) + (-0.4,0)$) (q0) {$q_1$};
	\node at ($0.5*(t2) + 0.5*(t1) + (0,-0.15)$) (q1) {$q_2$};
	\node at ($0.5*(t3) + 0.5*(t2) + (0,-0.15)$) (q2) {$q_2$};
	\node at ($0.5*(t4) + 0.5*(t3) + (0,-0.15)$) (q3) {$q_3$};
	\node at ($0.5*(t5) + 0.5*(t4) + (0,-0.15)$) (q4) {$q_3$};
	\node at ($0.5*(t6) + 0.5*(t5) + (0,-0.15)$) (q5) {$q_3$};
	\node at ($0.5*(t7) + 0.5*(t6) + (0,-0.15)$) (q6) {$q_3$};
	\node at ($0.5*(t8) + 0.5*(t7) + (0,-0.15)$) (q7) {$q_4$};
	\node at ($(q0) + (-1,0.05)$) {Run 2:};

	\node[below of=t1, node distance=0.4cm] (t1) {$\strans{\amark}$};
	\node[below of=t2, node distance=0.4cm] (t2) {$\strans{\umark}$};
	\node[below of=t3, node distance=0.4cm] (t3) {$\strans{\umark}$};
	\node[below of=t4, node distance=0.4cm] (t4) {$\strans{\umark}$};
	\node[below of=t5, node distance=0.4cm] (t5) {$\strans{\umark}$};
	\node[below of=t6, node distance=0.4cm] (t6) {$\strans{\amark}$};
	\node[below of=t7, node distance=0.4cm] (t7) {$\strans{\amark}$};
	\node[below of=t8, node distance=0.4cm,white] (t8) {$\strans{\amark}$};

	\node at ($(t1) + (0,-0.15) + (-0.4,0)$) (q0) {$q_1$};
	\node at ($0.5*(t2) + 0.5*(t1) + (0,-0.15)$) (q1) {$q_2$};
	\node at ($0.5*(t3) + 0.5*(t2) + (0,-0.15)$) (q2) {$q_2$};
	\node at ($0.5*(t4) + 0.5*(t3) + (0,-0.15)$) (q3) {$q_2$};
	\node at ($0.5*(t5) + 0.5*(t4) + (0,-0.15)$) (q4) {$q_2$};
	\node at ($0.5*(t6) + 0.5*(t5) + (0,-0.15)$) (q5) {$q_2$};
	\node at ($0.5*(t7) + 0.5*(t6) + (0,-0.15)$) (q6) {$q_3$};
	\node at ($0.5*(t8) + 0.5*(t7) + (0,-0.15)$) (q7) {$q_4$};
	\node at ($(q0) + (-1,0.05)$) {Run 3:};

	\node[below of=t1, node distance=0.4cm, white] (t1) {$\strans{\umark}$};
	\node[below of=t2, node distance=0.4cm] (t2) {$\strans{\amark}$};
	\node[below of=t3, node distance=0.4cm] (t3) {$\strans{\amark}$};
	\node[below of=t4, node distance=0.4cm] (t4) {$\strans{\umark}$};
	\node[below of=t5, node distance=0.4cm] (t5) {$\strans{\amark}$};
	\node[below of=t6, node distance=0.4cm, white] (t6) {$\strans{\umark}$};
	\node[below of=t7, node distance=0.4cm, white] (t7) {$\strans{\umark}$};
	\node[below of=t8, node distance=0.4cm,white] (t8) {$\strans{\amark}$};

	\node[white] at ($(t1) + (0,-0.15) + (-0.4,0)$) (q0) {$q_1$};
	\node at ($0.5*(t2) + 0.5*(t1) + (0,-0.15)$) (q1) {$q_1$};
	\node at ($0.5*(t3) + 0.5*(t2) + (0,-0.15)$) (q2) {$q_2$};
	\node at ($0.5*(t4) + 0.5*(t3) + (0,-0.15)$) (q3) {$q_3$};
	\node at ($0.5*(t5) + 0.5*(t4) + (0,-0.15)$) (q4) {$q_3$};
	\node at ($0.5*(t6) + 0.5*(t5) + (0,-0.15)$) (q5) {$q_4$};
	\node[white] at ($0.5*(t7) + 0.5*(t6) + (0,-0.15)$) (q6) {$q_3$};
	\node[white] at ($0.5*(t8) + 0.5*(t7) + (0,-0.15)$) (q7) {$q_4$};
	\node at ($(q0) + (-1,0.05)$) {Run 4:};

	\node[below of=t1, node distance=0.4cm, white] (t1) {$\strans{\umark}$};
	\node[below of=t2, node distance=0.4cm] (t2) {$\strans{\amark}$};
	\node[below of=t3, node distance=0.4cm] (t3) {$\strans{\amark}$};
	\node[below of=t4, node distance=0.4cm] (t4) {$\strans{\umark}$};
	\node[below of=t5, node distance=0.4cm] (t5) {$\strans{\umark}$};
	\node[below of=t6, node distance=0.4cm] (t6) {$\strans{\umark}$};
	\node[below of=t7, node distance=0.4cm] (t7) {$\strans{\amark}$};
	\node[below of=t8, node distance=0.4cm,white] (t8) {$\strans{\amark}$};

	\node[white] at ($(t1) + (0,-0.15) + (-0.4,0)$) (q0) {$q_1$};
	\node at ($0.5*(t2) + 0.5*(t1) + (0,-0.15)$) (q1) {$q_1$};
	\node at ($0.5*(t3) + 0.5*(t2) + (0,-0.15)$) (q2) {$q_2$};
	\node at ($0.5*(t4) + 0.5*(t3) + (0,-0.15)$) (q3) {$q_3$};
	\node at ($0.5*(t5) + 0.5*(t4) + (0,-0.15)$) (q4) {$q_3$};
	\node at ($0.5*(t6) + 0.5*(t5) + (0,-0.15)$) (q5) {$q_3$};
	\node at ($0.5*(t7) + 0.5*(t6) + (0,-0.15)$) (q6) {$q_3$};
	\node at ($0.5*(t8) + 0.5*(t7) + (0,-0.15)$) (q7) {$q_4$};
	\node at ($(q0) + (-1,0.05)$) {Run 5:};

	\node[below of=t1, node distance=0.4cm, white] (t1) {$\strans{\umark}$};
	\node[below of=t2, node distance=0.4cm] (t2) {$\strans{\amark}$};
	\node[below of=t3, node distance=0.4cm] (t3) {$\strans{\umark}$};
	\node[below of=t4, node distance=0.4cm] (t4) {$\strans{\umark}$};
	\node[below of=t5, node distance=0.4cm] (t5) {$\strans{\umark}$};
	\node[below of=t6, node distance=0.4cm] (t6) {$\strans{\amark}$};
	\node[below of=t7, node distance=0.4cm] (t7) {$\strans{\amark}$};
	\node[below of=t8, node distance=0.4cm,white] (t8) {$\strans{\amark}$};

	\node[white] at ($(t1) + (0,-0.15) + (-0.4,0)$) (q0) {$q_1$};
	\node at ($0.5*(t2) + 0.5*(t1) + (0,-0.15)$) (q1) {$q_1$};
	\node at ($0.5*(t3) + 0.5*(t2) + (0,-0.15)$) (q2) {$q_2$};
	\node at ($0.5*(t4) + 0.5*(t3) + (0,-0.15)$) (q3) {$q_2$};
	\node at ($0.5*(t5) + 0.5*(t4) + (0,-0.15)$) (q4) {$q_2$};
	\node at ($0.5*(t6) + 0.5*(t5) + (0,-0.15)$) (q5) {$q_2$};
	\node at ($0.5*(t7) + 0.5*(t6) + (0,-0.15)$) (q6) {$q_3$};
	\node at ($0.5*(t8) + 0.5*(t7) + (0,-0.15)$) (q7) {$q_4$};
	\node at ($(q0) + (-1,0.05)$) {Run 6:};

      \end{scope}

		\end{tikzpicture}
	\end{center}
  \caption{A CEA representing $Q_1$ from Figure~\ref{fig:ceql-queries} and some of its runs on an example stream.}
  \label{fig:example-cea}
\end{figure}

\begin{example}
  In Figure~\ref{fig:example-cea} we show a \CEA{} $\cA$ that
  represents  query $Q_1$ from Figure~\ref{fig:ceql-queries}. There, we
  depict predicates by listing, in array notation, the event type, the requested value of the
  \textsf{name} attribute, and the constraint on the \textsf{price}
  attribute. The initial state is $q_1$ and there is only one final state:
  $q_4$. The figure also shows an example stream $S$, and several runs of
  $\cA$ on $S$. Every run shown is accepting (i.e., ends in an accepting state
  of the automaton), and as such returns a complex event. The time of this
  complex event is the interval $[i,j]$ with $i$ the position where the run
  starts, and $j$ the position where the run ends. The data of this complex
  event consists of all positions marked by the run. For example, the complex
  event $C_1$ output by run $1$ is $([0,4], \{0,2,4\})$; the complex event $C_2$
  output by run $2$ is $([0,6], \{0,2,6\})$, and so on.
\end{example}

Formally, a \emph{Complex Event Automaton (\CEA)} is a tuple
$\cA = (Q, \Delta, q_0, F)$ where $Q$ is a finite set of states, $\Delta \subseteq Q\times \pset \times \{\amark, \umark\} \times (Q \setminus \{q_0\})$ is a finite transition relation, $q_0 \in Q$ is the initial state, and $F \subseteq Q$ is the set of final states. We will denote transitions in $\Delta$ by $q \trans{P/m} q'$. A \emph{run of $\cA$ over stream $S$ from positions $i$ to $j$}
is a sequence
$
\rho \ := \  q_{i} \, \trans{P_{i}/m_i} \, q_{i+1} \, \trans{P_{i+1}/m_{i+1}} \, \ldots \, \trans{P_j/m_j} \, q_{j+1}
$
such that $q_i$ is the initial state of $\cA$ and for every $k \in [i,j]$  it holds that $q_{k} \trans{P_{k/m_k}} q_{k+1} \in \Delta$ and $t_k \models P_k$. A run $\rho$  is \emph{accepting} if $q_{j+1} \in F$.
An accepting run $\rho$ of $\cA$ over $S$ from $i$ to $j$ naturally defines the complex event
$
C_{\rho} \ := \ ([i,j], \{k \mid i \leq k \leq j \wedge m_k = \amark\}).
$
If position $i$ and $j$ are clear from the context, we say that $\rho$ is a run of $\cA$ over $S$. 
Finally, we define the semantics of $\cA$ over a stream $S$ as 
$
\sem{\cA}(S) := \{C_\rho \mid \text{$\rho$ is an accepting run of $\cA$ over $S$} \}.
$

We note that in this paper, because we consider complex events with time
intervals, a run may start at an arbitrary position $i$ in the stream, which
differs from the semantics of \CEA{} considered in~\cite{ICDT2019,CELJOURNAL} where complex events do not have time
intervals and runs always start at the beginning of the stream. It is also
important to note that in the definition above no transition can re-enter the
initial state $q_0$; this will be important for defining the time-interval of
the output complex events in Section~\ref{sec:evaluation}. This requirement on
the initial state is without loss of generality, since any incoming transitions into the initial state $q_0$ may be removed without modifying semantics by  making a copy $q'_0 $ of $q_0$ (also copying its outgoing
transitions) and rewrite any transition into $q_0$ to go to $q'_0$ instead.

The usefulness of \CEA{} comes from the fact that CEL can be translated into CEA~\cite{ICDT2019,CELJOURNAL}. Because the \textsf{SELECT-WHERE} part of a CEQL query is in essence a CEL formula, this reduces the evaluation problem of the \textsf{SELECT-WHERE-WITHIN} part of CEQL query into the evaluation problem for CEA{}, in the following sense.\footnote{We remark that any selection policy mentioned in the \textsf{SELECT} clause can also be expressed using CEA, see ~\cite{ICDT2019,CELJOURNAL}.}
\begin{restatable}{theorem}{theoceltocea}\label{theo:CEL-to-CEA}
	For every CEL formula $\varphi$ we can construct a \CEA{} $\cA$ of size linear in $\varphi$ such that for every~$\epsilon$:
	$$
	\sem{\varphi \within \epsilon}(S) \ = \ \{C \mid C \in \sem{\cA}(S) \wedge C(\cend) - C(\cstart) \leq \epsilon\}.
	$$
\end{restatable}
Our evaluation algorithm will compute the right-hand side in this equation. It requires, however, that the input \CEA $\cA$ is \emph{I/O-deterministic}: for every pair of transitions $q \trans{P_1/m_1} q_1$ and $q \trans{P_2/m_2} q_2$ from the same state $q$, if $P_1 \cap P_2 \neq \emptyset$ then $m_1 \not = m_2$. In other words,  an event $t$ may trigger both transitions at the same time  (i.e., $t \models P_1$ and $t \models P_2$) only if one transition marks the event, but the other does not. In~\cite{ICDT2019,CELJOURNAL}, it was shown that any CEA can be
I/O-determinized. The determinization method we use is based on the classical
subset construction of finite state automata, thus possibly adding an
exponential blow-up in the number of states. To avoid this exponential blow-up
in practice, in CORE we determinize the \CEA{} not all at once, but on the
fly while the stream is being processed. Importantly, we cache the previous
states that we have computed. In Section~\ref{sec:evaluation:implementation} we discuss the
internal implementation of CORE and how this exponential factor impacts system performance.


%% file: sections/algorithm.tex

In this section, we present an efficient evaluation algorithm that, given a \CEA{} $\cA$, time window $\epsilon$, and stream $S$, computes the set
\[ 
  \sem{\cA}^\epsilon(S) \ := \ \{C \mid C \in \sem{\cA}(S) \wedge C(\cend) -
  C(\cstart) \leq \epsilon\}. \] In fact, our algorithm will compute this set
\emph{incrementally}: at  every position $j$ in the stream, it outputs the set
\[\sem{\cA}^\epsilon_j(S) \ := \ \{C \in \sem{\cA}^\epsilon \mid C(\cend) = j\}.\]
The algorithm works by incrementally maintaining a data structure that
\emph{compactly} represents partial outputs (i.e., fragments of $S$ that later
may cause a complex event to be output). Whenever a new tuple arrives, it takes
\emph{constant time} (in data complexity~\cite{DBLP:conf/stoc/Vardi82}) to
update the data structure. Furthermore, from the data structure, we may at each
position $j$ enumerate the complex events of $\sem{\cA}^\epsilon_j(S)$ one by
one, without duplicates, and with \emph{output-linear delay}~\cite{Segoufin13,CELJOURNAL}. This means
that the time required to print the first complex event of
$\sem{\cA}^\epsilon_j(S)$ from the data structure, or any of the following ones,
is linear in the size of complex event being printed. 
Note in particular
that the data complexity of our algorithm is asymptotically optimal: any
evaluation algorithm needs to at least inspect every input tuple and list the
query answers. Also note that, because it takes constant time to update the data structure with a new input event, the size of our data structure is at most linear in the number of seen events.

We first define the data structure in
Section~\ref{sec:evaluation:data-structure}, and operations on it in
Section~\ref{sec:evaluation-structure}. The evaluation algorithm is given in
Section~\ref{sec:evaluation:algo} and aspects of its implementation in Section~\ref{sec:evaluation:implementation}.

\subsection{The data structure}
\label{sec:evaluation:data-structure}

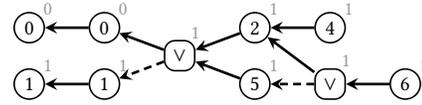
\begin{figure}
  \small
		\begin{tikzpicture}[->,>=stealth, semithick, auto, initial text= {}, initial distance= {3mm}, accepting distance= {4mm}, node distance=3cm, semithick]

                  \tikzstyle{bnode}=[draw=black,text=black,inner sep=0pt, minimum size=4mm, circle]
                  \tikzstyle{onode}=[draw=black,text=black,inner sep=0pt, minimum size=4mm, circle]
                  \tikzstyle{unode}=[draw=black, text=black,inner sep=0pt, minimum size=4mm,rounded corners]
                  \tikzstyle{mnode}=[text=black!40]

		\tikzstyle{onext}=[->, ultra thick, >=stealth, line width=1pt]

		\tikzstyle{ulist}=[->, ultra thick, >=stealth, line width=1pt, blue]

                \tikzstyle{uleft}=[->, densely dashed, ultra thick, >=stealth, line width=1pt]
		
                \tikzstyle{uright}=[->, ultra thick, >=stealth, line width=1pt]

                \tikzstyle{qnode}=[text=blue,color=blue, inner sep=0pt]

                \begin{scope}[rotate=-90]
                  \node[bnode] (bot0) at (0,0) {$0$};
                  \node[onode] (pos0) at (0,1) {$0$};
                  \draw[onext] (pos0) to (bot0);

                  \node[bnode] (bot1) at (0.75,0) {$1$};
                  \node[onode] (pos1) at (0.75,1) {$1$};
                  \draw[onext] (pos1) to (bot1);

                  \node[unode] (u01) at (0.3725,2) {$\vee$};
                  \node[onode] (pos2) at (0.0,3) {$2$};
                  \draw[onext] (pos2) to (u01);
                  \draw[uleft] (u01) to (pos1);
                  \draw[uright] (u01) to (pos0);

                  \node[onode] (pos4) at (0.0,4) {$4$};                  
                  \draw[onext] (pos4) to (pos2);

                  \node[onode] (pos5) at (0.75,3) {$5$};                  
                  \draw[onext] (pos5) to (u01);

                  \node[unode] (u25)  at (0.75,4) {$\vee$};
                  \draw[uleft] (u25) to (pos5);
                  \draw[uright] (u25) to (pos2);

                  \node[onode] (pos6) at (0.75, 5) {$6$};
                  \draw[onext] (pos6) to (u25);
                  
                  \node[mnode] at ($(bot0.north east)+(-0.1,0.1)$) {\footnotesize{$0$}};
                  \node[mnode] at ($(pos0.north east)+(-0.1,0.1)$) {\footnotesize{$0$}};
                  \node[mnode] at ($(bot1.north east)+(-0.1,0.1)$) {\footnotesize{$1$}};
                  \node[mnode] at ($(pos1.north east)+(-0.1,0.1)$) {\footnotesize{$1$}};
                  \node[mnode] at ($(u01.north east)+(-0.1,0)$) {\footnotesize{$1$}};
                  \node[mnode] at ($(pos2.north east)+(-0.1,0.1)$) {\footnotesize{$1$}};
                  \node[mnode] at ($(pos4.north east)+(-0.1,0.1)$) {\footnotesize{$1$}};
                  \node[mnode] at ($(pos5.north east)+(-0.1,0.1)$) {\footnotesize{$1$}};
                  \node[mnode] at ($(u25.north east)+(-0.1,0)$) {\footnotesize{$1$}};
                  \node[mnode] at ($(pos6.north east)+(-0.1,0.1)$) {\footnotesize{$1$}};

                \end{scope}
		\end{tikzpicture}
  \caption{An example tECS. Union nodes are labeled by $\vee$ while non-union nodes are depicted as circles. Left and right children of union nodes are indicated by dashed and solid edges, respectively. The maximum-start of each node is at its top-right in grey.}
  \label{fig:tecs}
\end{figure}

Our data structure is called a \emph{timed Enumerable Compact Set}
(tECS). Figure~\ref{fig:tecs} gives an example.  Specifically, a tECS is a directed
acyclic graph (DAG) $\ecs$  with two kinds of nodes: union nodes and non-union
nodes.  Every union node
$\unode$ has exactly two children, the left child $\uleft(\unode)$ and the right child
$\uright(\unode)$, which are depicted by dashed and solid edges in Figure~\ref{fig:tecs},
respectively. Every non-union node $\nnode$ is labeled by a stream position (an
element of $\bbN$) and has at most one child. If non-union node $\nnode$ has no child it is called a
\emph{bottom node}, otherwise it is an \emph{output node}. We write $\opos(\nnode)$
for the label of non-union node $\nnode$ and $\onext(\onode)$ for the unique child of output node $\onode$. To simplify presentation in what follows, we will range over nodes of any kind by  $\nnode$; over bottom, output, and union nodes by  $\bnode$, $\onode$, and $\unode$, respectively.

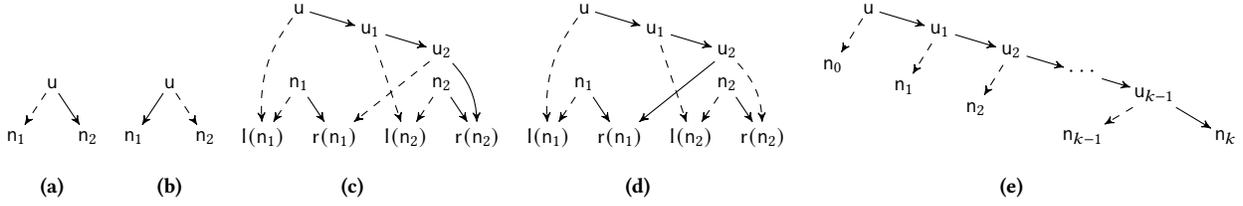
\begin{figure*}[t]
	\centering
        \small
	\begin{varwidth}[t]{0.4\textwidth}
		\begin{tikzpicture}[->,>=stealth',roundnode/.style={circle,inner sep=1pt},squarednode/.style={rectangle,inner sep=2pt}, scale=0.95]
		
		\node at (1, -0.7) {\textbf{(a)}};
		\node[squarednode] (0) at (0.5, 0) {$\nnode_1$};
		\node[squarednode] (1) at (1.5, 0) {$\nnode_2$};
		\node[roundnode] (2) at (1, 0.75) {$\unode$};
		\draw[dashed] (2) edge (0);
		\draw (2) to (1);
		\end{tikzpicture}
	\end{varwidth} \ \ 
	\begin{varwidth}[t]{0.4\textwidth}
		\begin{tikzpicture}[->,>=stealth',roundnode/.style={circle,inner sep=1pt},squarednode/.style={rectangle,inner sep=2pt}, scale=0.95]
		
		\node at (1, -0.7) {\textbf{(b)}};
		\node[squarednode] (0) at (0.5, 0) {$\nnode_1$};
		\node[squarednode] (1) at (1.5, 0) {$\nnode_2$};
		\node[roundnode] (2) at (1, 0.75) {$\unode$};
		\draw (2) edge (0);
		\draw[dashed] (2) to (1);
		\end{tikzpicture}
	\end{varwidth} \ \ 
	\begin{varwidth}[t]{0.4\textwidth}
	\begin{tikzpicture}[->,>=stealth',roundnode/.style={circle,inner sep=1pt},squarednode/.style={rectangle,inner sep=2pt}, scale=0.95]
	
	\node at (1.75, -0.7) {\textbf{(c)}};
	
	\node[squarednode] (0) at (0.5, 0) {$\uleftcomp(\nnode_1)$};
	\node[squarednode] (1) at (1.5, 0) {$\urightcomp(\nnode_1)$};
	\node[roundnode] (2) at (1, 0.75) {$\nnode_1$};
	\node[squarednode] (3) at (2.5, 0) {$\uleftcomp(\nnode_2)$};
	\node[squarednode] (4) at (3.5, 0) {$\urightcomp(\nnode_2)$};
	\node[roundnode] (5) at (3, 0.75) {$\nnode_2$};
	\node[roundnode] (6) at (3, 1.2) {$\unode_2$};
	\node[roundnode] (7) at (2, 1.5) {$\unode_1$};
	\node[roundnode] (8) at (1, 1.8) {$\unode$};
	\draw[dashed] (2) edge (0);
	\draw (2) to (1);
	\draw[dashed] (5) to (3);
	\draw (5) to (4);
	\draw (6) edge[bend left=20] (4);
	\draw[dashed] (6) to (1);
	\draw (7) to (6);
	\draw[dashed] (7) to (3);
	\draw (8) to (7);
	\draw[dashed] (8) edge[bend right=20] (0);
	\end{tikzpicture}
	\end{varwidth} \ \ 
	\begin{varwidth}[t]{0.4\textwidth}
		\begin{tikzpicture}[->,>=stealth',roundnode/.style={circle,inner sep=1pt},squarednode/.style={rectangle,inner sep=2pt}, scale=0.95]
		\node at (1.75, -0.7) {\textbf{(d)}};
		\node[squarednode] (0) at (0.5, 0) {$\uleftcomp(\nnode_1)$};
		\node[squarednode] (1) at (1.5, 0) {$\urightcomp(\nnode_1)$};
		\node[roundnode] (2) at (1, 0.75) {$\nnode_1$};
		\node[squarednode] (3) at (2.5, 0) {$\uleftcomp(\nnode_2)$};
		\node[squarednode] (4) at (3.5, 0) {$\urightcomp(\nnode_2)$};
		\node[roundnode] (5) at (3, 0.75) {$\nnode_2$};
		\node[roundnode] (6) at (3, 1.2) {$\unode_2$};
		\node[roundnode] (7) at (2, 1.5) {$\unode_1$};
		\node[roundnode] (8) at (1, 1.8) {$\unode$};
		\draw[dashed] (2) edge (0);
		\draw (2) to (1);
		\draw[dashed] (5) to (3);
		\draw (5) to (4);
		\draw[dashed] (6) edge[bend left=20] (4);
		\draw (6) to (1);
		\draw (7) to (6);
		\draw[dashed] (7) to (3);
		\draw (8) to (7);
		\draw[dashed] (8) edge[bend right=20] (0);
		\end{tikzpicture}
	\end{varwidth}
	\ \ \ \ 
	\begin{varwidth}[t]{0.4\textwidth}
		\begin{tikzpicture}[->,>=stealth',roundnode/.style={circle,inner sep=1pt},squarednode/.style={rectangle,inner sep=2pt}, scale=0.95]
		\node at (3, -0.7) {\textbf{(e)}};
		\node[roundnode] (0) at (1, 1.8) {$\unode$};
		\node[roundnode] (1) at (2, 1.5) {$\unode_1$};
		\node[roundnode] (2) at (3, 1.2) {$\unode_2$};
		\node[roundnode] (3) at (4, 0.9) {$\cdots$};
		\node[roundnode] (4) at (5, 0.6) {$\unode_{k-1}$};
		\node[roundnode] (5) at (0.5, 1) {$\nnode_0$};
		\node[roundnode] (6) at (1.5, 0.7) {$\nnode_1$};
		\node[roundnode] (7) at (2.5, 0.4) {$\nnode_2$};
		\node[roundnode] (8) at (4, 0) {$\nnode_{k-1}$};
		\node[roundnode] (9) at (6, 0) {$\nnode_{k}$};
		
		\draw (0) to (1);
		\draw (1) to (2);
		\draw (2) to (3);
		\draw (3) to (4);
		
		\draw[dashed] (0) edge (5);
		\draw[dashed] (1) edge (6);
		\draw[dashed] (2) edge (7);
		\draw[dashed] (4) edge (8);
		\draw (4) edge (9);
		\end{tikzpicture}
	\end{varwidth}
	\caption{Gadgets for the implementation of the $\dunion$ method. The $\unode$ nodes are union nodes, where the dashed and bold arrows symbolize the left and right nodes, respectively.}
	\label{fig:union-gadgets}
	\vspace{-3mm}
      \end{figure*}

A tECS  represents  sets of complex events or, more precisely, sets of \emph{open} complex events. An \emph{open complex event} is a pair $(i, D)$ where $i \in \bbN$ and $D$ is a finite subset of $\{i, i+1, \ldots\}$. An open complex event is almost a complex event, with a start time $i$ and set of positions $D$, but where the end  time  is missing: if we choose $j \geq \max(D)$, then $([i, j], D)$ is a complex event. Intuitively, when processing a stream, the  open complex events represented by a tECS are partial results that may later become full complex events.

The representation is as follows. A \emph{full-path} in $\ecs$ is a path 
$\seq{p} = \nnode_1, \nnode_2, \dots, \nnode_k$ such that
$\nnode_k$ is a bottom node. 
Each full-path
$\seq{p}$ represents the open complex event $\outce{\seq{p}} = (i, D)$ where
$i = \opos(n_k)$ is the label of the bottom node $n_k$, and $D$ is the set of
labels of the other non-union nodes in $\seq{p}$. For instance, for the
full-path $\seq{p} = 4, 2, \vee, 1, 1$ in Figure~\ref{fig:tecs}, we have
$\outce{\seq{p}} = (1, \{1,2,4\})$.  Given a node $\nnode$, the
set $\outce{\nnode}$ of open complex events represented by $\nnode$ consists
of all open complex events $\outce{\seq{p}}$ with $\seq{p}$ a full-path in
$\ecs$ starting at $\nnode$.

 \begin{example}
In Figure~\ref{fig:tecs} we have $\outce{4} = \{ (0, \{0,2,4\}), \allowbreak (1, \{1,2,4\})$ and $\outce{6} = \{ (0, \{0,2,6\})$, $(0, \{0,5,6\}), (1, \{1,2,6\})$, $(1,\{1,5,6\})$. 
 \end{example}


Remember that our purpose in constructing $\ecs$ is to be able to enumerate the
set $\sem{\cA}^\epsilon_j(S)$ at every  $j$. To that end, it will be necessary to enumerate, for certain nodes $\nnode$ in $\ecs$, the set
\[
\outce{\nnode}^\epsilon(j) \ := \  \{\, ([i, j], D) \mid (i, D) \in \outce{\nnode} \wedge j-i \leq \epsilon\, \},
\]
i.e., all open complex events represented by $\nnode$ that, when
closed with $j$, are within a time window of size $\epsilon$.

\begin{example}
  The set of all complex events output by the accepting runs of CEA $\cA$ in Figure~\ref{fig:example-cea} can be retrieved from the tECS of Figure~\ref{fig:tecs} by enumerating $\outce{4}^6(4)$ and $\outce{6}^6(6)$,  which contains all complex events output at position $4$ and $6$, respectively.
\end{example}

A straightforward algorithm for enumerating $\outce{\nnode}^\epsilon(j)$ is to
perform a depth-first search (DFS) starting at $\nnode$. During the search we maintain
the full-path $\seq{p}$ from $\nnode$ to the currently visited node
$\textsf{m}$. Every time we reach a bottom node, we check whether $\outce{\seq{p}}$ satisfies the time window, and, if so, output it. There are two problems with this algorithm. First,  it does not satisfy our delay requirements: the DFS may spend unbounded time before reaching a full-path $\seq{p}$ that satisfies the time window and  actually generates output. Second, it may enumerate the same complex event multiple times. This happens if there are multiple full-paths from $n$ that represent the same open complex event. We therefore impose three restrictions on the structure of a tECS.

The first restriction is that $\ecs$ needs to be time-ordered, which is defined
as follows. For a node $\nnode$ define its \emph{maximum-start}, denoted
$\maxstart(\nnode)$, as
$\maxstart(\nnode) = \max\big(\, \{i \mid (i, D) \in \outce{\nnode}\} \,
\big).$ A tECS is \emph{time-ordered} if (1) every node $\nnode$ carries
$\max(\nnode)$ as an extra label (so that it can be retrieved in $O(1)$ time)
and (2) for every union node $\unode$ it holds that
$\max(\uleft(\unode)) \geq \max(\uright(\unode))$. For instance, the tECS of
Figure~\ref{fig:tecs} is time-ordered. 
The DFS-based enumeration algorithm
described above can be modified to avoid needless searching on a time-ordered
tECS: before starting the search, first check that $j - \max(n) \leq
\epsilon$. If so, $\outce{\nnode}^\epsilon(j)$ is non-empty and we perform the
DFS-based enumeration. However, when we traverse a union node $\unode$ we always
visit $\uleft(\unode)$ before $\uright(\unode)$. Moreover, we only visit
$\uright(\unode)$ if $j - \max(\uright(\unode)) \leq \epsilon$. Otherwise,
$\uright(\unode)$ and its descendants do not contribute to
$\outce{\nnode}^\epsilon(j)$, and can be skipped.

The second restriction is that $\ecs$ needs to be \emph{$k$-bounded} for some
$k \in \bbN$, which is defined as follows. 
Define the \emph{(left) output-depth} $\odepth(\nnode)$ of a node $\nnode$ recursively as follows:
if $\nnode$ is a non-union node, then $\odepth(\nnode) = 0$; otherwise,
$\odepth(\nnode) = \odepth(\uleft(\nnode))+1$.  The output depth tell us how
many union nodes we need to traverse to the left before we find a non-union node
that, therefore, produces part of the output.  Then, 
$\ecs$ is \emph{$k$-bounded} if $\odepth(\nnode) \leq k$ for every node
$\nnode$. For instance, the tECS of Figure~\ref{fig:tecs} is $1$-bounded. The $k$-boundedness restriction is necessary because even though we know that, on a time-ordered $\ecs$, we will find a complex event to output by consistently visiting left children of union nodes, starting from $\nnode$,  there may be an unbounded number of union nodes to visit before reaching a bottom node. In that case, the length of the corresponding full-path $\seq{p}$ risks being significantly bigger than the size of $\outce{\seq{p}}$, violating the output-linear delay.

The third restriction on $\ecs$, needed to ensure that we may enumerate without duplicates, is for it to be \emph{duplicate-free}. Here, $\ecs$ is duplicate-free if all of its nodes are duplicate-free, and a node  $\nnode$ is duplicate-free if for  every pair of distinct full-paths $\seq{p}$ and $\seq{q}$ that start at $\nnode$ we have  $\outce{\seq{p}} \not = \outce{\seq{q}}$.


\begin{restatable}{theorem}{theotimeordered}\label{theo:time-ordered}
  Fix $k$. For every $k$-bounded and time-ordered tECS $\ecs$, and for every
  duplicate-free node $\nnode$ of $\ecs$, time-window bound $\epsilon$, and position $j$, the
  set $\outce{\nnode}^\epsilon(j)$ can be enumerated with output-linear delay and without duplicates.
      \end{restatable}

      \inFullVersion{Appendix~\ref{sec:proof:theo:time-ordered} gives the pseudocode of the enumeration algorithm, and shows its correctness.}

\subsection{Methods for managing the data structure}\label{sec:evaluation-structure}

The evaluation algorithm will build the tECS $\ecs$ incrementally: it starts
from the empty tECS and, whenever a new tuple arrives on the stream $S$,
it modifies $\ecs$ to correctly represent the relevant open complex events. To
ensure that we may enumerate $\sem{\cA}^\epsilon_j(S)$ from $\ecs$ by use of
Theorem~\ref{theo:time-ordered}, $\ecs$ will always be time-ordered, $k$-bounded
for $k = 3$, and duplicate-free.  We next discuss the operations for modifying a
tECS $\ecs$ required by the evaluation algorithm.

It is important to remark that, in order to ensure that newly created nodes are
$3$-bounded, many of these operations expect their argument nodes to be
\emph{safe}. Here, a node is \emph{safe} if it is a non-union node or if both
$\odepth(\nnode) = 1$ and $\odepth(\uright(\nnode)) \leq 2$. All of our
operations themselves return safe nodes, as we will see.

\paragraph*{Operations on tECS.}
We consider the following three operations:
$$
\bnode \leftarrow \dbottom(i) \ \  \ \ \ \ \onode \leftarrow \dextend(\nnode, j) \ \ \ \ \ \  \unode \leftarrow \dunion(\nnode_1, \nnode_2)
$$ 
where $i, j \in \bbN$, $\nnode$, $\nnode_1$ and $\nnode_2$ are nodes in $\ecs$, and $\bnode$, $\onode$, and $\unode$ are the bottom, output, and union nodes, respectively, created by these methods.  
The first method, $\dbottom(i)$ simply adds a new bottom node $\bnode$ labeled by $i$ to $\ecs$. The second method, $\dextend(\nnode, j)$ adds a new output node $\onode$ to $\ecs$ with $\opos(\onode) = j$ and $\onext(\onode) = \nnode$. The third method, $\dunion(\nnode_1, \nnode_2)$ returns a node $\unode$ such that $\outce{\unode} = \outce{\nnode_1} \cup \outce{\nnode_2}$. This method requires a more detailed discussion.

Specifically, $\dunion$ requires that its inputs $\nnode_1$ and $\nnode_2$ are
safe and that $\maxstart(\nnode_1) = \maxstart(\nnode_2)$.
Under these requirements, $\dunion(\nnode_1,\nnode_2)$ operates as follows. If $\nnode_1$ is non-union then a new union node $\unode$ is created which is connected to $\nnode_1$ and $\nnode_2$ as shown in Figure~\ref{fig:union-gadgets}(a). If $\nnode_2$ is non-union, then $\unode$ is created as shown in Figure~\ref{fig:union-gadgets}(b). When $\nnode_1$ and $\nnode_2$ are both union nodes we distinguish two cases. If  $\maxstart(\uright(\nnode_1)) \geq \maxstart(\uright(\nnode_2))$, three new union nodes, $\unode$, $\unode_1$, and $\unode_2$ are added, and connected as shown in Figure~\ref{fig:union-gadgets}(c). Finally, if $\maxstart(\uright(\nnode_1)) < \maxstart(\uright(\nnode_2))$, three new union nodes are added, but connected as we show in Figure~\ref{fig:union-gadgets}(d). In all cases, $\outce{\unode} = \outce{\nnode_1} \cup \outce{\nnode_2}$ and the newly created union nodes are time-ordered. Furthermore, the reader is invited to check that, because $\nnode_1$ and $\nnode_2$ are safe, $\unode$ is also safe, and the output-depth of the newly created union nodes is a most $3$.

Because also the nodes created by $\dbottom$ and $\dextend$ are safe, time-ordered, and have output-depth at most $3$, it follows that any tECS that is created using only these three methods is time-ordered and $3$-bounded. Moreover, all of these methods output safe nodes and take constant time. 


\let\oldReturn\Return
\renewcommand{\Return}{\State\oldReturn}

\begin{algorithm*}
	\caption{Evaluation of an I/O-deterministic CEA $\cA = (Q,\Delta,q_0,F)$  over a stream $S$ given a time-bound $\epsilon$.}\label{alg:evaluation}
	\smallskip
	\begin{varwidth}[t]{0.66\textwidth}
		\begin{algorithmic}[1]
			\Procedure{Evaluation}{$\cA,S, \epsilon$}
			\State $j \gets -1$
			\State $\ctable \gets \emptyset$
			\While{$t \gets \yield(S)$}\label{line:evaluation-while}
			\State $j \gets j+1$
			\State $\ctable' \gets \emptyset$
			\State $\ulist \gets \linit(\dbottom(j))$\label{line:evaluation-linit1}
			\State $\textproc{ExecTrans}(q_0, \ulist, t, j)$
			\For{$p \in \cordkeys(\ctable)$} \label{line:teval-forq}
			\State $\textproc{ExecTrans}(p, \ctable[p], t, j)$
			\EndFor
			\State $\ctable \gets \ctable'$
			\State $\textproc{Output}(j, \epsilon)$
			\EndWhile
			\EndProcedure
			\algstore{myalg}
		\end{algorithmic}
	\end{varwidth} \ \ \ \ \ \ \
	\begin{varwidth}[t]{0.66\textwidth}
		\begin{algorithmic}[1]
			\algrestore{myalg}
			\Procedure{ExecTrans}{$p, \ulist, t, j$}
			\State $\nnode \gets \lmerge(\ulist)$\label{line:evaluation-lmerge1}
			\If{$q \gets \Delta(p, t, \amark)$}
			\State $\nnode' \gets \dextend(\nnode, j)$
			\State $\ulist' \gets \linit(\nnode')$ \label{line:evaluation-linit2}
			\State $\textproc{Add}(q, \nnode', \ulist')$
			\EndIf 
			\If{$q \gets \Delta(p, t, \umark)$}
			\State $\textproc{Add}(q, \nnode, \ulist)$
			\EndIf 
			\Return
			\EndProcedure
			\algstore{myalg}
		\end{algorithmic}
	\end{varwidth} \ \ \ \ \ \ \
	\begin{varwidth}[t]{0.66\textwidth}
		\begin{algorithmic}[1]
			\algrestore{myalg}
			\Procedure{Add}{$q, \nnode, \ulist$}
			\If{$q \in \ckeys(\ctable')$}
			\State $\ctable'[q] \gets \linsert(\ctable'[q], \nnode)$ \label{line:evaluation-linsert}
			\Else
			\State $\ctable'[q] \gets \ulist$
			\EndIf
			\Return
			\EndProcedure
			\State
			\Procedure{Output}{$j, \epsilon$}
			\For{$p \in \ckeys(\ctable)$}
			\If{$p \in F$}
			\State $\nnode \gets \lmerge(\ctable[p])$\label{line:evaluation-lmerge2}
			\State $\textproc{Enumerate}(\nnode, j)$
			\EndIf
			\EndFor
			\EndProcedure
		\end{algorithmic}
	\end{varwidth}
	\smallskip
      \end{algorithm*}

\paragraph*{Union-lists and their operations.} To incrementally maintain $\ecs$, the evaluation algorithm will also need to manipulate \emph{union-lists}. A union-list is a non-empty sequence $\ulist$ of safe nodes of the form $
\ulist =  \nnode_0, \nnode_1, \ldots, \nnode_k
$ such that (1) $\nnode_0$ is non-union, (2) $\maxstart(\nnode_0) \geq \maxstart(\nnode_i)$ and (3) $\maxstart(\nnode_{j}) > \maxstart(\nnode_{j+1})$, for every $i \leq k$ and $1 \leq j < k$. 
In other words, a union-list is a non-empty sequence of safe nodes sorted decreasingly by maximum-start.

We require three operations on union-lists, all of which take safe nodes as arguments. The first method, $\linit(\nnode)$, creates a new union-list containing the single non-union node $\nnode$. The second method, $\linsert(\ulist, \nnode)$, mutates union-list $\ulist = \nnode_0, \ldots, \nnode_k$ in-place by inserting a safe node $\nnode$ such that $\maxstart(\nnode) \leq \maxstart(n_0)$.  Specifically, if there is $i>0$ such that $\maxstart(\nnode_i) = \maxstart(\nnode)$, then it replaces $\nnode_i$ in $\ulist$ by the result of calling  $\dunion(\nnode_i, \nnode)$. This hence also updates~$\ecs$.  Otherwise, we discern two cases. If $\maxstart(\nnode) = \maxstart(\nnode_0)$, then $\nnode$ is inserted at position $1$ in $\ulist$. Otherwise, $\nnode$ is inserted between $\nnode_i$ and $\nnode_{i+1}$ with $i > 0$ such that $\maxstart(\nnode_i) > \maxstart(\nnode) > \maxstart(\nnode_{i+1})$. 
The last method, $\lmerge(\ulist)$, takes a union-list $\ulist$ and returns a node $\unode$ such that  $\outce{\unode} = \outce{\nnode_0} \cup \ldots \cup \outce{\nnode_k}$. Specifically, if $k = 0$, then  $\unode = \nnode_0$. Otherwise, we add  $k$ union nodes $\unode, \unode_1, \ldots, \unode_{k-1}$ to $\ecs$, and  connect them as shown in Figure~\ref{fig:union-gadgets}~(e).  It is important to observe that, because $\nnode_0$ is a non-union node, $\odepth(\unode) \leq 1$. Moreover, because all $\nnode_i$ are safe, $\odepth(\unode_i) \leq 2$. As a result, $\unode$ is safe. 
Furthermore, all of the new union nodes are time-ordered and are $3$-bounded.
This, combined with the  properties of $\dbottom$, $\dextend$, and $\dunion$ described above implies that any tECS that is created using only these three methods plus $\lmerge$ is time-ordered and $3$-bounded.
Furthermore, all of these methods retrieve safe nodes, and their outputs are hence valid inputs to further calls.
Finally,  we remark that all methods on union-lists take time linear in the length of $\ulist$.

\paragraph*{Hash tables.} In order to incrementally maintain $\ecs$, the evaluation algorithm will also need to manipulate hash tables that map CEA states to union-lists of nodes. If $\ctable$ is such a hash table, then we  write $\ctable[q]$ for the union-list associated to state $q$ and $\ctable[q] \gets \ulist$ for inserting or updating it with a union-list $\ulist$. We use the method $\ckeys(\ctable)$ to iterate through all the current states of $\ctable$ and write $q \in \ckeys(\ctable)$ for checking if $q$ is already a key in $\ctable$ or not. For technical reasons, we also consider a method called $\cordkeys(\ctable)$ that  iterates over keys of $T$ \emph{in the order in which they have been inserted into} $T$. If a key is inserted and then later is inserted again (i.e., an update), then it is the time of first insertion that counts for the iteration order. One can easily implement  $\cordkeys(\ctable)$ by maintaining a traditional hash table together with a linked list that stores  keys sorted in insertion order. Compatible with the RAM model of computation~\cite{aho1974design}, we assume that hash table lookups and insertion take constant time, while iteration over their keys by means of $\ckeys(\ctable)$ and $\cordkeys(\ctable)$ is with constant delay.

\subsection{The evaluation algorithm}
\label{sec:evaluation:algo}

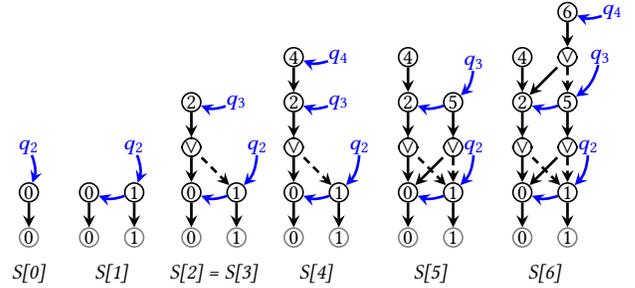
\begin{figure}[tpb]
  \small
		\begin{tikzpicture}[->,>=stealth, semithick, auto, initial text= {}, initial distance= {3mm}, accepting distance= {4mm}, node distance=3cm, semithick,scale=0.8]

                  \tikzstyle{bnode}=[draw=gray,text=black,inner sep=0pt, minimum size=2.5mm, circle]
                  \tikzstyle{onode}=[draw=black,text=black,inner sep=0pt, minimum size=2.5mm, circle]
                  \tikzstyle{unode}=[draw=black, text=black,inner sep=0pt, minimum size=2.5mm,rounded corners]

		\tikzstyle{onext}=[->, ultra thick, >=stealth, line width=1pt]

		\tikzstyle{ulist}=[->, ultra thick, >=stealth, line width=1pt, blue]

                \tikzstyle{uleft}=[->, densely dashed, ultra thick, >=stealth, line width=1pt]
		
                \tikzstyle{uright}=[->, ultra thick, >=stealth, line width=1pt]

                \tikzstyle{qnode}=[text=blue,color=blue, inner sep=0pt]

                \begin{scope}
                  \node (lbl) at (0,-0.6) {\textit{S[0]}};
                  \node[bnode] (bot0) at (0,0) {$0$};
                  \node[onode] (pos0) at (0,0.75) {$0$};

                  \draw[onext] (pos0) to (bot0);

                  \node[qnode] (q2) at (0,1.5) {$q_2$};
                  \draw[ulist] (q2) to[bend left=20]  (pos0);
                \end{scope}

                \begin{scope}[xshift=1cm]
                  \node (lbl) at (0.375,-0.6) {\textit{S[1]}};

                  \node[bnode] (bot0) at (0,0) {$0$};
                  \node[onode] (pos0) at (0,0.75) {$0$};
                  \draw[onext] (pos0) to (bot0);

                  \node[bnode] (bot1) at (0.75,0) {$1$};
                  \node[onode] (pos1) at (0.75,0.75) {$1$};
                  \draw[onext] (pos1) to (bot1);

                  \node[qnode] (q2) at (0.75,1.5) {$q_2$};
                  \draw[ulist] (q2) to[bend left=20]  (pos1);
                  \draw[ulist] (pos1) to[bend left=20] (pos0);
                \end{scope}

                \begin{scope}[xshift=2.7cm]
                  \node (lbl) at (0.375,-0.6) {\textit{S[2] = S[3]}};

                  \node[bnode] (bot0) at (0,0) {$0$};
                  \node[onode] (pos0) at (0,0.75) {$0$};
                  \draw[onext] (pos0) to (bot0);

                  \node[bnode] (bot1) at (0.75,0) {$1$};
                  \node[onode] (pos1) at (0.75,0.75) {$1$};
                  \draw[onext] (pos1) to (bot1);

                  \node[unode] (u01) at (0.0,1.5) {$\vee$};
                  \node[onode] (pos2) at (0.0,2.25) {$2$};
                  \draw[onext] (pos2) to (u01);
                  \draw[uleft] (u01) to (pos1);
                  \draw[uright] (u01) to (pos0);

                  \node[qnode] (q2) at (1.1,1.5) {$q_2$};
                  \draw[ulist] (q2) to[bend left=20]  (pos1);
                  \draw[ulist] (pos1) to[bend left=20] (pos0);

                  \node[qnode] (q3) at (0.75,2.25) {$q_3$};
                  \draw[ulist] (q3) to[bend left=20]  (pos2);
                \end{scope}

                \begin{scope}[xshift=4.4cm]
                  \node (lbl) at (0.375,-0.6) {\textit{S[4]}};

                  \node[bnode] (bot0) at (0,0) {$0$};
                  \node[onode] (pos0) at (0,0.75) {$0$};
                  \draw[onext] (pos0) to (bot0);

                  \node[bnode] (bot1) at (0.75,0) {$1$};
                  \node[onode] (pos1) at (0.75,0.75) {$1$};
                  \draw[onext] (pos1) to (bot1);

                  \node[unode] (u01) at (0.0,1.5) {$\vee$};
                  \node[onode] (pos2) at (0.0,2.25) {$2$};
                  \draw[onext] (pos2) to (u01);
                  \draw[uleft] (u01) to (pos1);
                  \draw[uright] (u01) to (pos0);

                  \node[onode] (pos4) at (0.0,3.0) {$4$};                  
                  \draw[onext] (pos4) to (pos2);

                  \node[qnode] (q2) at (1.1,1.5) {$q_2$};
                  \draw[ulist] (q2) to[bend left=20]  (pos1);
                  \draw[ulist] (pos1) to[bend left=20] (pos0);

                  \node[qnode] (q3) at (0.75,2.25) {$q_3$};
                  \draw[ulist] (q3) to[bend left=20]  (pos2);

                  \node[qnode] (q4) at (0.75,3) {$q_4$};
                  \draw[ulist] (q4) to[bend left=20]  (pos4);
                \end{scope}

                \begin{scope}[xshift=6.3cm]
                  \node (lbl) at (0.375,-0.6) {\textit{S[5]}};

                  \node[bnode] (bot0) at (0,0) {$0$};
                  \node[onode] (pos0) at (0,0.75) {$0$};
                  \draw[onext] (pos0) to (bot0);

                  \node[bnode] (bot1) at (0.75,0) {$1$};
                  \node[onode] (pos1) at (0.75,0.75) {$1$};
                  \draw[onext] (pos1) to (bot1);

                  \node[unode] (u01) at (0.0,1.5) {$\vee$};
                  \node[onode] (pos2) at (0.0,2.25) {$2$};
                  \draw[onext] (pos2) to (u01);
                  \draw[uleft] (u01) to (pos1);
                  \draw[uright] (u01) to (pos0);

                  \node[onode] (pos4) at (0.0,3.0) {$4$};                  
                  \draw[onext] (pos4) to (pos2);

                  \node[unode] (u01v2) at (0.75,1.5) {$\vee$};
                  \draw[uleft] (u01v2) to (pos1);
                  \draw[uright] (u01v2) to (pos0);

                  \node[onode] (pos5) at (0.75,2.25) {$5$};                  
                  \draw[onext] (pos5) to (u01v2);

                  \node[qnode] (q2) at (1.1,1.5) {$q_2$};
                  \draw[ulist] (q2) to[bend left=20]  (pos1);
                  \draw[ulist] (pos1) to[bend left=20] (pos0);

                  \node[qnode] (q3) at (1.1,2.9) {$q_3$};
                  \draw[ulist] (q3) to[bend left=20]  (pos5);
                  \draw[ulist] (pos5) to[bend left=20]  (pos2);
                \end{scope}

                \begin{scope}[xshift=8.2cm]
                  \node (lbl) at (0.375,-0.6) {\textit{S[6]}};

                  \node[bnode] (bot0) at (0,0) {$0$};
                  \node[onode] (pos0) at (0,0.75) {$0$};
                  \draw[onext] (pos0) to (bot0);

                  \node[bnode] (bot1) at (0.75,0) {$1$};
                  \node[onode] (pos1) at (0.75,0.75) {$1$};
                  \draw[onext] (pos1) to (bot1);

                  \node[unode] (u01) at (0.0,1.5) {$\vee$};
                  \node[onode] (pos2) at (0.0,2.25) {$2$};
                  \draw[onext] (pos2) to (u01);
                  \draw[uleft] (u01) to (pos1);
                  \draw[uright] (u01) to (pos0);

                  \node[onode] (pos4) at (0.0,3.0) {$4$};                  
                  \draw[onext] (pos4) to (pos2);

                  \node[unode] (u01v2) at (0.75,1.5) {$\vee$};
                  \draw[uleft] (u01v2) to (pos1);
                  \draw[uright] (u01v2) to (pos0);

                  \node[onode] (pos5) at (0.75,2.25) {$5$};                  
                  \draw[onext] (pos5) to (u01v2);

                  \node[unode] (u25)  at (0.75,3) {$\vee$};
                  \draw[uleft] (u25) to (pos5);
                  \draw[uright] (u25) to (pos2);

                  \node[onode] (pos6) at (0.75, 3.75) {$6$};
                  \draw[onext] (pos6) to (u25);

                  \node[qnode] (q2) at (1.1,1.5) {$q_2$};
                  \draw[ulist] (q2) to[bend left=20]  (pos1);
                  \draw[ulist] (pos1) to[bend left=20] (pos0);

                  \node[qnode] (q3) at (1.3,3) {$q_3$};
                  \draw[ulist] (q3) to[bend left=20]  (pos5);
                  \draw[ulist] (pos5) to[bend left=20]  (pos2);

                  \node[qnode] (q4) at (1.5,3.75) {$q_4$};
                  \draw[ulist] (q4) to[bend left=20]  (pos6);
                \end{scope}

              \end{tikzpicture}
              \vspace{-3mm}
  \caption{Illustration of Algorithm~\ref{alg:evaluation} on the CEA $\cA$ and stream $S$ of Figure~\ref{fig:example-cea}. 
  }
  \label{fig:update-example}
  \label{fig:update-run}
\end{figure}

\systemname's main evaluation algorithm is presented in
Algorithm~\ref{alg:evaluation}. It receives as input an I/O deterministic CEA
$\cA = (Q,\Delta,q_0,F)$, a stream $S$, and a time-bound $\epsilon$. As already
mentioned at the beginning of Section~\ref{sec:evaluation}, its goal is to
enumerate, at every position $j$ in the stream, the set
$\sem{\cA}_j^\epsilon(S)$ of complex events produced by accepting runs
terminating at $j$ that satisfy the time-bound $\varepsilon$.  It does so by
maintaining (1) a tECS $\ecs$ to represents all open complex events up to the
current position $j$ and (2) the set of \emph{active states} of $\cA$. Here, a
state $q \in Q$ is \emph{active} at stream position $j$ if there is some (not
necessarily accepting) run of $\cA$ that starts at position $i \leq j$ which is
in state $q$ at position $j$.  Specifically, in order to incrementally maintain
the tECS, Algorithm~\ref{alg:evaluation} will link active states to the set
of open complex events that they generate. Towards that goal, it uses a hash
table $T$ that maps active states of $Q$ to a union-list of nodes.

We next explain how Algorithm~\ref{alg:evaluation} works. During our discussion, the reader may find it helpful to refer to Figure~\ref{fig:update-run}, which  illustrates Algorithm~\ref{alg:evaluation} as it evaluates the CEA of Figure~\ref{fig:example-cea} over the stream $S$ of Figure~\ref{fig:example-cea}. Each subfigure depicts the state after processing $S[j]$. The tECS is denoted in black, while the hash table $T$ that links the  active states to union-lists is illustrated in blue.  For each position, the set $\cordkeys(T)$ will be ordered top down. (E.g., for $S[4]$ this will be $q_4, q_3, q_2$ while for $S[3]$ this will be $q_3,q_2$.)

Algorithm~\ref{alg:evaluation} consists of four procedures, of which
$\textproc{Evaluation}$ is the main one. It starts by initializing the current
stream position $j$ to $-1$ and the hash table $T$ to empty (lines 2--3). Then,
for every tuple in the stream, it executes the while loop in lines 4--12. Here,
we assume that $\yield(S)$ returns the next unprocessed tuple~$t$ from $S$.  For
every such tuple, $j$ is updated, and the hash table $\ctable'$ is initialized
to empty (lines 5-6). Intuitively, in lines 6--10, the hash table $T$
will hold the states that are active at position $j-1$ (plus corresponding
union-lists) while $T'$ will hold the states that are active at position $j$. In
particular, $T'$ is computed from $T$ in lines 7--10. Specifically, lines 7--8 take into account that a new run may start at any
position in the stream, and hence in particular at the current position $j$. For
this purpose, the algorithm creates a new union-list starting at position $j$
(line 7) and executes all transitions of initial state $q_0$ by calling
$\textproc{ExecTrans}$ (line 8), whose operation is explained
below. Subsequently, lines 9--10 take into account that a state $q$ is active at position $j$ if there is a state $p$ active at position $j-1$ and a 
transition $p \trans{P/m} q$ of $\cA$ with $t \models P$. As such, we iterate through all active states of $T$ and execute
all of their transitions (line 9-10). Once this is done, we swap the content of
$T$ with $T'$ to prepare of the next iteration. We also call the
$\textproc{Output}$ method (lines 29-33) who is in charge of enumerating all
complex events in $\sem{\cA}_j^\epsilon(S)$, and whose operation is explained
below.

The procedure $\textproc{ExecTrans}$ is the workhorse of
Algorithm~\ref{alg:evaluation}. It receives an active state $p$, a union-list
$\ulist$, the current tuple $t$, and the current position $j$.  The union-list
$\ulist$ encodes all open complex events of runs that have reached $p$.
$\textproc{ExecTrans}$ first merges $\ulist$ into a single node $\nnode$ (line
14). Then it executes the marking (line 15) and non-marking transitions (line
19) that can read $t$ while in state~$p$. Specifically, we write
$q \gets \Delta(p, t, m)$ to indicate that there is $p \trans{P/m} q$ in
$\Delta$ with $t \models P$. Given that $\cA$ is I/O-deterministic, there exists
at most one such state $q$ and, if there is none, we interpret
$q \gets \Delta(p, t, m)$ to be \textsf{false}.  In lines 15--18, if there is a marking
transition reaching $q$ from $p$, then we extend all open complex events
represented by $\nnode$ with the new position $j$ (line 16) and add them to
$\ctable'[q]$.  In lines 19--20, if there is a non-marking transition reaching $q$ from $p$, we
add $\nnode$ directly to $\ctable'[q]$ without extending it.  To add the open
complex events to $\ctable'[q]$, we use the method $\textproc{Add}$ (lines
22-27). This method checks whether it is the first time that we reach $q$ on the
$j$-th iteration or not. Specifically, if $q \in \ckeys(\ctable')$, then we have
already reached $q$ on the $j$-th iteration and therefore we insert $\nnode$ in
the list $\ctable'[q]$ (lines 23-24). Instead, if it is the first time that we
reach $q$ on the $j$-iteration, then we initialize the $q$ entry of $\ctable'$
with the union-list representation of~$\nnode$.

The $\textproc{Output}$ procedure (lines 29-33) is in charge of enumerating all
complex events in $\sem{\cA}_j^\epsilon(S)$. Given that, when it is called,
$\ctable$ contains all active states at position $j$, it suffices to iterate over
$p \in \ckeys(\ctable)$ and check whether $p$ is a final state or not (lines
30-31). If $p$ is final, then we merge the union-list at $\ctable[p]$ into a
node $\nnode$ and call $\textproc{Enumerate}(\nnode, j)$, where $\textproc{Enumerate}$ is the enumeration algorithm of Theorem~\ref{theo:time-ordered}.

Recall that by Theorem~\ref{theo:time-ordered} if the tECS is $k$-bounded,
time-ordered, and duplicate-free then the set $\outce{\nnode}(j)$ can be
enumerated with output-linear delay.  Because Algorithm~\ref{alg:evaluation} builds $\ecs$ only through the methods of Section~\ref{sec:evaluation-structure}, we are guaranteed it is $3$-bounded and time-ordered. Moreover, we can show that, because $\cA$ is I/O-deterministic, $\ecs$ will also be duplicate-free. From this, we can derive the following correctness statement of Algorithm~\ref{alg:evaluation}.

\begin{restatable}{theorem}{theoalgocorrect}\label{theo:algo-correct}

  After the $j$-th iteration of $\textproc{Evaluation}$, the $\textproc{Output}$
  method enumerates the set $\sem{\cA}_j^\epsilon(S)$ with output-linear delay.
\end{restatable}

We note that the order in which we iterate over
active states and execute transitions in lines 7--11 is important for the
correctness of the algorithm\inFullVersion{ (as explained in detail in the Appendix)}. In
particular, because we first execute transitions of initial state $q_0$ and then
process all states according to their insertion-order, we can prove that states
are processed following a decreasing order of the max-start of active
states. From this, we also derive that every call to $\textsf{insert}(T'[q],n)$
in line 24 is legal: when it is called we have that
$\maxstart(T'[q]) \geq \maxstart(n)$, as is required by the definition of
\textsf{insert}. 

Let us now analyze the update-time. When a new tuple arrives, lines 5--11 of Algorithm~\ref{sec:evaluation:algo} update $T$, $T'$, and $\ecs$  by means of the methods of Section~\ref{sec:evaluation-structure}. All of these  either take constant time, or time linear in the size of the union list being manipulated. We can show\inFullVersion{ (see the Appendix)} that, for every position $j$,  the length of every union list is bounded by the number of active states (i.e., the number of keys in $\ctable$). Then, because in each invocation of lines 5--11 we iterate over all transitions in the worst case, and because executing a transition takes time proportional to the length of union-list, which is at most the number of states, we may conclude that the time for processing a new tuple is $\cO(|Q| \cdot |\Delta|)$. This is constant in data complexity. 

\vspace{-1ex}
\subsection{Implementation aspects of \systemname}\label{sec:evaluation:implementation}

\input{sections/implementation}

%% file: sections/implementation.tex
We review here some implementation aspects of CORE that we did not cover by the algorithm or previous sections. 

The system receives a CEQL query and a stream, reading it tuple by tuple. From the query, CORE collects all atomic predicates (e.g., \texttt{price > 100}) into a list, call it $P_1, \ldots, P_k$. For each tuple $t$ of the stream, the system evaluates $t$ over $P_1, \ldots, P_k$ by building a bit vector $\vec{v}_t$ of $k$ entries such that $\vec{v}_t[i] = 1$ if, and only if, $t \models P_i$, for every $i \leq k$. Then, CORE uses $\vec{v}_t$ as the internal representation of $t$ for optimizing the evaluation of complex predicates (e.g., conjunctions or disjunctions of atomic predicates) and for the determinization procedure (see below). Furthermore, CORE evaluates each predicate once, improving the performance over costly attributes (e.g., text).

As already mentioned, CORE compiles the CEQL query into a non-deterministic CEA $\cA$ (Theorem~\ref{theo:CEL-to-CEA}). For the evaluation of $\cA$ with Algorithm~\ref{alg:evaluation}, CORE runs a determinization procedure \emph{on-the-fly}: for a state $p$ in the determinization of $\cA$ and the bit vector $\vec{v}$, the states $q_\amark:= \Delta(p, \vec{v}, \amark)$ and $q_\umark:= \Delta(p, \vec{v}, \umark)$ are computed in linear time over $|\cA|$. Moreover, we cache $q_\amark$ and $q_\umark$ in main memory and use a fast-index to recover $\Delta(p, \vec{v}, \amark)$ and $\Delta(p, \vec{v}, \umark)$ whenever is needed again. Although the determinization of $\cA$ could be of exponential size in the worst case, this rarely happens in practice. Note that the determinization process depends on the selection strategy which can also computed on-the-fly by following the constructions in~\cite{CELJOURNAL}.

For reducing memory usage when dealing with time windows, the system manages the memory itself with the help of Java weak references. Nodes in the tECS data structure are weakly referenced, while the strong references are stored in a list, ordered by creation time. When the system goes too long without any outputs, it will remove the strong references from nodes that are now outside the time window, allowing Java's garbage collector to reclaim that memory without the need to modify the tECS data structure. Although this memory management could break the constant time update and output-linear delay, it takes constant \emph{amortized} time and works well in practice (see Section~\ref{sec:experiments}).

For evaluating the \textsf{PARTITION BY} clause, CORE partitions the stream by the corresponding \textsf{PARTITION BY} attributes, running one instance of the algorithm for each partition. This process is done by hashing the corresponding attribute values, assigning them their own runs, or creating new ones if they don't exist.

We implemented CORE in Java. Its code is open-source and available at~\cite{system} under the GNU GPLv3 license.

%% file: sections/experiments.tex
\newcommand{\T}{\texttt{T}}

\newtext{In this section, we compare \systemname\xspace against four leading CER systems: SASE~\cite{SASE}, Esper~\cite{EsperTech}, FlinkCEP~\cite{FlinkCEP}, and OpenCEP~\cite{OpenCEP}.}
These all provide a CER query language with features like pattern matching, windowing, and partition-by based correlation, whose semantics is comparable to CORE. \newtext{We have surveyed the literature for other systems to compare against but found ourselves limited to these baselines, as explained \altVersion{in Appendix~\ref{sec:app-other-competitors}}{in the online appendix~\cite{core-full-paper-version}}}.

\smallskip
\noindent\textbf{Setup}. \newtext{We compare against SASE v.1.0, Esper v.8.7.0, FlinkCEP v.1.12.2, and OpenCEP (commit \texttt{e320ad8}). All systems are implemented in Java except for OpenCEP which uses Python 3.9.0.}
We run experiments on a server equipped with
an 8-core AMD Ryzen 7 5800X processor running
at 3.8GHz, 64GB of RAM, Windows 10 operating system, OpenJDK Runtime
17+35-2724, and the OpenJDK 64-Bit Server Virtual
Machine build 17+35-2724. 
\newtext{Java and Python virtual machines are restarted with freshly allocated memory before each run.}

\newtext{We compare systems on throughput and memory consumption. All reported
  numbers are averages over 3 runs.  We measure throughput, expressed as the
  number of events processed per second (e/s), as follows. We first load the
  input stream completely in main memory to avoid measuring the data loading
  time. We then start the timer and allow systems to read and process events as fast as they can. After 30 seconds, we disallow reading further events and stop the timer when the last read event has been fully processed, with a timeout of 1 minute.  We report the average throughput, expressed as the
  total number of processed events divided by the total running time. Runs that time-out have a value of ``aborted'' for the average, and will not be
  plotted. Recognized complex events are logged to main memory.
We adopt the consumption policy~\cite{DBLP:journals/vldb/GiatrakosAADG20,cugola2012} that forgets all events read so far when a complex event is found. We adopt this policy for all systems because it is the only one supported by Esper and SASE.}
\newtext{%
We measure  memory consumption every 10000 events, and report the average value. Before measuring memory consumption, we always first call the garbage collector. The experiments that measure memory consumption are run separately from the experiments that measure throughput.}



For the sake of consistency, we have verified that all systems produced the same set of complex events. When this was not the case, we explicitly mention this difference below.

CORE and SASE are single-core, sequential programs. To ensure fair comparison, all of the systems are therefore run in a single-core, sequential setup. \newtext{Esper, FlinkCEP, and OpenCEP} may exploit parallelism in a multi-core setup and support work in a distributed environment. While this may improve their performance, we stress that in many of the experiments below, CORE outperforms \newtext{the competition} by orders of magnitude. As such, even if we assume that these systems have perfect linear scaling in the number of added processors (which is unlikely in practice), they would need  orders of magnitude more processors before meeting COREs throughput. 
 Therefore, we do not consider a setup with parallelization.

All the experiments are reproducible. The data and scripts can be found in~\cite{system}.

\smallskip
\noindent\textbf{Datasets.} \newtext{We run our experiments over three real datasets: (1) \emph{the stock market dataset}~\cite{stockmarket} containing buy and sell events of stocks in a single market day; (2) \emph{the smart homes dataset}~\cite{smarthomes} containing power measurements of smart plugs deployed in different households; and (3) \emph{the taxi trips dataset}~\cite{taxitrips} recording taxi trips events in New York. Each original dataset includes several millions of real-world events and have already been  used in the past to compare  CER systems (e.g.~\cite{PoppeLAR17,PoppeLRM17,DBLP:conf/sigmod/PoppeLR019,PoppeLMRR21}).
For each dataset, we run experiments on a prefix of the full stream consisting of about one million events. 
\altVersion{Full details on the datasets, the prefix considered, and the queries that we run are given in Appendix~\ref{sec:app-seq-stocks} until \ref{sec:app-seq-taxis}.}{Full details on the datasets, the prefix considered, and the queries that we run are given in the online appendix~\cite{system}.}
}

\smallskip
\noindent \textbf{Sequence queries with output.} \newtext{We start by considering sequence queries, which have been used for benchmarking in CER before (see, for example, \cite{SASE,SASEcomplexity,cayuga,WPI,SAPesp}). Specifically, for each dataset we fix a workload of queries $P_n$ of the following form, each detecting sequences of $n$ events.}
{\small
\begin{verbatim}
	Pn := SELECT * FROM Dataset
	      WHERE A1 ; A2 ; ... ; An
	      FILTER A1[filter_1] AND ... AND An[filter_n] 
	      WITHIN T
\end{verbatim}}
\noindent  
\newtext{Here, \texttt{A1}, $\ldots$, \texttt{An} and \texttt{filter\_1},
  $\ldots$, \texttt{filter\_n} are dataset-dependent, \altVersion{as detailed in
    Appendix~\ref{sec:app-seq-stocks}--\ref{sec:app-seq-taxis}}{as detailed in
    the online appendix}. We consider sequences of length $n=3$, $6$, $9$, $12$,
  and $24$. In this experiment, we use a fixed time window $\texttt{T}$ of 10
  seconds for the stock market and smart homes dataset, and of 2.7 hours for the
  taxi trip dataset, which is enough to find several outputs. While $P_{12}$
  and $P_{24}$ usually do not produce outputs, they are nevertheless useful
  to see how systems scale to long sequence queries.  }

\begin{figure*}[t]
  \centering
  \includegraphics[width=17.5cm]{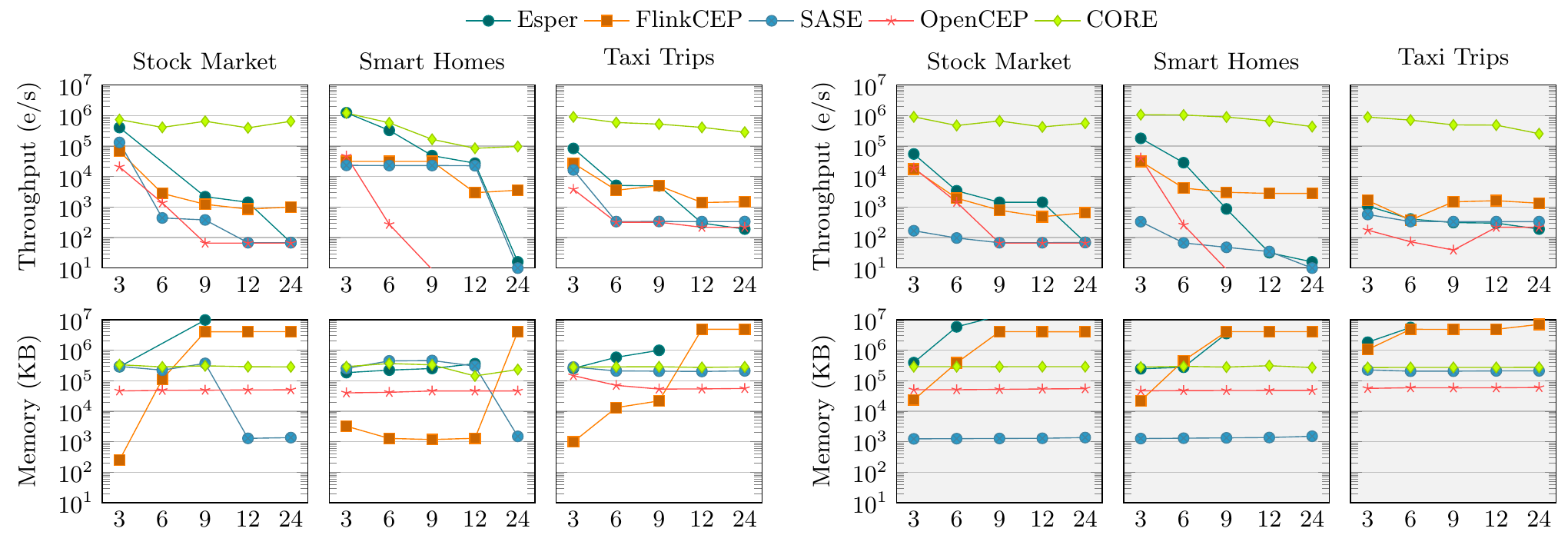}
  \vspace{-3mm}
  \caption{Throughput (higher is better) and memory consumption (lower is better) as a function of sequence
    length $n$, on queries with output (left) and queries without outputs
    (right, in gray).}
  \label{plot:seq}
\end{figure*}

\newtext{In the following analysis, when the processing of an input event
  triggers one or more complex events to be recognized, we will refer to
  this triggering input event as a \emph{pattern occurrence}. Each pattern
  occurrence may generate multiple complex event outputs. In our experiments,
  when a pattern occurrence is found, we only enumerate the first thousand
  corresponding complex events outputs.  An exception is FlinkCEP where, for implementation reasons, we only generate the first such output. Note
  that this favors FlinkCEP since it needs to produce less
  output. For OpenCEP we use the so-called ANY evaluation strategy, which is the only one that allows enumerating all matches. Unfortunately, it does not produce the same outputs as CORE, SASE, and FlinkCEP. Despite this, the number of results produced by OpenCEP is similar to other systems, making the comparison reasonable.}



\newcommand{\oom}{OOM\xspace}

\newtext{Figure~\ref{plot:seq} (left), plots the measured throughput and memory
  consumption (in log scale) as a function of the sequence length
  $n$. 
  \systemname's throughput is in the range of $[10^5,10^6]$ e/s. This throughput
  fluctuates due to variations in the number of complex events found, which must
  all be reported before an event can be declared fully processed. For example,
  for the smart homes dataset, queries $P_9$ and $P_{12}$ are very
  output intensive (i.e., around 1 million pattern occurrences in total), and the
  fluctuations are most noticeable for this dataset. Instead, the stocks and taxi
  trips datasets have less pattern occurrences (i.e., around a few thousand) and we can see that \systemname's throughput is more stable. Note in particular that \systemname's throughput is only linearly affected by $n$ in these cases.}

\newtext{We next compare to the other systems. For $n = 3$, the throughput of
  most baselines is one order of magnitude (\oom) lower than \systemname,
  except for Esper which has comparable throughput on stock and smart homes.
  However, as $n$ grows, the throughput of all systems except CORE degrades
  exponentially. For the smart homes dataset, the throughput of FlinkCEP is more stable, although up to 2 \oom
  lower than \systemname. Recall, however, that FlinkCEP produces only a
  single complex event per pattern occurrence, while all other systems enumerate up to one
  thousand complex events per occurrence. In contrast to the baselines,
  \systemname's throughput is stable, only affected by the high number of
  complex events found, and degrades only linearly in $n$ on stocks and taxis. 
As a consequence, on these datasets,  \systemname's throughput is 1 to 5 \oom higher than the
  baselines for large
  values of $n$. }

\newtext{
  The memory used by \systemname\xspace is high ($\sim$300MB) but stable in $n$. By contrast, the memory consumption of Esper and FlinkCEP can grow exponentially in $n$. OpenCEP and SASE\footnote{The memory consumption of SASE drops for $P_{12}$ and $P_{24}$. This is because in these cases the number of events that SASE can successfully process in full is  significantly less than for smaller values of $n$, and also less than other systems .
  } are special cases whose memory consumption is stable and comparable to \systemname. }

\smallskip
\noindent \textbf{Sequence queries without output.}  \newtext{In practice, we
  may expect CER systems to look for unusual patterns in the sense that 
  the number of complex events found is small.  Our next experiment captures
  this setting. For each query $P_n$ we create a variant $P_n'$ by adding, to
  the sequence pattern of $P_n$, an additional event that never occurs in the
  stream.  These variant queries hence never produce outputs. Since systems
  do not know this, however, they must inherently look for partial matches that satisfy
  the original sequence query $P_n$. Because no time is spent enumerating
  complex events when processing $P_n'$, this experiment can hence also be viewed as a
  way of measuring only the update performance of $P_n$. Note that, even though
  no complex events are found, systems may still materialize a large number of
  partial matches.}


\newtext{Figure~\ref{plot:seq} (right) plots the results (in log scale) on
  $P_n'$ for $n= 3$, $6$, $9$, $12$, $24$. We see that \systemname's throughput
  is comparable to the throughput on $P_n$ (Figure~\ref{plot:seq}, left), except
  for the smart homes dataset, where the throughput has
  improved. 
  Overall, \systemname's throughput is on the order of $10^6$ e/s, decreasing
  mostly linearly with $n$. By contrast, other systems have lower throughput, by
  1 to 5 \oom, and this throughput decreases much more rapidly in $n$.  This is
  most observable for Esper and OpenCEP, whose performance drops exponentially in
  $n$. For the taxi dataset, the throughput of all systems reduces to less
  $10^4$ e/s compared to the original $P_n$ setting, most probably because of a
  high number of partial matches that are maintained but never
  completed. \systemname\xspace is not affected by the number of partial matches, which explains its stable performance.}

\newtext{Notice that CORE does not use \emph{lazy evaluation}~\cite{mei2009zstream,DBLP:conf/debs/KolchinskySS15,DBLP:journals/pvldb/KolchinskyS18a} as some other systems do: it eagerly updates its internal data structure on each input event.
A CER system that uses lazy evaluation could perform well when there is no output (i.e., for $P_n'$) but badly when several pattern occurrences appear (i.e., for $P_n$). One can see CORE as the best of both worlds. It does work event after event, and results are ready for enumeration at each pattern occurrence.}


\newtext{
  Memory consumption of most systems remains high, with the memory usage of Esper and FlinkCEP  increases exponentially with $n$.  \systemname's memory consumption is stable, like the consumption of SASE\footnote{Memory consumption for SASE is lower compared to Figure~\ref{plot:seq} (left) because the number of events that SASE can successfully process in full is  significantly less.} and OpenCEP.}

\smallskip
\noindent \textbf{Varying the time-window size.} \newtext{We next measure the effect of varying the time window size on processing performance. 
We do so by 
fixing 
query $P'_3$,  and varying the window size from \texttt{T} to \texttt{4T} where \texttt{T} is the original window size. 
 We fix $P_3'$ because this allows us to measure the effect of update processing only, and because performance on $P_n'$ is highest for all systems when $n=3$. 
}

\begin{figure}[t]
	\centering
	\includegraphics[width=8.3cm]{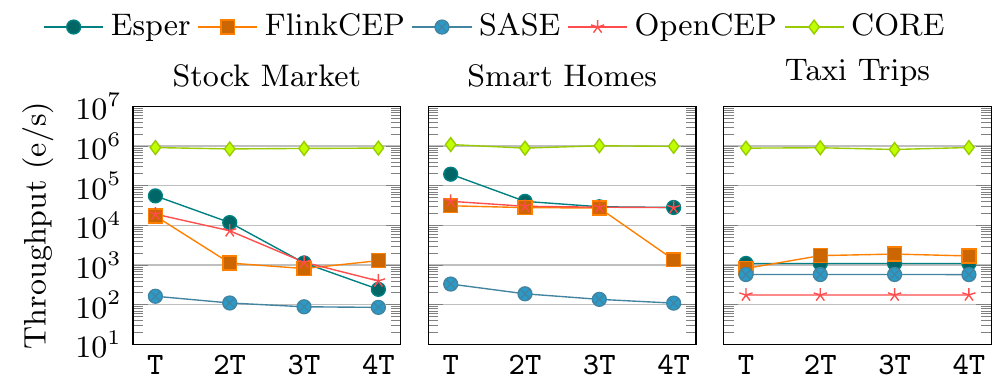}
	\vspace{-3mm}
	\caption{Throughput as a function of time-window size.}
	\label{plot:seq-windows}
		\vspace{-3mm}
\end{figure} 

\newtext{The results (log scale) are in Figure~\ref{plot:seq-windows}. We see that \systemname\xspace outperforms other systems by at least one \oom for size \texttt{T} and by 2 to 4 \oom for \texttt{4T}. 
  Note that for stock market and smart homes dataset we are still using relatively small time windows of a few hundreds events; in practice windows may be significantly larger (e.g., taxi trips dataset). We also observe that the throughput of other systems may degrade exponentially as the size of the time window grows. Indeed, this is clear for Esper, FlinkCEP, and OpenCEP on stock market dataset, where the throughput is around $10^5$ e/s at \texttt{T} but less than $10^3$ e/s at \texttt{4T}.  In contrast, CORE consistently maintains its performance at $10^6$ e/s, and it is not affected by the window size, as the theoretical analysis predicted.}

\newtext{
\altVersion{In Appendix~\ref{sec:app-seq-strategies}, }{In the online appendix~\cite{core-full-paper-version}, } we report similar experimental results  even when all systems are allowed to use a selection strategy heuristic to aid in faster processing. 
}


\smallskip
\noindent \textbf{Other operators.} \newtext{For the last experiment, we consider a diverse workload of queries where other operators like disjunction, iteration, or partition-by are used. For these queries, we report only on the stock market dataset. 
Given space restrictions, we present the full CEQL definition of each query in the online appendix~\cite{system}, and limit ourselves here to the the following simplified description.}
\begin{center}
\small
	\begin{tabular}{|l|}
		\hline
		$Q_1$ \ := \ \texttt{SELL} ; \texttt{BUY} ; \texttt{BUY} ; \texttt{SELL} \\ \hline
		$Q_2$ \ := \  $Q_1$ + \texttt{FILTER} \\ \hline
		$Q_3$ \ := \  $Q_1$  + \texttt{PARTITION BY} \\ \hline
		$Q_4$ \ := \  \texttt{SELL} ; \texttt{(BUY OR SELL)} ; \texttt{(BUY OR SELL)} ; \texttt{SELL} \\ \hline
		$Q_5$ \ := \  $Q_4$  + \texttt{FILTER} \\ \hline
		$Q_6$ \ := \  $Q_4$  + \texttt{PARTITION BY} \\ \hline
		$Q_7$ \ := \  \texttt{SELL} ; \texttt{(BUY OR SELL)}+ ; \texttt{SELL} \\
		\hline
	\end{tabular}
\end{center}

\begin{figure}[t]
	\centering
	\includegraphics[width=8.3cm]{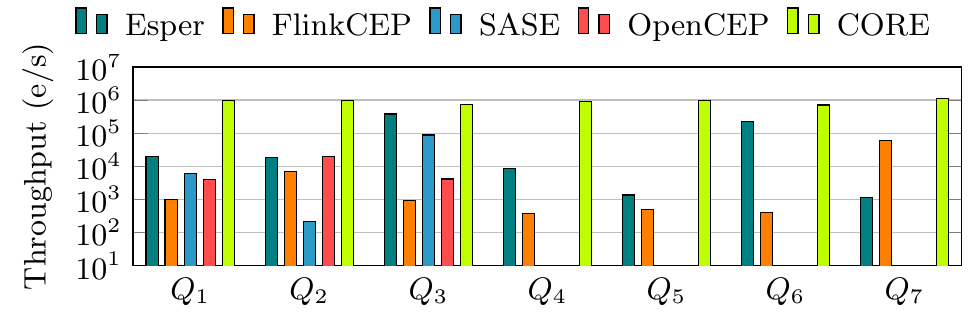}
	\vspace{-3mm}
	\caption{Throughput of evaluating queries $Q_1$ to $Q_7$, which include non-sequence operators, over stock market data.}
	\label{plot:stock-queries}
	\vspace{-3mm}
\end{figure} 

\newtext{SASE and OpenCEP do not support disjunction, and  we hence omit $Q_4$--$Q_7$ for them. Queries $Q_3$ and $Q_6$ use the partition-by clause. Unfortunately, every system gave different outputs when we tried partition-by queries. Therefore, for $Q_3$ and $Q_6$ we cannot guarantee query equivalence for all systems. 
In all other cases, the results provided by each system are the same.  }

\newtext{In Figure~\ref{plot:stock-queries} we show the throughput (log-scale),
  grouped per query. The results confirm our observations from the previous
  experiments. 
  In particular, \systemname's throughput is stable over all queries (i.e.,
  $10^6$ e/s), in contrast to the baselines, which are not stable. In particular
  for every baseline system there is at least one query where \systemname' exhibits 2
  \oom or more higher throughput.}

\newtext{$Q_2$ and $Q_5$ add filter clauses to $Q_1$ and $Q_4$ respectively. If we
  contrast system performance on $Q_2$ and $Q_5$ with that on $Q_1$ and $Q_4$,
  respectively, then we see that adding such filters reduces performance of some baselines, most notably SASE and Esper. CORE does not suffer from these problems due to its evaluation algorithm.}

\newtext{$Q_3$ and $Q_6$ add a partition-by clause to $Q_1$ and $Q_4$, respectively.
  If we contrast system performance on $Q_3$ and $Q_6$ with that on $Q_1$ and
  $Q_4$, respectively, then we see that partition-by aids the performance of
  systems like Esper and SASE but slightly decreases the throughput of CORE. This is because \systemname~ evaluate the partition-by clause by running several instances of the main algorithm, one for each partition, which slightly diminishes the throughput. Nevertheless,  CORE still outperforms the baselines.}


\smallskip
\noindent \textbf{Limitations of CORE.} \newtext{As the previous experiments show, \systemname\xspace outperforms other systems by several orders of magnitudes on different query and data workloads. We finish this section by a discussion of the limitations of \systemname.}

\newtext{In particular, we see two limitations compared to other approaches. First, \systemname's asymptotically optimal performance comes from representing partial matches succinctly in a compact structure,  and enumerating complex events from this structure whenever a pattern occurrence is found. The enumeration is with output-linear delay, meaning that the time spent to enumerate complex event $C$ is asymptotically $\mathcal{O}(|C|)$, i.e., linear in $C$. This asymptotic analysis hides a constant factor. While our experiments show that this constant is negligible in practice, enumeration from the data structure may take longer than directly fetching the complex event from an uncompressed representation, as most other systems do. This difference may be important in situations where a user wants to access outputs multiple times. }

\newtext{Second, \systemname\xspace compiles \systemql\xspace queries into
  non-deterministic CEA, which need to be (on-the-fly) determinized before
  execution. In the worst case, this determinized CEA can be of size exponential
  in the size of the query. Since the time needed to process a new event tuple,
  while constant in data complexity, does depends on the CEA size
  (Section~\ref{sec:evaluation:algo}), this may also affect processing
  performance. While we did not observe this blowup on the queries in this
  section, \systemql\xspace queries with complicated nesting of iteration and
  disjunction can theoretically exhibit such behavior. Unfortunately, no
  baseline system (except \systemname) currently allows such nesting, and we are
  therefore unable to experimentally compare \systemname\xspace in such a
  setting. 
}


%% file: sections/conclusions.tex
We introduced CORE, a CER system whose evaluation algorithm guarantees constant time per event, followed by output-linear delay enumeration. To the best of our knowledge, CORE is the first system that supports both guarantees. We showed through experiments that the evaluation algorithm provides stable performance for CORE, which is not affected by the size of the stream, query, or time window. This property means a throughput up to five orders of magnitudes higher than leading systems in the area. 

CORE provides a novel query evaluation approach; however, there is space for several improvements. A natural problem is to extend CEQL to allow time windows or partition-by operators inside the \textsf{WHERE} clause, which will increase the expressive power of CEQL. We currently do not know how to extend the evaluation algorithm for such queries while maintaining the performance guarantees. Other relevant features to include in CORE are aggregation, integration of non-event data sources, or the algorithm's parallelization, among others, which we leave as future work.


%% file: appendix/ceql-semantics.tex

\subsection{Proof of Theorem~\ref{theo:CEL-to-CEA}}
\label{sec:proof:theo:CEL-to-CEA}

\theoceltocea*

\begin{proof}
	This result follows almost directly from the proof for Theorem 6.2 in~\cite{CELJOURNAL}, which states that for every (unary) CEL formula $\varphi$ there is an equivalent CEA $\cA$.
	However, there are certain subtleties that need to be addressed so that this result translates into ours.
	For completeness, we present here the construction of~\cite{CELJOURNAL}, modified to fit our setting with intervals.
	However, it is worth noting that the only three differences in the construction are in the cases $\varphi = R$, $\varphi = \psi_1 \sq \psi_2$ and $\varphi = \psi \ks$, in order to handle the fact that now a run may start at an arbitrary position $i$ in the stream.
	
	In~\cite{CELJOURNAL}, they use an extended model of CEA, which replaces the $\amark, \umark$ marks in the transitions with subsets of variables.
	Formally, a \emph{valuation complex event automaton} (VCEA) is a tuple $\cA = (Q,\Delta,I,F)$, where $Q$ is the set of states, $I,F \subseteq Q$ are the set of initial and final states, respectivelly, and $\Delta$ is the transition relation $\Delta \subseteq Q \times \pset \times 2^\xset \times Q$.
	The definitions of run and accepting run remain the same as for CEA, i.e., a run is a  sequence
	$
	\rho \ := \  q_{i} \, \trans{P_{i}/L_i} \, q_{i+1} \, \trans{P_{i+1}/L_{i+1}} \, \ldots \, \trans{P_j/L_j} \, q_{j+1}
	$
	such that $q_i \in I$ and for every $k \in [i,j]$  it holds that $q_{k} \trans{P_k/L_k} q_{k+1} \in \Delta$ and $t_k \models P_k$; and $\rho$ is \emph{accepting} if $q_{j+1} \in F$.
	Now $\rho$ defines a valuation $V_{\rho} \ := \ ([i,j], \mu)$, where $\mu(X) = \{k \mid X \in L_k\}$, and we define the valuation semantics of $\cA$ over a stream $S$ as 
	$
	\semaux{\cA}(S) := \{V_\rho \mid \text{$\rho$ is an accepting run of $\cA$ over $S$} \}.
	$
	Now, we say that a CEL formula $\varphi$ is equivalent to a VCEA if $\semaux{\varphi}(S) = \semaux{\cA}(S)$ for every stream $S$.
	
	Consider any CEL formula $\varphi$.
	We can construct a VCEA $\cA_\varphi$ that is equivalent to $\varphi$ by induction on $\varphi$, as follows.
	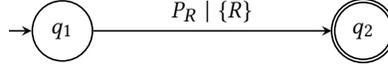
\begin{figure}
		\begin{center}
			\begin{tikzpicture}[->,>=stealth, semithick, auto, initial text= {}, initial distance= {3mm}, accepting distance= {4mm}, node distance=4cm, semithick]
				\tikzstyle{every state}=[draw=black,text=black,inner sep=0pt, minimum size=8mm]
				\node[initial,state]	(1) 				{$q_1$};
				\node[accepting,state]	(2) [right of=1]	{$q_2$};
				\path
				(1)
				edge 				node {$P_R \mid \{R\}$}	(2);
			\end{tikzpicture}
			\caption{A VCEA for the atomic formula $R$. Here, $P_R = \{t \mid t(\type) = R\}$.}\label{fig:R_ucea}
		\end{center}
	\end{figure}
	\begin{itemize}
		\item If $\varphi = R$, then $\cA_\varphi$ is defined as depicted in Figure~\ref{fig:R_ucea}, i.e. $\cA_\varphi = (\{q_1,q_2\},\{(q_1,P_R,\{R\},q_2)\},\{q_1\},\{q_2\})$, where $P_R$ is the predicate containing all tuples with type $R$, namely, $P_R := \{t \mid t(\type) = R\}$.
		
		\item If $\varphi = \psi \IN X$, then $\cA_\varphi = (Q_\psi, \Delta_\varphi,I_\psi,F_\psi)$ where $\Delta_\varphi$ is the result of adding variable $X$ to all marking transitions of $\Delta_\psi$, i.e., $\Delta_\varphi = \{(p,P,L,q)\in\Delta_\psi \mid L = \emptyset\}\cup\{(p,P,L,q)\mid \exists L'\neq\emptyset \text{ such that } (p,P,L',q)\in \Delta_\psi \land L=L'\cup\{X\}\}$.
		
		\item If $\varphi = \psi \FILTER X\texttt{[}P\texttt{]}$ for some variable $X$ and predicate $P$, then $\cA_\varphi = (Q_\psi, \Delta_\varphi,I_\psi,F_\psi)$ where $\Delta_\varphi$ is defined as $\{(p,P',L,q)\in\Delta_\psi \mid X \notin L\}\cup\{(p,P \wedge P',L,q)\mid (p,P',L,q)\in \Delta_\psi \land X \in L\}$.
		
		\item If $\varphi = \psi_1 \cor \psi_2$, then $\cA_\varphi$ is the automata union between $\cA_{\psi_1}$ and $\cA_{\psi_2}$: $\cA_\varphi = (Q_{\psi_1} \cup Q_{\psi_2}, \Delta_{\psi_1} \cup \Delta_{\psi_2}, I_{\psi_1} \cup I_{\psi_2}, F_{\psi_1} \cup F_{\psi_2})$. Here, we assume w.l.o.g. that $\cA_{\psi_1}$ and $\cA_{\psi_2}$ have disjoint sets of states.
		
		\item If $\varphi = \psi_1 \sq \psi_2$, then $\cA_\varphi = (Q_{\psi_1} \cup Q_{\psi_2}, \Delta_\varphi, I_{\psi_1}, F_{\psi_2})$ where $\Delta_\varphi = \Delta_{\psi_1} \cup \Delta_{\psi_2} \cup \{(p,\TRUE,\emptyset,p) \mid p \in I_{\psi_2}\} \cup \{(p,P,L,q) \mid q \in I_{\psi_2} \land \exists q' \in F_{\psi_1}.(p,P,L,q') \in \Delta_{\psi_1}\}$. Here, we assume w.l.o.g. that $\cA_{\psi_1}$ and $\cA_{\psi_2}$ have disjoint sets of states.
		
		\item If $\varphi = \psi \ks$, then $\cA_\varphi = (Q_\psi \cup \{q\}, \Delta_\varphi, I_\psi, F_\psi)$ where $q$ is a fresh state, $\Delta_\varphi = \Delta_\psi \cup \{(p,P,L,q) \mid \exists q' \in F_\psi.(p,P,L,q') \in \Delta_\psi\} \cup \{(q,P,L,p) \mid \exists q' \in I_\psi. (q',P,L,p) \in \Delta_\psi\}$.
		
		\item If $\varphi = \pi_L(\psi)$ for some $L \subseteq \lset$, then $\cA_\varphi = (Q_\psi, \Delta_\varphi,I_\psi,F_\psi)$ where $\Delta_\varphi$ is the result of intersecting the labels of each transition in $\Delta_\psi$ with $L$.
		Formally, that is $\Delta_\varphi = \{(p,P,L \cap L' ,q) \mid (p,P,L',q) \in \Delta_\psi\}$.
	\end{itemize}

	Note that in the VCEA model we replaced $q_0$ with a subset of states $I$.
	This is not a problem, since one can modify the final automaton to have only one initial state by adding a fresh state $q_0$ and replicating every transition coming from an initial state to go from $q_0$, i.e., adding transitions $\{(q_0,P,L,q) \mid \exists p \in I. (p,P,L,q) \in \Delta\}$.
	This also assures that $q_0$ has no incoming transitions.
	From now on, we assume that $\cA_\varphi$ has only one initial state $q_\varphi$.
	
	Now that we have a VCEA $\cA_\varphi = (Q_\varphi, \Delta_\varphi,q_\varphi,F_\varphi)$ that is equivalent to $\varphi$, we can construct a CEA $\cA$ by changing the sets in the transitions with $\amark, \umark$ marks.
	More precisely, we define $\cA = (Q_\varphi,\Delta,q_\varphi,F_\varphi)$ where $\Delta = \{(p,P,\amark,q) \mid (p,P,L,q) \in \Delta_\varphi \land L \neq \emptyset\} \cup \{(p,P,\umark,q) \mid (p,P,L,q) \in \Delta_\varphi \land L = \emptyset\}$.
	One can verify that this CEA satisfies the theorem.
\end{proof}


%% file: appendix/algorithm.tex

\subsection{Proof of Theorem~\ref{theo:time-ordered}}
\label{sec:proof:theo:time-ordered}

\theotimeordered*

\begin{proof}
	Consider a $k$-bounded and time-ordered tECD $\ecs$, a duplicate-free node $\nnode$ of $\ecs$, a time-window bound $\epsilon$ and a position $j$.
	We provide Algorithm~\ref{alg:enumeration} and show that: (1) it enumerates the set $\outce{\nnode}^\epsilon(j)$, and (2) it does so with output-linear delay.
	To simplify presentation in what follows, we denote the sets of bottom, output and union nodes by $\bnodes$, $\onodes$ and $\unodes$, respectively.
	
	\begin{algorithm*}
		\caption{Enumeration of $\outce{\nnode}^\epsilon(j)$.}\label{alg:enumeration}
		\begin{algorithmic}[1]
			\Procedure{Enumerate}{$\ecs, \nnode,\epsilon, j$}
				\State $\stack \gets \sinit()$
				\State $\tau \gets j - \epsilon$
				\If{$\maxstart(\nnode) \geq \tau$}
					\State $\push(\stack, (\nnode,\emptyset))$
				\EndIf
				\While{$(\nnode', P) \gets \pop(\stack)$}\label{line:enumeration-while1}
					\While{$\texttt{true}$}\label{line:enumeration-while2}
						\If{$\nnode' \in \bnodes$}\label{line:enumeration-bnodeif}
							\State $\texttt{output}([\opos(\nnode'),j], P)$\label{line:enumeration-output}
							\State $\textbf{break}$\label{line:enumeration-break}
						\ElsIf{$\nnode' \in \onodes$}\label{line:enumeration-onodeif}
							\State $P \gets P \cup \{\opos(\nnode')\}$
							\State $\nnode' \gets \onext(\nnode')$
						\ElsIf{$\nnode' \in \unodes$}\label{line:enumeration-unodeif}
							\If{$\maxstart(\uright(\nnode')) \geq \tau$}
								\State $\push(\stack,(\uright(\nnode'),P))$
							\EndIf
							\State $\nnode' \gets \uleft(\nnode')$
						\EndIf
					\EndWhile
				\EndWhile
			\EndProcedure
		\end{algorithmic}
	\end{algorithm*}

	\newcommand\tinyswarrow{\vcenter{\hbox{\scalebox{0.5}{$\swarrow$}}}}
	\newcommand{\leftmost}{\operatorname{\pi_{\tinyswarrow}}}
	\newcommand{\paths}{\operatorname{paths}}
	
	Algorithm~\ref{alg:enumeration} uses a stack $\stack$ with the typical stack operations: $\sinit()$ to initialize an empty stack, $\push(\stack, e)$ to add an element $e$ at the beginning of $\stack$, and $\pop(\stack)$ to get the first element in $\stack$.
	Moreover, when $\stack$ is empty, $e \gets \pop(\stack)$ is interpreted as \textsf{false}.
	We assume that each of the previous operations can be performed in constant time.
	
	Recall that $\ecs$ encodes a directed acyclic graph (DAG) $G_\ecs = (\nnodes, E)$ where $\nnodes$ are the vertices, and edges goes from any union node $\unode$ to $\uleft(\unode)$ and $\uright(\unode)$, and from any output node $\onode$ to $\onext(\onode)$.
	For every node $\nnode' \in \nnodes$ reachable from $\nnode$ and a threshold $\tau\geq 0$, let $\paths_{\geq \tau}(\nnode')$ be all paths of $G_\ecs$ that start at $\nnode'$ and end at some bottom node $\bnode$ with $\opos(\bnode) \geq \tau$.
	It is clear that there exists a bijection between $\outce{\nnode'}^\epsilon(j)$ and $\paths_{\geq j - \epsilon}(\nnode')$, namely, for every complex event within a time window of size $\epsilon$ there exists exactly one path that reaches a bottom node $\bnode$ with $\opos(\bnode) \geq j - \epsilon$ (this is true because, since $\nnode$ is duplicate-free, so must be $\nnode'$).
	Then, Algorithm~\ref{alg:enumeration} receives as input a tECD $\ecs$, a node $\nnode$, a time-window bound $\epsilon$ and a position$j$, and traverses $G_\ecs$ in a DFS manner and following the left-to-right order, i.e., for every union node $\unode$, the paths of $\uleft(\unode)$ are traversed before the ones of $\uright(\unode)$.
	Each iteration of the \textbf{while} of line~\ref{line:enumeration-while1} traverses a new path starting from the point it branches from the previous path.
	For this, the stack $\stack$ is used to store the node and partial complex event of that branching point.
	Then, the \textbf{while} of line~\ref{line:enumeration-while2} traverses through the nodes of the next path, following the left direction whenever a union node is reached and adding the right node to the stack whenever needed.
	Moreover, by checking for every node $\nnode'$ its value $\maxstart(\nnode')$ before adding it to the stack, it makes sure of only going through paths in $\paths_{\geq j - \epsilon}(\nnode')$.
	
	A simpler recursive algorithm could compute the same results.
	However, if done in a naive way, the output-linear delay might not be assured because when going back from a path to compute the next one, the number of backtracking steps might be as long as the longest path of $G_\ecs$.
	To avoid this, Algorithm~\ref{alg:enumeration} uses the stack $\stack$ which allows it to jump immediately to the node that continues the next path. Here we assume that storing $P$ in the stack takes constant time. We can realize this assumption by representing $P$ as a linked list of positions, where the list is ordered by the last element added and there is a link to the previous element. Then for storing $P$ we can save in $\stack$ a pointer to the last element added.
	
	We first prove that Algorithm~\ref{alg:enumeration} enumerates the set $\outce{\nnode}^\epsilon(j)$, provided that $\ecs$ is time-ordered and $\nnode$ a duplicate-free node.
	\begin{lemma}\label{lem:enumeration-correctness}
		Let $\ecs$ be a time-ordered tECD, $\nnode$ a duplicate-free node of $\ecs$ and $\epsilon$ a time-window.
		Then, Algorithm~\ref{alg:enumeration} enumerates $\outce{\nnode}^\epsilon(j)$ without duplicates.
	\end{lemma}
	\begin{proof}
		For the rest of this proof fix a tECD $\ecs$ and a threshold $\tau = j-\epsilon$.
		For every node $\mnode$ of tECD, let $l_\mnode$ be the longest path from $\mnode$ to a bottom node.
		We say a pair $(\mnode,R)$ of a node and a set of positions is \emph{visited} if at some iteration of the \textbf{while} in line~\ref{line:enumeration-while2}, variables $\nnode'$ and $P$ take the values $\mnode$ and $R$, respectively.
		Moreover, we say $(\mnode,R)$ is \emph{left} when the pair $(\mnode',R')$ is popped from the stack, where $(\mnode',R')$ is the top of the stack when $(\mnode,R)$ was visited (or when the algorithm ends, if the stack was empty).
		The \emph{enumeration period} of $(\mnode,R)$ is the period between it being visited and it being left.
		Intuitively, in the enumeration period of $(\mnode,R)$ is when the subtree rooted at $\mnode$ is enumerated.

		Now, we prove that for every visited pair $(\mnode,R)$, the set $[\mnode](R,j) = \{([i,j],D \cup R) \mid (i,D) \in \outce{\mnode} \land i \geq \tau \}$ is output in its enumeration period without duplicates.
		If this is the case, then it is clear that Lemma~\ref{lem:enumeration-correctness} holds, since, upon calling $\textproc{Enumerate}(\ecs, \nnode, \epsilon, j)$ the pair $(\nnode,\emptyset)$ is the first one visited, and the enumerated set corresponds to $[\nnode](\emptyset,j) = \{([i,j],D) \mid (i,D) \in \outce{\nnode} \land i \geq \tau\} = \outce{\nnode}^\epsilon(j)$.
		
		We prove the above by strong induction over $l_\mnode$.
		The base case is when $l_\mnode = 0$, which means that $\mnode$ must be itself a bottom node.
		In that case, it enters in the \textbf{if} of line~\ref{line:enumeration-bnodeif}, outputs the corresponding complex event $([\opos(\mnode),j],R)$, clearly without duplicates.
		Note that $\outce{\mnode} = \{\opos(\mnode),\emptyset\}$ and that $\maxstart(\mnode) \geq \tau$, so $\opos(\mnode) \geq \tau$ and the statement holds.
		
		Now for the inductive step, consider $l_\mnode > 0$, which means that $\mnode$ is either an output node or a union node.
		In the former case, after entering the \textbf{if} of line~\ref{line:enumeration-onodeif}, $P$ and $\nnode'$ are updated to $R \cup \{\opos(\mnode)\}$ and $\onext(\mnode)$, respectively, and the new iteration of the \textbf{while} begins with those values.
		Then, since $l_{\onext(\mnode)} = l_\mnode - 1$, by induction hypothesis the set $[\onext(\mnode)](R\cup\{\opos(\mnode)\},j)$ is enumerated, but by definition $\outce{\mnode} = \{ (i, D' \cup \{\opos(\mnode)\}) \mid (i, D') \in \outce{\onext(\mnode)} \}$, so
		\[
		[\onext(\mnode)](R\cup\{\opos(\mnode)\},j) \; = \; \{([i,j],D' \cup R\cup\{\opos(\mnode)\}) \mid (i,D') \in \outce{\onext(\mnode)} \land i \geq \tau \} \; = \; [\mnode](R,j).
		\]
		Therefore, the enumerated set corresponds to $[\mnode](R,j)$.
		Also, by induction hypothesis this set is enumerated without duplicates and the statement holds.
		
		If $\mnode$ is a union node, the \textbf{if} of line~\ref{line:enumeration-unodeif} is entered, which first pushes the pair $(\uright(\mnode),R)$ if $\maxstart(\uright(\mnode)) \geq \tau$ and then updates $\nnode'$ to $\uleft(\mnode)$ before continuing with the next iteration.
		Since $l_{\uleft(\mnode)}$ and $l_{\uright(\mnode)}$ are not greater than $l_\mnode -1$, by induction hypothesis, both sets $[\uleft(\mnode)](R,j)$, $[\uright(\mnode)](R,j)$ are enumerated without duplicates (note that after enumerating $[\uleft(\mnode)](R,j)$, the pair $(\uright(\mnode),R)$ is popped).
		Because $\outce{\unode} = \outce{\uleft(\unode)} \uplus \outce{\uright(\unode)}$, then this enumerated set corresponds to $[\uleft(\mnode)](R,j) \cup [\uright(\mnode)](R,j) = [\mnode](R,j)$.
		Also, since $\nnode'$ is duplicate-free, this set is enumerated without duplicates and the statement holds.
		
		It has been proved that the set $\outce{\nnode}^\epsilon(j)$ is enumerated without duplicates.
	\end{proof}

	Now that the correctness of the algorithm has been proved, we proceed to show that the enumeration is performed with output-linear delay.

	\begin{lemma}\label{lem:enumeration-delay}
		Fix $k$. Let $\ecs$ be a $k$-bounded and time-ordered tECD, $\nnode$ a node of $\ecs$ and $\epsilon$ a time-window.
		Then, Algorithm~\ref{alg:enumeration} enumerates $\outce{\nnode}^\epsilon(j)$ with output-linear delay.
	\end{lemma}

	\begin{proof}
		Fix $\ecs$ and $\tau$.
		We showed that Algorithm~\ref{alg:enumeration} traverses all paths of $\paths_{\geq \tau}(\nnode)$.
		Moreover, the order in which paths are traversed is completely determined by the order of the union nodes: for each union node $\unode$, the paths to its left are traversed first, and then the ones to its right.
		Formally, for every node $\nnode'$ define the leftmost path from $\nnode'$ as $\leftmost(\nnode') := \nnode_0 \rightarrow \nnode_1 \rightarrow \ldots \rightarrow \nnode_l$ such that $\nnode_0 = \nnode'$ and, for every $i \leq l$:
		\begin{itemize}
			\item if $\nnode_i \in \bnodes$, then $i = l$,
			\item if $\nnode_i \in \onodes$, then $\nnode_{i+1} = \onext(\nnode_i)$, and
			\item if $\nnode_i \in \unodes$, then $\nnode_{i+1} = \uleft(\nnode_i)$.
		\end{itemize}
		Consider a path $\pi := \nnode_0 \rightarrow \nnode_1 \rightarrow \ldots \rightarrow \nnode_l$, and let $j \leq l$ be the last position such that $\nnode_j$ is a union node, $\nnode_{j+1} = \uleft(\nnode_j)$ and $\maxstart(\uright(\nnode_j)) \geq \tau$.
		Then, let $\pi^u$ be the path $\pi$ up to position $j$, i.e., that stops at such union node.
	
		Let $P = \{\pi_1, \pi_2, \ldots, \pi_m\}$ be the set of paths enumerated by Algorithm~\ref{alg:enumeration} in that order.
		Then, by analysing Algorithm~\ref{alg:enumeration}, one can see that $\pi_1 = \leftmost(\nnode)$ and, for every $i \leq m$, $\pi_i = \pi_{i-1}^u \cdot \leftmost(\uright(\unode))$, where $\unode$ is the last node of $\pi_{i-1}^u$.
		In other words, it performs a greedy DFS from left to right: the first path to enumerate is $\pi_1 = \leftmost(\nnode)$, then each $\pi_i$ is the path of $\paths_{\geq \tau}(\nnode)$ that branches from $\pi_{i-1}$ to the right at the deepest level ($\unode$) and from there follows the leftmost path.
		Moreover, to jump from $\pi_{i-1}$ to $\pi_i$, the node popped by the stack is exactly $\unode$, that is, the last node of $\pi_{i-1}^u$.
		
		To show that the enumeration is done with output-linear delay, we study how long it takes between enumerating the complex events of $\pi_{i-1}$ and $\pi_i$.
		Consider that $\pi_{i-1}$ was just traversed and its complex event was output by line~\ref{line:enumeration-output}.
		Then, the \textbf{break} of line~\ref{line:enumeration-break} is executed, breaking the \textbf{while} of line~\ref{line:enumeration-while2}.
		Afterwards, either the stack is empty and the algorithm ends, or a pair $(\mnode,R)$ is popped from the stack, where $\mnode$ corresponds to the last node of $\pi_{i-1}^u$.
		From that point, it is straightforward to see that the number of iterations of the while of line~\ref{line:enumeration-while2} (each taking constant time) is equal to the number of nodes $l$ in $\leftmost(\mnode)$, so those nodes are traversed and the complex event of the path $\pi_i$ is output.
		But, because $\ecs$ is $k$-bounded, then $l \leq k \cdot |O|$, where $O$ is the complex event of $\pi_i$.
		Finally, the time taken is bounded by the size of the output, and the enumeration is performed with output-linear delay.
	\end{proof}

	By Lemmas~\ref{lem:enumeration-correctness} and ~\ref{lem:enumeration-delay}, Theorem~\ref{theo:time-ordered} immediately holds.
\end{proof}

\subsection{Proof of Theorem~\ref{theo:algo-correct}}
\label{sec:proof:theo:algo-correct}

\theoalgocorrect*

\begin{proof}
	Fix a CEA $\cA = (Q, \Delta, q_0, F)$ and a stream $S = t_0 t_2 \ldots$.
	Given a run of $\cA$ over $S$ of the form $\rho:= q_i \trans{P_i/m_i}q_{i+1} \trans{P_{i+1}/m_{i+1}} \ldots \trans{P_j/m_j} q_{j+1}$, let $o(\rho) = (i,D)$ where $D = \{k \mid i \leq k \leq j \land m_k = \amark\}$ be the open complex event associated to $\rho$.
	For any position $j$ and a state $q \in Q$, let $R_j^q$ be the set of runs that reach $q$ after reading position $j$ of the stream, that is, $R_j^q = \{\rho \mid \exists i.\, \text{$\rho$ is a run of $\cA$ over $S$ from positions $i$ to $j$} \}$.
	Further, let $CE_j^q = \{o(\rho) \mid \rho \in R_j^q\}$ be the set of corresponding open complex events.
	
	For a union-list $\ulist = \nnode_0, \nnode_1, \ldots, \nnode_k$, of a tECD $\ecs$ define its set of open complex events as expected: $o(\ulist) = \outce{\nnode_0} \cup \ldots \cup \outce{\nnode_k}$.
	Then, for a position $j$ and a state $q \in Q$ let $ACE_j^q = o(\ctable[q])$ be the set of open complex events stored by $\ctable[q]$ after reading position $j$ of the stream, i.e., after the $(j+1)$-th iteration of the \textbf{while} of line~\ref{line:evaluation-while} of Algorithm~\ref{alg:evaluation}.
	
	Note that $\sem{\cA}^\epsilon_j(S) = \{([i,j],D) \mid (i,D) \in \bigcup_{q \in F}CE_j^q \land j-i \leq \epsilon \}$.
	One can see that in Algorithm~\ref{alg:evaluation}, if after the $j$-th iteration the underlying tECS $\ecs$ is time-ordered and $k$-bounded for some constant $k$ and all nodes of $\ecs$ are duplicate-free, then by Theorem~\ref{theo:time-ordered} procedure $\textproc{Output}$ enumerates the set $\{([i,j],D) \mid (i,D) \in \bigcup_{q \in F}ACE_j^q \land j-i \leq \epsilon \}$ with output-linear delay.
	Therefore, if we prove that, for every $j$ and $q$, (1) $CE_j^q = ACE_j^q$, (2) $\ecs$ is time-ordered and $k$-bounded and (3) all nodes in $\ecs$ are duplicate-free, then the theorem holds.
	
	Before we start proving (1)-(3), let us first analyse procedure $\textproc{ExecTrans}$.
	At the beginning of iteration $j$, every $ACE^q_j$ starts empty.
	Then, every time $\textproc{ExecTrans}(p,\ulist,t,j)$ is called, it takes all open complex events in $ACE^p_{j-1}$ and:
	\begin{itemize}
		\item for $q_\umark = \Delta(p,t,\umark)$, it makes $ACE^{q_\umark}_j \gets ACE^{q_\umark}_j \cup ACE^p_{j-1}$; and
		\item for $q_\amark = \Delta(p,t,\amark)$, it makes $ACE^{q_\amark}_j \gets ACE^{q_\amark}_j \cup (ACE^p_{j-1} \times \{j\})$, where $ACE^p_{j-1} \times \{j\} = \{(i,D \cup \{j\}) \mid (i,D) \in ACE^p_{j-1} \}$.
	\end{itemize}
	The two cases in $\textproc{Add}$ just handle whether $ACE^{q_\amark}_j$ or $ACE^{q_\umark}_j$ are empty at the moment or not.
	
	We will prove (1) by simple induction over $j$.
	For the base case, consider the state before the first iteration, namely $j = -1$ and $\ctable = \emptyset$.
	It is easy to see that $CE_{-1}^q = ACE_{-1}^q = \emptyset$, so the property holds.
	Now for the inductive case, consider that $CE_j^q = ACE_j^q$.
	First, a new list with only a bottom node is created and stored in $\ulist$ and then $\textproc{ExecTrans}(q_0,\ulist,t,j)$ is called.
	This starts possibly two runs from the initial state $q_0$ with either a $\umark$ or $\amark$-transition.
	Then, $\textproc{ExecTrans}(q_0,\ctable[p],t,j)$ is called for every state that is active before reading position $j$, i.e., for every $p$ with $ACE_j^p \neq \emptyset$.
	Let us consider any $(i,D) \in ACE_{j+1}^q$.
	If it came from the first call to $\textproc{ExecTrans}$, then it means that $i = j$ and, either $D = \{j\}$ and there is a run (from $j+1$ to $j+1$) of the form $\rho = q_0 \trans{P/\amark}q$, or $D = \emptyset$ and there is a run (from $j+1$ to $j+1$) $\rho = q_0 \trans{P/\umark}q$.
	Either way, the corresponding run assures that $(i,D) \in CE_{j+1}^q$.
	Now, if $(i,D)$ came from one of the subsequent calls to $\textproc{ExecTrans}$, it means that there was some $(i,D') \in ACE_j^p$ such that either $D = D'$ and $q = \Delta(p,t_j,\umark)$, or $D = D' \cup \{j\}$ and $q = \Delta(p,t_j,\amark)$.
	Call such transition $e$.
	By induction hypothesis, there is a run from $i$ to $j$ $\rho = q_i \trans{P_i/m_i} q_{i+1} \trans{P_{i+1}/m_{i+1}} \ldots \trans{P_j/m_j} p$ with $o(\rho) = (i,D')$.
	Therefore, there is a run $\rho'$ that extends $\rho$ with $e$ such that $o(\rho') = (i,D)$ and $\rho' \in R^q_{j+1}$, therefore $(i,D) \in ACE^q_{j+1}$.
	Then, $CE^q_{j+1} \subseteq ACE^q_{j+1}$.
	We can use the same analysis to prove that $ACE^q_{j+1} \subseteq CE^q_{j+1}$, thus $CE^q_{j+1} = ACE^q_{j+1}$ and the induction holds.
	
	Now, we focus on proving (2).
	It has already been shown that every operation over tECS $\ecs$ used by Algorithm~\ref{alg:evaluation} preserves the time-ordered and $k$-boundedness for $k = 3$, provided that every time we call $\dunion(\nnode_1,\nnode_2)$, both $\nnode_1$ and $\nnode_2$ are safe nodes.
	In particular, since we only manage unions using union-lists, we only need to check that every time $\linit(\nnode)$ or $\linsert(\ulist, \nnode)$ are called, the input node $\nnode$ is a safe one.
	This is easily verifiably by looking at lines \ref{line:evaluation-linit1}, \ref{line:evaluation-linit2} and \ref{line:evaluation-linsert} of Algorithm~\ref{alg:evaluation} (for \ref{line:evaluation-linit2} and \ref{line:evaluation-linsert}, recall that $\lmerge(\ulist)$ returns a safe node).
	Note that $\lmerge(\ulist)$ does not have any requirement, since its correct behaviour is ensured by using the other operations correctly.
	Therefore, at any moment in the evaluation, the underlying tECS $\ecs$ is time-ordered and 3-bounded.
	Note that $\linsert(\ulist,\nnode)$ has another precondition: that $\maxstart(\nnode_0) \geq \maxstart(\nnode)$, where $\nnode_0$ is the first node of $\ulist$.
	Since this does not relate directly with the correctness of the output of Algorithm~\ref{alg:evaluation}, but with an overall correct behaviour of the algorithm, this property is discussed in the next section, along with another property that is needed to ensure constant update-time.
	
	Next, we prove (3).
	By taking a look at Algorithm~\ref{alg:evaluation}, one can see that the only ways it could add to $\ecs$ nodes with duplicates are with methods $\lmerge$ and $\linsert$, particularly in lines~\ref{line:evaluation-lmerge1}, \ref{line:evaluation-linsert} and \ref{line:evaluation-lmerge2}.
	Fix a position $j$.
	Let $\operatorname{All-N}$ be the set all nodes in the union-list of some state at position $j$, i.e., $\operatorname{All-N} = \{\nnode \mid \exists q. \exists i. T[q] = \nnode_0,\ldots,\nnode_l \land \nnode_i = \nnode\}$, where $T[q]$ is its value at iteration $j$.
	If, for every two $\nnode_1,\nnode_2 \in \operatorname{All-N}$ with $\nnode_1 \neq \nnode_2$ it holds that $\outce{\nnode_i} \cap \outce{\nnode_j} = \emptyset$, then all cases are solved and the algorithm creates no nodes with duplicates.
	One can see that each $\outce{\nnode}$ actually stores the open complex events for a disjoint set of runs.
	Then, if there were some $\nnode_1,\nnode_2 \in \operatorname{All-N}$ with $\outce{\nnode_1} \cap \outce{\nnode_2} \neq \emptyset$, it would mean that there are two different runs defining the same output, thus contradicting the fact that $\cA$ is I/O-deterministic.
	Therefore, all nodes must define disjoint sets and Algorithm~\ref{alg:evaluation} only adds duplicate-free nodes to $\ecs$.
	
	Finally, because (1)-(3) hold, also does the theorem.
\end{proof}
\subsection{Special invariants of union-lists in Algorithm~\ref{alg:evaluation}}

In this section we discuss in more detail two invariants maintained by Algorithm~\ref{alg:evaluation} which are needed for its correct behaviour.
Specifically, we prove that: 
\begin{enumerate}
	\item[(1)] every time a node $\nnode$ is added to some union-list $\ulist = \nnode_0, \ldots, \nnode_k$, it is the case that $\maxstart(\nnode_0) \geq \maxstart(\nnode)$; and 
	\item[(2)] at any moment, the length of every union-list $\ctable[q]$ is smaller than the number of states of the automaton $\cA$.
\end{enumerate}
The former assures that the first node $\nnode_0$ always remains a non-union node, while the latter is needed so that operations $\lmerge$ and $\linsert$ take constant time (when $\cA$ is fixed).

First, we focus on (1).
This comes directly from the order in which we update the union-lists.
As notation, for a non-empty union-list $\ulist = \nnode_0, \ldots, \nnode_k$, we write $\maxstart(\ulist) = \maxstart(\nnode_0)$.
Consider any iteration $j \geq 0$ and let $q_1,\ldots,q_l$ be the order given by $\cordkeys(\ctable)$.
One can prove that this order actually preserves the following property: $\maxstart(\ctable[q_i]) \geq \maxstart(\ctable[q_{i+1}])$ for every $i < l$.
One can prove this by simple induction over $j$.
As consequence, if $\textproc{ExecTrans}(q_i,\ctable[q_i],t,j)$ is called and we need to add some node $\nnode$ at some union-list $\ulist$, it must be that $\ulist$ was the result of some previous call $\textproc{ExecTrans}(q_k,\ctable[q_k],t,j)$ with $k < i$, and so $\maxstart(\ulist) \geq \maxstart(\ctable[q_k]) \geq \maxstart(\nnode)$.

Now we prove (2).
First, we show a property about the possible maximum-start values at any iteration of the algorithm.
For $j \geq 0$, we define the set of maximum-starts at iteration $j$ as $M_j = \{ \maxstart(\ctable[q]) \mid q \in \ckeys(\ctable) \}$.
The property we use is that the set $M_j$ is closely restricted by the set $M_{j-1}$ of the previous iteration.
Specifically, one can see that $M_j$ must always be a subset of $M_{j-1}$, with the possibility of adding the maximum-start $j$ if a new run begins at that iteration, i.e., $M_j \subseteq (M_{j-1} \cup \{j\})$.
This is because the maximum-start of every union-list $\ctable[q]$ at some iteration came either from extending the already existing runs of the previous iteration, therefore maintaining its maximum-start, or from starting a new run with maximum-start $j$.

Consider any union-list $\ctable[q]$ created at iteration $j$.
From (1), we know that we will only add to $\ctable[q]$ nodes $\nnode$ with $\maxstart(\nnode) \leq \maxstart(\ctable[q])$.
Moreover, it is always the case that $\maxstart(\nnode) \in M_j$.
This is clear for iteration $j$, but it actually holds for the following ones.
Indeed, $\ctable[q]$ could actually receive insertions in some future iteration, call it $k$.
However, since $\maxstart(\ctable[q])$ does not change and the new maximum-start values in $M_k$ are higher than $j \geq \maxstart(\ctable[q])$, the actual maximum-start values that could be added to $\ctable[q]$ are still in $M_j \cap M_k \subseteq M_j$.
Therefore, the size of $\ctable[q]$ is bounded by $|M_j| \leq |Q|$.


%% file: appendix/app-other-competitors.tex
We have surveyed the literature for other systems to compare against, but have chosen to limit our experiments to SASE~\cite{SASE}, Esper~\cite{EsperTech}, FlinkCEP~\cite{FlinkCEP}, and OpenCEP~\cite{OpenCEP}, for the following reasons.

First, we cannot compare against systems that, while published in the
literature, do not publish an accompanying implementation (e.g.,
ZStream~\cite{mei2009zstream}, NextCEP~\cite{nextCEP}, DistCED~\cite{distCED},
SAP ESP~\cite{SAPesp}\footnote{SAP offers 30-day trial versions of its products,
  however, SAP ESP itself is no longer available from its website.}) or whose
implementation was discontinued in the last decade (e.g.,
Cayuga~\cite{cayuga}). Second, we cannot compare against proposals that do not
offer a CER query language or whose implementation does not allow running the
queries considered here (e.g.~CET~\cite{PoppeLAR17},
GRETA~\cite{PoppeLRM17}). We do not consider comparison against system whose
optimization focus is on additional CER
features~\cite{DBLP:conf/sigmod/0019ANNW21,hirzel2012partition,DBLP:conf/icde/ZhaoHW20},
such as load shedding~\cite{DBLP:conf/icde/ZhaoHW20} or interacting with
external, non-stream data sources~\cite{DBLP:conf/sigmod/0019ANNW21}, for which
the comparison would not be fair.  Finally, we do not compare against streaming
SQL systems like Spark and Flink SQL whose SQL-based query languages do not
natively support basic CER operators, like sequencing and iteration.

We \emph{have} run experiments to compare against TESLA/TRex
\cite{TESLA,Cugola:2012} and
Siddhi~\cite{suhothayan2011siddhi,Siddhi}. Unfortunately, their event
recognition features differ from \systemname's; in particular, they do not have
dedicated support for count-based time windows.\footnote{Although Siddhi has an
  operator for time windows, on can only use it in the presence of
  aggregation. As such, one cannot retrieve the complex events found, only
  aggregations thereof.} As such, they are not optimized for the use cases that
we aim to test here.  When we expressed simple sequence queries with time
windows by means of other features in their query languages (e.g., by using
inequality predicates), performance was so limited (i.e., a throughput of less
than ten events per second), that we decided not to further report the results
of TESLA/TRex and Siddhi.


%% file: appendix/app-seq-stocks.tex
\noindent \textbf{Dataset.} We reuse the stock market dataset from
\cite{stockmarket} also used in related
work~\cite{PoppeLMRR21,DBLP:conf/sigmod/PoppeLR019,PoppeLAR17,PoppeLRM17} to
evaluate the effectiveness of CER processing strategies. This dataset contains
approximately 225k transaction records of 10 stocks from NYSE in a single day,
listing stock name, transaction ID, volume, price, time, and transaction type
(SELL/BUY). We use the transaction type to divide the dataset into two streams: one stream of SELL events, and one steam of BUY events.

\medskip
\noindent \textbf{Queries.} The workload of sequence queries with output over the stock market dataset are the following queries of length $3$, $6$, $9$, $12$ and $24$, respectively:
	\begin{verbatim}
		S3 = SELECT * FROM S
		     WHERE (SELL as T1; BUY as T2; BUY as T3)
		     FILTER T1[name = 'INTC'] AND T2[name = 'RIMM'] AND T3[name = 'QQQ'] 
		     WITHIN 10000 [stock_time]
	
		S6 = SELECT * FROM S
		     WHERE (SELL as T1; BUY as T2; BUY as T3; 
		            SELL as T4; BUY as T5; BUY as T6)
		     FILTER T1[name = 'INTC'] AND T2[name = 'RIMM'] AND T3[name = 'QQQ'] AND
		            T4[name = 'IPIX'] AND T5[name = 'AMAT'] AND T6[name = 'CSCO']
		     WITHIN 10000 [stock_time]
		
		S9 = SELECT * FROM S
		     WHERE (SELL as T1; BUY as T2; BUY as T3;
		            SELL as T4; BUY as T5; BUY as T6;
		            SELL as T7; BUY as T8; BUY as T9)
		     FILTER T1[name = 'INTC'] AND T2[name = 'RIMM'] AND T3[name = 'QQQ'] AND
		            T4[name = 'IPIX'] AND T5[name = 'AMAT'] AND T6[name = 'CSCO'] AND
		            T7[name = 'YHOO'] AND T8[name = 'DELL'] AND T9[name = 'ORCL']
		     WITHIN 10000 [stock_time]
		
		S12 = SELECT * FROM S
		      WHERE (SELL as T1; BUY as T2; BUY as T3;
		             SELL as T4; BUY as T5; BUY as T6;
		             SELL as T7; BUY as T8; BUY as T9;
		             SELL as T10; BUY as T11; BUY as T12)
		      FILTER T1[name = 'INTC'] AND T2[name = 'RIMM'] AND T3[name = 'QQQ'] AND
		             T4[name = 'IPIX'] AND T5[name = 'AMAT'] AND T6[name = 'CSCO'] AND
		             T7[name = 'YHOO'] AND T8[name = 'DELL'] AND T9[name = 'ORCL'] AND
		             T10[name = 'MSFT'] AND T11[name = 'INTC'] AND T12[name = 'RIMM']
		      WITHIN 10000 [stock_time]
		
		S24 = SELECT * FROM S
		      WHERE (SELL as T1; BUY as T2; BUY as T3;
		             SELL as T4; BUY as T5; BUY as T6;
		             SELL as T7; BUY as T8; BUY as T9;
		             SELL as T10; BUY as T11; BUY as T12;
		             SELL as T13; BUY as T14; BUY as T15;
		             SELL as T16; BUY as T17; BUY as T18;
		             SELL as T19; BUY as T20; BUY as T21;
		             SELL as T22; BUY as T23; BUY as T24)
		      FILTER T1[name = 'INTC'] AND T2[name = 'RIMM'] AND T3[name = 'QQQ'] AND
		             T4[name = 'IPIX'] AND T5[name = 'AMAT'] AND T6[name = 'CSCO'] AND
		             T7[name = 'YHOO'] AND T8[name = 'DELL'] AND T9[name = 'ORCL'] AND
		             T10[name = 'MSFT'] AND T11[name = 'INTC'] AND T12[name = 'RIMM'] AND
		             T13[name = 'INTC'] AND T14[name = 'RIMM'] AND T15[name = 'QQQ'] AND
		             T16[name = 'IPIX'] AND T17[name = 'AMAT'] AND T18[name = 'CSCO'] AND
		             T19[name = 'YHOO'] AND T20[name = 'DELL'] AND T21[name = 'ORCL'] AND
		             T22[name = 'MSFT'] AND T23[name = 'INTC'] AND T24[name = 'RIMM']
		      WITHIN 10000 [stock_time]
	\end{verbatim}
For the workload of sequence queries without output, we extend each query of the previous workload with a ``non-existing'' event. Queries \texttt{S3'}, \texttt{S6'}, \texttt{S9'}, \texttt{S12'}, and \texttt{S24'} are the aforementioned extension of \texttt{S3}, \texttt{S6}, \texttt{S9}, \texttt{S12}, and \texttt{S24}, respectively :
\begin{verbatim}
	S3' = SELECT * FROM S
	      WHERE (SELL as T1; BUY as T2; BUY as T3; BUY as NE)
	      FILTER T1[name = 'INTC'] AND T2[name = 'RIMM'] AND T3[name = 'QQQ'] AND NE[name = 'NotExists']
	      WITHIN 10000 [stock_time]
	
	S6' = SELECT * FROM S
	      WHERE (SELL as T1; BUY as T2; BUY as T3; 
	             SELL as T4; BUY as T5; BUY as T6; BUY AS NE)
	      FILTER T1[name = 'INTC'] AND T2[name = 'RIMM'] AND T3[name = 'QQQ'] AND
	             T4[name = 'IPIX'] AND T5[name = 'AMAT'] AND T6[name = 'CSCO'] AND NE[name = 'NotExists']
	      WITHIN 10000 [stock_time]
	
	S9' = SELECT * FROM S
	      WHERE (SELL as T1; BUY as T2; BUY as T3;
	             SELL as T4; BUY as T5; BUY as T6;
	             SELL as T7; BUY as T8; BUY as T9; BUY AS NE)
	      FILTER T1[name = 'INTC'] AND T2[name = 'RIMM'] AND T3[name = 'QQQ'] AND
	             T4[name = 'IPIX'] AND T5[name = 'AMAT'] AND T6[name = 'CSCO'] AND
	             T7[name = 'YHOO'] AND T8[name = 'DELL'] AND T9[name = 'ORCL'] AND NE[name = 'NotExists']
	      WITHIN 10000 [stock_time]
	
	S12' = SELECT * FROM S
	       WHERE (SELL as T1; BUY as T2; BUY as T3;
	              SELL as T4; BUY as T5; BUY as T6;
	              SELL as T7; BUY as T8; BUY as T9;
	              SELL as T10; BUY as T11; BUY as T12; BUY AS NE)
	       FILTER T1[name = 'INTC'] AND T2[name = 'RIMM'] AND T3[name = 'QQQ'] AND
	              T4[name = 'IPIX'] AND T5[name = 'AMAT'] AND T6[name = 'CSCO'] AND
	              T7[name = 'YHOO'] AND T8[name = 'DELL'] AND T9[name = 'ORCL'] AND
	              T10[name = 'MSFT'] AND T11[name = 'INTC'] AND T12[name = 'RIMM'] AND NE[name = 'NotExists']
	       WITHIN 10000 [stock_time]
	
	S24' = SELECT * FROM S
	       WHERE (SELL as T1; BUY as T2; BUY as T3;
	              SELL as T4; BUY as T5; BUY as T6;
	              SELL as T7; BUY as T8; BUY as T9;
	              SELL as T10; BUY as T11; BUY as T12;
	              SELL as T13; BUY as T14; BUY as T15;
	              SELL as T16; BUY as T17; BUY as T18;
	              SELL as T19; BUY as T20; BUY as T21;
	              SELL as T22; BUY as T23; BUY as T24; BUY AS NE)
	       FILTER T1[name = 'INTC'] AND T2[name = 'RIMM'] AND T3[name = 'QQQ'] AND
	              T4[name = 'IPIX'] AND T5[name = 'AMAT'] AND T6[name = 'CSCO'] AND
	              T7[name = 'YHOO'] AND T8[name = 'DELL'] AND T9[name = 'ORCL'] AND
	              T10[name = 'MSFT'] AND T11[name = 'INTC'] AND T12[name = 'RIMM'] AND
	              T13[name = 'INTC'] AND T14[name = 'RIMM'] AND T15[name = 'QQQ'] AND
	              T16[name = 'IPIX'] AND T17[name = 'AMAT'] AND T18[name = 'CSCO'] AND
	              T19[name = 'YHOO'] AND T20[name = 'DELL'] AND T21[name = 'ORCL'] AND
	              T22[name = 'MSFT'] AND T23[name = 'INTC'] AND T24[name = 'RIMM'] AND NE[name = 'NotExists']
	       WITHIN 10000 [stock_time]
\end{verbatim}
For the workload of sequence queries of length 3 with increasing time windows and without output, we use query \texttt{S3'} and modify the size of the time window by $20$, $30$, and $40$ seconds as follow:
\begin{verbatim}
	S3' = SELECT * FROM S
	      WHERE (SELL as T1; BUY as T2; BUY as T3; BUY as NE)
	      FILTER T1[name = 'INTC'] AND T2[name = 'RIMM'] AND T3[name = 'QQQ'] AND NE[name = 'NotExists']
	      WITHIN 10000 [stock_time]
	
	S3'x2 = SELECT * FROM S
	        WHERE (SELL as T1; BUY as T2; BUY as T3; BUY as NE)
	        FILTER T1[name = 'INTC'] AND T2[name = 'RIMM'] AND T3[name = 'QQQ'] AND NE[name = 'NotExists']
	        WITHIN 20000 [stock_time]
	
	S3'x3 = SELECT * FROM S
	        WHERE (SELL as T1; BUY as T2; BUY as T3; BUY as NE)
	        FILTER T1[name = 'INTC'] AND T2[name = 'RIMM'] AND T3[name = 'QQQ'] AND NE[name = 'NotExists']
	        WITHIN 30000 [stock_time]
	
	S3'x4 = SELECT * FROM S
	        WHERE (SELL as T1; BUY as T2; BUY as T3; BUY as NE)
	        FILTER T1[name = 'INTC'] AND T2[name = 'RIMM'] AND T3[name = 'QQQ'] AND NE[name = 'NotExists']
	        WITHIN 40000 [stock_time]
\end{verbatim}
Finally, for the last workout of sequence queries with selection strategies, we consider queries \texttt{S3'}, \texttt{S3'x2}, \texttt{S3'x3}, and \texttt{S3'x4} and use the selection strategy provided by each system.


%% file: appendix/app-seq-homes.tex
\noindent \textbf{Dataset.} We reuse the smart homes dataset from the DEBS 2014
Grand Challenge~\cite{smarthomes}, which contains events originating from smart
plugs deployed in private households. Each smart plug acts as a
proxy between the wall power outlet and the device connected to it, and is
equipped with sensors to measure power load and power work at a resolution of
one measurement per second. The full dataset consists of over 4055 million
measurements for 2125 plugs distributed across 40 houses. We select the
substream of events of a single house, with house id 28.  This substream
contains data from 85 plugs distributed over 17 households in the stream. We
process the first million events, focusing on load measurements only. Each event
consists of an event id, the timestamp, the measured load value, the plug id,
and the household id. 

\medskip
\noindent \textbf{Queries.} All queries look for sequences of measurements in different household that are above the dataset average  load of 76 Watts, in a time window of $10$ seconds (note that the plug timestamp is in seconds).
The workload of sequence queries with output over the smart homes dataset are the following queries of length $3$, $6$, $9$, $12$ and $24$, respectively.\footnote{We note that, because \texttt{L} is bound to all events in the sequence, the filter \texttt{L[value > 76]} requires all events in the sequence to have a load value above 76.}

	\begin{verbatim}
		H3 = SELECT * FROM S
		     WHERE (LOAD as L1; LOAD as L2; LOAD as L3) as L
		     FILTER L[value > 76] AND 
		            L1[household_id = 0] AND L2[household_id = 2] AND L3[household_id = 4]
		     WITHIN 10 [plug_timestamp]
		
		H6 = SELECT * FROM S
		     WHERE (LOAD as L1; LOAD as L2; LOAD as L3; 
		            LOAD as L4; LOAD as L5; LOAD as L6) as L
		     FILTER L[value > 76] AND 
		            L1[household_id = 0] AND L2[household_id = 2] AND L3[household_id = 4] AND 
		            L4[household_id = 6] AND L5[household_id = 9] AND L6[household_id = 10]
		     WITHIN 10 [plug_timestamp]

		H9 = SELECT * FROM S
		     WHERE (LOAD as L1; LOAD as L2; LOAD as L3; 
		            LOAD as L4; LOAD as L5; LOAD as L6;
		            LOAD as L7; LOAD as L8; LOAD as L9) as L
		     FILTER L[value > 76] AND 
		            L1[household_id = 0] AND L2[household_id = 2] AND L3[household_id = 4] AND
		            L4[household_id = 6] AND L5[household_id = 9] AND L6[household_id = 10] AND
		            L7[household_id = 12] AND L8[household_id = 14] AND L9[household_id = 15]
		     WITHIN 10 [plug_timestamp]
		     
		H12 = SELECT * FROM S 
		      WHERE (LOAD as L1; LOAD as L2; LOAD as L3; 
		             LOAD as L4; LOAD as L5; LOAD as L6;
		             LOAD as L7; LOAD as L8; LOAD as L9;
		             LOAD as L10; LOAD as L11; LOAD as L12) as L
		      FILTER L[value > 76] AND 
		             L1[household_id = 0] AND L2[household_id = 2] AND L3[household_id = 4] AND
		             L4[household_id = 6] AND L5[household_id = 9] AND L6[household_id = 10] AND
		             L7[household_id = 12] AND L8[household_id = 14] AND L9[household_id = 15] AND
		             L10[household_id = 4] AND L11[household_id = 9] AND L12[household_id = 10]
		      WITHIN 10 [plug_timestamp]
		      
		H24 = SELECT * FROM S 
		      WHERE (LOAD as L1; LOAD as L2; LOAD as L3; 
		             LOAD as L4; LOAD as L5; LOAD as L6;
		             LOAD as L7; LOAD as L8; LOAD as L9;
		             LOAD as L10; LOAD as L11; LOAD as L12;
		             LOAD as L13; LOAD as L14; LOAD as L15; 
		             LOAD as L16; LOAD as L17; LOAD as L18;
		             LOAD as L19; LOAD as L20; LOAD as L21;
		             LOAD as L22; LOAD as L23; LOAD as L24) as L
		      FILTER L[value > 76] AND 
		             L1[household_id = 0] AND L2[household_id = 2] AND L3[household_id = 4] AND
		             L4[household_id = 6] AND L5[household_id = 9] AND L6[household_id = 10] AND
		             L7[household_id = 12] AND L8[household_id = 14] AND L9[household_id = 15] AND
		             L10[household_id = 4] AND L11[household_id = 9] AND L12[household_id = 10] AND
		             L13[household_id = 0] AND L14[household_id = 2] AND L15[household_id = 4] AND
		             L16[household_id = 6] AND L17[household_id = 9] AND L18[household_id = 10] AND
		             L19[household_id = 12] AND L20[household_id = 14] AND L21[household_id = 15] AND
		             L22[household_id = 4] AND L23[household_id = 9] AND L24[household_id = 10]
		      WITHIN 10 [plug_timestamp]
	\end{verbatim}
Similarly, we extend each query of the previous workload with a ``non-existing'' event. The queries \texttt{H3'}, \texttt{H6'}, \texttt{H9'}, \texttt{H12'}, and \texttt{H24'} are the corresponding extensions of \texttt{H3}, \texttt{H6}, \texttt{H9}, \texttt{H12}, and \texttt{H24}:
\begin{verbatim}
	H3' = SELECT * FROM S
	      WHERE (LOAD as L1; LOAD as L2; LOAD as L3; LOAD as NE) as L
	      FILTER L[value > 76] AND 
	             L1[household_id = 0] AND L2[household_id = 2] AND L3[household_id = 4] AND 
	             NE[household_id = 1000]
	      WITHIN 10 [plug_timestamp]
	
	H6' = SELECT * FROM S
	      WHERE (LOAD as L1; LOAD as L2; LOAD as L3; 
	             LOAD as L4; LOAD as L5; LOAD as L6; LOAD as NE) as L
	      FILTER L[value > 76] AND 
	             L1[household_id = 0] AND L2[household_id = 2] AND L3[household_id = 4] AND
	             L4[household_id = 6] AND L5[household_id = 9] AND L6[household_id = 10] AND 
	             NE[household_id = 1000]
	      WITHIN 10 [plug_timestamp]
	
	H9' = SELECT * FROM S
	      WHERE (LOAD as L1; LOAD as L2; LOAD as L3; 
	             LOAD as L4; LOAD as L5; LOAD as L6;
	             LOAD as L7; LOAD as L8; LOAD as L9; LOAD as NE) as L
	      FILTER L[value > 76] AND 
	             L1[household_id = 0] AND L2[household_id = 2] AND L3[household_id = 4] AND
	             L4[household_id = 6] AND L5[household_id = 9] AND L6[household_id = 10] AND
	             L7[household_id = 12] AND L8[household_id = 14] AND L9[household_id = 15] AND
	             NE[household_id = 1000]
	      WITHIN 10 [plug_timestamp]
	
	H12' = SELECT * FROM S 
	       WHERE (LOAD as L1; LOAD as L2; LOAD as L3; 
	              LOAD as L4; LOAD as L5; LOAD as L6;
	              LOAD as L7; LOAD as L8; LOAD as L9;
	              LOAD as L10; LOAD as L11; LOAD as L12; LOAD as NE) as L
	       FILTER L[value > 76] AND
	              L1[household_id = 0] AND L2[household_id = 2] AND L3[household_id = 4] AND
	              L4[household_id = 6] AND L5[household_id = 9] AND L6[household_id = 10] AND
	              L7[household_id = 12] AND L8[household_id = 14] AND L9[household_id = 15] AND
	              L10[household_id = 4] AND L11[household_id = 9] AND L12[household_id = 10] AND 
	              NE[household_id = 1000]
	       WITHIN 10 [plug_timestamp]
	       
	H24' = SELECT * FROM S 
	       WHERE (LOAD as L1; LOAD as L2; LOAD as L3; 
	              LOAD as L4; LOAD as L5; LOAD as L6;
	              LOAD as L7; LOAD as L8; LOAD as L9;
	              LOAD as L10; LOAD as L11; LOAD as L12;
	              LOAD as L13; LOAD as L14; LOAD as L15; 
	              LOAD as L16; LOAD as L17; LOAD as L18;
	              LOAD as L19; LOAD as L20; LOAD as L21;
	              LOAD as L22; LOAD as L23; LOAD as L24; LOAD as NE) as L
	       FILTER L[value > 76] AND 
	              L1[household_id = 0] AND L2[household_id = 2] AND L3[household_id = 4] AND
	              L4[household_id = 6] AND L5[household_id = 9] AND L6[household_id = 10] AND
	              L7[household_id = 12] AND L8[household_id = 14] AND L9[household_id = 15] AND
	              L10[household_id = 4] AND L11[household_id = 9] AND L12[household_id = 10] AND
	              L13[household_id = 0] AND L14[household_id = 2] AND L15[household_id = 4] AND
	              L16[household_id = 6] AND L17[household_id = 9] AND L18[household_id = 10] AND
	              L19[household_id = 12] AND L20[household_id = 14] AND L21[household_id = 15] AND
	              L22[household_id = 4] AND L23[household_id = 9] AND L24[household_id = 10] AND
	              NE[household_id = 1000]
	       WITHIN 10 [plug_timestamp]
\end{verbatim}
For the workload of sequence queries of length 3 with increasing time windows and without output, we use query \texttt{H3'} and modify the size of the time window by $20$, $30$, and $40$ seconds:
\begin{verbatim}
	H3' = SELECT * FROM S
	      WHERE (LOAD as L1; LOAD as L2; LOAD as L3; LOAD as NE) as L
	      FILTER L[value > 76] AND 
	             L1[household_id = 0] AND L2[household_id = 2] AND L3[household_id = 4] AND 
	             NE[household_id = 1000]
	      WITHIN 10 [plug_timestamp]
	
	H3'x2 = SELECT * FROM S
	        WHERE (LOAD as L1; LOAD as L2; LOAD as L3; LOAD as NE) as L
	        FILTER L[value > 76] AND 
	               L1[household_id = 0] AND L2[household_id = 2] AND L3[household_id = 4] AND 
	               NE[household_id = 1000]
	        WITHIN 20 [plug_timestamp]
	
	H3'x3 = SELECT * FROM S
	        WHERE (LOAD as L1; LOAD as L2; LOAD as L3; LOAD as NE) as L
	        FILTER L[value > 76] AND 
	               L1[household_id = 0] AND L2[household_id = 2] AND L3[household_id = 4] AND 
	               NE[household_id = 1000]
	        WITHIN 30 [plug_timestamp]
	
	H3'x4 = SELECT * FROM S
	        WHERE (LOAD as L1; LOAD as L2; LOAD as L3; LOAD as NE) as L
	        FILTER L[value > 76] AND 
	               L1[household_id = 0] AND L2[household_id = 2] AND L3[household_id = 4] AND 
	               NE[household_id = 1000]
	        WITHIN 40 [plug_timestamp]
\end{verbatim}
For the workout of sequence queries with selection strategies, \texttt{H3'}, \texttt{H3'x2}, \texttt{H3'x3}, and \texttt{H3'x4} are executed with the selection strategy provided by each system.


%% file: appendix/app-seq-taxis.tex
\noindent \textbf{Dataset.} We reuse the taxi trips dataset from the DEBS 2015
Grand Challenge~\cite{taxitrips}, which contains events reporting on taxi trips
in New York, listing the trip's pickup and drop-off location, timestamps,
information on the taxi, and information related to the payment. Data are
reported at the end of the trip, i.e., upon arrival in the order of the drop-off
timestamps. The full dataset consists of 173 million events, of which we process
the first million. The original dataset contains absolute (latitude, longitue)
coordinates of pickup and dropoff locations. We preprocess events and use
reverse geo-location to convert these coordinates into corresponding zone (E.g.,
East Harlem, Midwood, \dots) using the official definition of New York City
zones available at 
\url{https://www1.nyc.gov/site/tlc/about/tlc-trip-record-data.page}.

\medskip
\noindent \textbf{Queries.} Queries look for sequences of events where the dropoff location of one event equals the pickup location of the next event. We search for this pattern in a time window of 2.7 hours (i.e., 10000 seconds).
Concretely, the workload of sequence queries with output over the NY taxi trips are the following queries of length $3$, $6$, $9$, $12$ and $24$, respectively:
	\begin{verbatim}
		T3 = SELECT * FROM S
		     WHERE (TRIP as loc1; TRIP as loc2; TRIP as loc3)
		     FILTER loc1[pickup_loc = 'East Harlem North' and dropoff_loc = 'Midwood'] AND 
		            loc2[pickup_loc = 'Midwood' AND dropoff_loc = 'Gravesend'] AND
		            loc3[pickup_loc = 'Gravesend' AND dropoff_loc = 'West Brighton']
		     WITHIN 10000 [dropoff_datetime]
		
		T6 = SELECT * FROM S
		     WHERE (TRIP as loc1; TRIP as loc2; TRIP as loc3; 
		            TRIP as loc4; TRIP as loc5; TRIP as loc6)
		     FILTER loc1[pickup_loc = 'East Harlem North' and dropoff_loc = 'Midwood'] AND 
		            loc2[pickup_loc = 'Midwood' AND dropoff_loc = 'Gravesend'] AND
		            loc3[pickup_loc = 'Gravesend' AND dropoff_loc = 'West Brighton'] AND
		            loc4[pickup_loc = 'West Brighton' AND dropoff_loc = 'Lincoln Square West'] AND
		            loc5[pickup_loc = 'Lincoln Square West' AND dropoff_loc = 'Sutton Place/Turtle Bay North'] AND
		            loc6[pickup_loc = 'Sutton Place/Turtle Bay North' AND
		                 dropoff_loc = 'East Concourse/Concourse Village']
		     WITHIN 10000 [dropoff_datetime]
		     
		T9 = SELECT * FROM S
		     WHERE (TRIP as loc1; TRIP as loc2; TRIP as loc3; 
		            TRIP as loc4; TRIP as loc5; TRIP as loc6; 
		            TRIP as loc7; TRIP as loc8; TRIP as loc9)
		     FILTER loc1[pickup_loc = 'East Harlem North' and dropoff_loc = 'Midwood'] AND 
		            loc2[pickup_loc = 'Midwood' AND dropoff_loc = 'Gravesend'] AND 
		            loc3[pickup_loc = 'Gravesend' AND dropoff_loc = 'West Brighton'] AND
		            loc4[pickup_loc = 'West Brighton' AND dropoff_loc = 'Lincoln Square West'] AND
		            loc5[pickup_loc = 'Lincoln Square West' AND dropoff_loc = 'Sutton Place/Turtle Bay North'] AND
		            loc6[pickup_loc = 'Sutton Place/Turtle Bay North' AND
		                 dropoff_loc = 'East Concourse/Concourse Village'] AND
		            loc7[pickup_loc = 'East Concourse/Concourse Village' AND dropoff_loc = 'East Harlem North'] AND
		            loc8[pickup_loc = 'East Harlem North' AND dropoff_loc = 'East Harlem North'] AND
		            loc9[pickup_loc = 'East Harlem North' AND dropoff_loc = 'Gravesend']
		     WITHIN 10000 [dropoff_datetime]
		     
		T12 = SELECT * FROM S
		      WHERE (TRIP as loc1; TRIP as loc2; TRIP as loc3; 
		             TRIP as loc4; TRIP as loc5; TRIP as loc6; 
		             TRIP as loc7; TRIP as loc8; TRIP as loc9; 
		             TRIP as loc10; TRIP as loc11; TRIP as loc12)
		      FILTER loc1[pickup_loc = 'East Harlem North' and dropoff_loc = 'Midwood'] AND 
		             loc2[pickup_loc = 'Midwood' AND dropoff_loc = 'Gravesend'] AND
		             loc3[pickup_loc = 'Gravesend' AND dropoff_loc = 'West Brighton'] AND
		             loc4[pickup_loc = 'West Brighton' AND dropoff_loc = 'Lincoln Square West'] AND
		             loc5[pickup_loc = 'Lincoln Square West' AND dropoff_loc = 'Sutton Place/Turtle Bay North'] AND
		             loc6[pickup_loc = 'Sutton Place/Turtle Bay North' AND
		                  dropoff_loc = 'East Concourse/Concourse Village'] AND
		             loc7[pickup_loc = 'East Concourse/Concourse Village' AND dropoff_loc = 'East Harlem North'] AND
		             loc8[pickup_loc = 'East Harlem North' AND dropoff_loc = 'East Harlem North'] AND
		             loc9[pickup_loc = 'East Harlem North' AND dropoff_loc = 'Gravesend'] AND
		             loc10[pickup_loc = 'Gravesend' AND dropoff_loc = 'Midwood'] AND
		             loc11[pickup_loc = 'Midwood' AND dropoff_loc = 'Midwood'] AND
		             loc12[pickup_loc = 'Midwood' AND dropoff_loc = 'Gravesend']
		      WITHIN 10000 [dropoff_datetime]
		      
	T24 = SELECT * FROM S
	      WHERE (TRIP as loc1; TRIP as loc2; TRIP as loc3; 
	             TRIP as loc4; TRIP as loc5; TRIP as loc6; 
	             TRIP as loc7; TRIP as loc8; TRIP as loc9; 
	             TRIP as loc10; TRIP as loc11; TRIP as loc12; 
	             TRIP as loc13; TRIP as loc14; TRIP as loc15; 
	             TRIP as loc16; TRIP as loc17; TRIP as loc18; 
	             TRIP as loc19; TRIP as loc20; TRIP as loc21; 
	             TRIP as loc22; TRIP as loc23; TRIP as loc24)
	      FILTER loc1[pickup_loc = 'East Harlem North' and dropoff_loc = 'Midwood'] AND 
	             loc2[pickup_loc = 'Midwood' AND dropoff_loc = 'Gravesend'] AND
	             loc3[pickup_loc = 'Gravesend' AND dropoff_loc = 'West Brighton'] AND
	             loc4[pickup_loc = 'West Brighton' AND dropoff_loc = 'Lincoln Square West'] AND
	             loc5[pickup_loc = 'Lincoln Square West' AND dropoff_loc = 'Sutton Place/Turtle Bay North'] AND
	             loc6[pickup_loc = 'Sutton Place/Turtle Bay North' AND
	                  dropoff_loc = 'East Concourse/Concourse Village'] AND
	             loc7[pickup_loc = 'East Concourse/Concourse Village' AND dropoff_loc = 'East Harlem North'] AND
	             loc8[pickup_loc = 'East Harlem North' AND dropoff_loc = 'East Harlem North'] AND
	             loc9[pickup_loc = 'East Harlem North' AND dropoff_loc = 'Gravesend'] AND
	             loc10[pickup_loc = 'Gravesend' AND dropoff_loc = 'Midwood'] AND
	             loc11[pickup_loc = 'Midwood' AND dropoff_loc = 'Midwood'] AND
	             loc12[pickup_loc = 'Midwood' AND dropoff_loc = 'Gravesend'] AND
	             loc13[pickup_loc = 'Gravesend' and dropoff_loc = 'Midwood'] AND 
	             loc14[pickup_loc = 'Midwood' AND dropoff_loc = 'Gravesend'] AND
	             loc15[pickup_loc = 'Gravesend' AND dropoff_loc = 'West Brighton'] AND
	             loc16[pickup_loc = 'West Brighton' AND dropoff_loc = 'Lincoln Square West'] AND
	             loc17[pickup_loc = 'Lincoln Square West' AND dropoff_loc = 'Sutton Place/Turtle Bay North'] AND
	             loc18[pickup_loc = 'Sutton Place/Turtle Bay North' AND
	                   dropoff_loc = 'East Concourse/Concourse Village'] AND
	             loc19[pickup_loc = 'East Concourse/Concourse Village' AND dropoff_loc = 'East Harlem North'] AND
	             loc20[pickup_loc = 'East Harlem North' AND dropoff_loc = 'East Harlem North'] AND
	             loc21[pickup_loc = 'East Harlem North' AND dropoff_loc = 'Gravesend'] AND
	             loc22[pickup_loc = 'Gravesend' AND dropoff_loc = 'Midwood'] AND
	             loc23[pickup_loc = 'Midwood' AND dropoff_loc = 'Midwood'] AND
	             loc24[pickup_loc = 'Midwood' AND dropoff_loc = 'Gravesend']
	      WITHIN 10000 [dropoff_datetime]
	\end{verbatim}
The workload of sequence queries without output extends each query of the previous workload with a ``non-existing'' event.
\begin{verbatim}
	T3' = SELECT * FROM S
	     WHERE (TRIP as loc1; TRIP as loc2; TRIP as loc3; TRIP as NE)
	     FILTER loc1[pickup_loc = 'East Harlem North' and dropoff_loc = 'Midwood'] AND 
	            loc2[pickup_loc = 'Midwood' AND dropoff_loc = 'Gravesend'] AND
	            loc3[pickup_loc = 'Gravesend' AND dropoff_loc = 'West Brighton'] AND
	            NE[pickup_loc = 'NotExists']
	     WITHIN 10000 [dropoff_datetime]
	
	T6' = SELECT * FROM S
	      WHERE (TRIP as loc1; TRIP as loc2; TRIP as loc3; 
	            TRIP as loc4; TRIP as loc5; TRIP as loc6; TRIP as NE)
	      FILTER loc1[pickup_loc = 'East Harlem North' and dropoff_loc = 'Midwood'] AND 
	             loc2[pickup_loc = 'Midwood' AND dropoff_loc = 'Gravesend'] AND
	             loc3[pickup_loc = 'Gravesend' AND dropoff_loc = 'West Brighton'] AND
	             loc4[pickup_loc = 'West Brighton' AND dropoff_loc = 'Lincoln Square West'] AND
	             loc5[pickup_loc = 'Lincoln Square West' AND dropoff_loc = 'Sutton Place/Turtle Bay North'] AND
	             loc6[pickup_loc = 'Sutton Place/Turtle Bay North' AND 
	                  dropoff_loc = 'East Concourse/Concourse Village'] AND
	             NE[pickup_loc = 'NotExists']
	      WITHIN 10000 [dropoff_datetime]
	
	T9' = SELECT * FROM S
	      WHERE (TRIP as loc1; TRIP as loc2; TRIP as loc3; 
	             TRIP as loc4; TRIP as loc5; TRIP as loc6; 
	             TRIP as loc7; TRIP as loc8; TRIP as loc9; TRIP as NE)
	      FILTER loc1[pickup_loc = 'East Harlem North' and dropoff_loc = 'Midwood'] AND 
	             loc2[pickup_loc = 'Midwood' AND dropoff_loc = 'Gravesend'] AND 
	             loc3[pickup_loc = 'Gravesend' AND dropoff_loc = 'West Brighton'] AND
	             loc4[pickup_loc = 'West Brighton' AND dropoff_loc = 'Lincoln Square West'] AND
	             loc5[pickup_loc = 'Lincoln Square West' AND dropoff_loc = 'Sutton Place/Turtle Bay North'] AND
	             loc6[pickup_loc = 'Sutton Place/Turtle Bay North' AND
	                  dropoff_loc = 'East Concourse/Concourse Village'] AND
	             loc7[pickup_loc = 'East Concourse/Concourse Village' AND dropoff_loc = 'East Harlem North'] AND
	             loc8[pickup_loc = 'East Harlem North' AND dropoff_loc = 'East Harlem North'] AND
	             loc9[pickup_loc = 'East Harlem North' AND dropoff_loc = 'Gravesend'] AND
	             NE[pickup_loc = 'NotExists']
	      WITHIN 10000 [dropoff_datetime]
	
	T12' = SELECT * FROM S
	       WHERE (TRIP as loc1; TRIP as loc2; TRIP as loc3; 
	              TRIP as loc4; TRIP as loc5; TRIP as loc6; 
	              TRIP as loc7; TRIP as loc8; TRIP as loc9; 
	              TRIP as loc10; TRIP as loc11; TRIP as loc12; TRIP as NE)
	       FILTER loc1[pickup_loc = 'East Harlem North' and dropoff_loc = 'Midwood'] AND 
	              loc2[pickup_loc = 'Midwood' AND dropoff_loc = 'Gravesend'] AND
	              loc3[pickup_loc = 'Gravesend' AND dropoff_loc = 'West Brighton'] AND
	              loc4[pickup_loc = 'West Brighton' AND dropoff_loc = 'Lincoln Square West'] AND
	              loc5[pickup_loc = 'Lincoln Square West' AND dropoff_loc = 'Sutton Place/Turtle Bay North'] AND
	              loc6[pickup_loc = 'Sutton Place/Turtle Bay North' AND
	                   dropoff_loc = 'East Concourse/Concourse Village'] AND
	              loc7[pickup_loc = 'East Concourse/Concourse Village' AND dropoff_loc = 'East Harlem North'] AND
	              loc8[pickup_loc = 'East Harlem North' AND dropoff_loc = 'East Harlem North'] AND
	              loc9[pickup_loc = 'East Harlem North' AND dropoff_loc = 'Gravesend'] AND
	              loc10[pickup_loc = 'Gravesend' AND dropoff_loc = 'Midwood'] AND
	              loc11[pickup_loc = 'Midwood' AND dropoff_loc = 'Midwood'] AND
	              loc12[pickup_loc = 'Midwood' AND dropoff_loc = 'Gravesend'] AND
	              NE[pickup_loc = 'NotExists']
	       WITHIN 10000 [dropoff_datetime]
	
	T24' = SELECT * FROM S
	       WHERE (TRIP as loc1; TRIP as loc2; TRIP as loc3; 
	              TRIP as loc4; TRIP as loc5; TRIP as loc6; 
	              TRIP as loc7; TRIP as loc8; TRIP as loc9; 
	              TRIP as loc10; TRIP as loc11; TRIP as loc12; 
	              TRIP as loc13; TRIP as loc14; TRIP as loc15; 
	              TRIP as loc16; TRIP as loc17; TRIP as loc18; 
	              TRIP as loc19; TRIP as loc20; TRIP as loc21; 
	              TRIP as loc22; TRIP as loc23; TRIP as loc24; TRIP as NE)
           FILTER loc1[pickup_loc = 'East Harlem North' and dropoff_loc = 'Midwood'] AND 
                  loc2[pickup_loc = 'Midwood' AND dropoff_loc = 'Gravesend'] AND
                  loc3[pickup_loc = 'Gravesend' AND dropoff_loc = 'West Brighton'] AND
                  loc4[pickup_loc = 'West Brighton' AND dropoff_loc = 'Lincoln Square West'] AND
                  loc5[pickup_loc = 'Lincoln Square West' AND dropoff_loc = 'Sutton Place/Turtle Bay North'] AND
                  loc6[pickup_loc = 'Sutton Place/Turtle Bay North' AND
                       dropoff_loc = 'East Concourse/Concourse Village'] AND
                  loc7[pickup_loc = 'East Concourse/Concourse Village' AND dropoff_loc = 'East Harlem North'] AND
                  loc8[pickup_loc = 'East Harlem North' AND dropoff_loc = 'East Harlem North'] AND
                  loc9[pickup_loc = 'East Harlem North' AND dropoff_loc = 'Gravesend'] AND
                  loc10[pickup_loc = 'Gravesend' AND dropoff_loc = 'Midwood'] AND
                  loc11[pickup_loc = 'Midwood' AND dropoff_loc = 'Midwood'] AND
                  loc12[pickup_loc = 'Midwood' AND dropoff_loc = 'Gravesend'] AND
                  loc13[pickup_loc = 'Gravesend' and dropoff_loc = 'Midwood'] AND 
                  loc14[pickup_loc = 'Midwood' AND dropoff_loc = 'Gravesend'] AND
                  loc15[pickup_loc = 'Gravesend' AND dropoff_loc = 'West Brighton'] AND
                  loc16[pickup_loc = 'West Brighton' AND dropoff_loc = 'Lincoln Square West'] AND
                  loc17[pickup_loc = 'Lincoln Square West' AND dropoff_loc = 'Sutton Place/Turtle Bay North'] AND
                  loc18[pickup_loc = 'Sutton Place/Turtle Bay North' AND
                        dropoff_loc = 'East Concourse/Concourse Village'] AND
                  loc19[pickup_loc = 'East Concourse/Concourse Village' AND dropoff_loc = 'East Harlem North'] AND
                  loc20[pickup_loc = 'East Harlem North' AND dropoff_loc = 'East Harlem North'] AND
                  loc21[pickup_loc = 'East Harlem North' AND dropoff_loc = 'Gravesend'] AND
                  loc22[pickup_loc = 'Gravesend' AND dropoff_loc = 'Midwood'] AND
                  loc23[pickup_loc = 'Midwood' AND dropoff_loc = 'Midwood'] AND
                  loc24[pickup_loc = 'Midwood' AND dropoff_loc = 'Gravesend'] AND
                  NE[pickup_loc = 'NotExists']
	       WITHIN 10000 [dropoff_datetime]
\end{verbatim}
For the workload of sequence queries of length 3 with increasing time windows and without output, we use query \texttt{T3'} and modify the size of the time window by $20$, $30$, and $40$ seconds as follow:
\begin{verbatim}
	T3' = SELECT * FROM S
	      WHERE (TRIP as loc1; TRIP as loc2; TRIP as loc3; TRIP as NE)
	      FILTER loc1[pickup_loc = 'East Harlem North' and dropoff_loc = 'Midwood'] AND 
	             loc2[pickup_loc = 'Midwood' AND dropoff_loc = 'Gravesend'] AND
	             loc3[pickup_loc = 'Gravesend' AND dropoff_loc = 'West Brighton'] AND
	             NE[pickup_loc = 'NotExists']
	      WITHIN 10000 [dropoff_datetime]
	
	T3'x2 = SELECT * FROM S
	        WHERE (TRIP as loc1; TRIP as loc2; TRIP as loc3; TRIP as NE)
	        FILTER loc1[pickup_loc = 'East Harlem North' and dropoff_loc = 'Midwood'] AND 
	               loc2[pickup_loc = 'Midwood' AND dropoff_loc = 'Gravesend'] AND
	               loc3[pickup_loc = 'Gravesend' AND dropoff_loc = 'West Brighton'] AND
	               NE[pickup_loc = 'NotExists']
	        WITHIN 20000 [dropoff_datetime]
	
	T3'x3 = SELECT * FROM S
	        WHERE (TRIP as loc1; TRIP as loc2; TRIP as loc3; TRIP as NE)
	        FILTER loc1[pickup_loc = 'East Harlem North' and dropoff_loc = 'Midwood'] AND 
	               loc2[pickup_loc = 'Midwood' AND dropoff_loc = 'Gravesend'] AND
	               loc3[pickup_loc = 'Gravesend' AND dropoff_loc = 'West Brighton'] AND
	               NE[pickup_loc = 'NotExists']
	        WITHIN 30000 [dropoff_datetime]
	
	T3'x4 = SELECT * FROM S
	        WHERE (TRIP as loc1; TRIP as loc2; TRIP as loc3; TRIP as NE)
	        FILTER loc1[pickup_loc = 'East Harlem North' and dropoff_loc = 'Midwood'] AND 
	               loc2[pickup_loc = 'Midwood' AND dropoff_loc = 'Gravesend'] AND
	               loc3[pickup_loc = 'Gravesend' AND dropoff_loc = 'West Brighton'] AND
	               NE[pickup_loc = 'NotExists']
	        WITHIN 40000 [dropoff_datetime]
\end{verbatim}
The last workout of sequence queries with selection strategies uses \texttt{T3'}, \texttt{T3'x2}, \texttt{T3'x3}, and \texttt{T3'x4} with the selection strategy provided by each system.


%% file: appendix/app-stock-queries.tex
The specification of the stock market queries $Q_1$ to $Q_7$ is the following:
\smallskip
	\begin{verbatim}
		Q1 = SELECT * FROM S
		     WHERE (SELL as msft; BUY as oracle; BUY as csco; SELL as amat)
		     FILTER msft[name = 'MSFT'] AND oracle[name = 'ORCL'] AND
		            csco[name = 'CSCO'] AND amat[name = 'AMAT']
		     WITHIN 30000 [stock_time]
		
		Q2 = SELECT * FROM S
		     WHERE (SELL as msft; BUY as oracle; BUY as csco; SELL as amat)
		     FILTER msft[name = 'MSFT'] AND msft[price > 26.0] AND 
		            oracle[name = 'ORCL'] AND oracle[price > 11.14] AND
		            csco[name = 'CSCO'] AND amat[name = 'AMAT'] AND amat[price >= 18.92]
		     WITHIN 30000 [stock_time]
		
		Q3 = SELECT * FROM S 
		     WHERE (SELL as msft; BUY as oracle; BUY as csco; SELL as amat) 
		     FILTER msft[name = 'MSFT'] AND oracle[name = 'ORCL'] AND 
		            csco[name = 'CSCO'] AND amat[name = 'AMAT'] 
		     PARTITION BY [volume] 
		     WITHIN 30000 [stock_time]
		     CONSUME BY ANY
		
		Q4 = SELECT * FROM S 
		     WHERE (SELL as msft; (BUY OR SELL) as oracle; (BUY OR SELL) as csco; SELL as amat) 
		     FILTER msft[name = 'MSFT'] AND oracle[name = 'ORCL'] AND 
		            csco[name = 'CSCO'] AND amat[name = 'AMAT'] 
		     WITHIN 30000 [stock_time]
		
		Q5 = SELECT * FROM S 
		     WHERE (SELL as msft; (BUY OR SELL) as oracle; (BUY OR SELL) as csco; SELL as amat) 
		     FILTER msft[name = 'MSFT'] AND msft[price > 26.0] AND 
		            oracle[name = 'ORCL'] AND oracle[price > 11.14] AND
		            csco[name = 'CSCO'] AND amat[name = 'AMAT'] AND amat[price >= 18.92] 
		     WITHIN 30000 [stock_time]
		
		Q6 = SELECT * FROM S 
		     WHERE (SELL as msft; (BUY OR SELL) as oracle; (BUY OR SELL) as csco; SELL as amat) 
		     FILTER msft[name = 'MSFT'] AND oracle[name = 'ORCL'] AND
		            csco[name = 'CSCO'] AND amat[name = 'AMAT'] 
		     PARTITION BY [volume] 
		     WITHIN 30000 [stock_time]
		     CONSUME BY ANY
		
		Q7 = SELECT * FROM S 
		     WHERE (SELL as msft; (BUY OR SELL)+ as qqq; SELL as amat) 
		     FILTER msft[name = 'MSFT'] AND qqq[name = 'QQQ'] and 
		            qqq[volume=4000] AND amat[name = 'AMAT'] 
		     WITHIN 30000 [stock_time]
	\end{verbatim}
\medskip

Next, we briefly explain each query. $Q_1$ is a sequence query looking for a sequence \texttt{SELL};\texttt{BUY};\texttt{BUY};\texttt{SELL} of four major tech companies. $Q_2$ restricts $Q_1$ by filtering the price of each stock over a certain threshold (similar than in Example~\ref{ex:cer-hard}). Instead, $Q_3$ is a restricted version $Q_1$ with a partition-by clause over the stock volume.
$Q_4$ is for testing disjunction by allowing $Q_1$ to test for $\texttt{BUY OR SELL}$ in the two middle events. $Q_5$ and $Q_6$ are the analogs of $Q_2$ and $Q_3$, extending now $Q_4$ with filters and partition-by, respectively. Indeed, $Q_5$ contains the same features as Example~\ref{ex:cer-hard}. Finally, $Q_7$ combines disjunction and iteration, searching for a pattern of the form \texttt{SELL};\texttt{(BUY OR SELL)}+;\texttt{SELL}.

For all queries, we used a time windows of 30 seconds (e.g., 30,000 milliseconds) over the \texttt{stock\_time} attribute. We measure the throughput of the stock market stream, and it gives 4,803 e/s. Thus, using a time window of 30 seconds, we will have approximately 100 active events in the window, which is comparable with the experiments that we performed over synthetic data.


%% file: appendix/app-sel-strategies.tex
Many CER systems offer so-called selection strategies~\cite{DBLP:journals/vldb/GiatrakosAADG20,cugola2012, SASE,SASEcomplexity}. 
A selection strategy can be seen as a heuristic for evaluating a query, where the system is asked to return only a specific subset of all matched complex event. Since this subset is often easier to recognize, it improves performance.

In the next experiment, we compare all systems in the presence of selection strategies. For this, we redo the experiments of fixing the sequence length to 3, adding a final non-existing event (i.e., no output), and varying the window length from \texttt{T} to \texttt{4T}, but this time we allow each system to use its own selection strategy.  
Unfortunately, each system has its own algorithm for selection strategy and, thus, it is not possible to guarantee that everyone generates the same outputs.
Given that there is no output in this experiment setup, we can argue that all systems are hence performing a similar task; namely, we test the case where the selection strategy found no result (i.e., an unusual event). Nevertheless, the systems are still free to adopt their performance-improving heuristics, consistent with their selection strategy.

For each system we use the selection strategy that gives better performance for this experiment (except for FlinkCEP that was already using a selection strategy in all experiments). Instead, CORE implements four different selection strategies: ALL (no selection strategy), NEXT, LAST, and MAX (see~\cite{CELJOURNAL} for the semantics of each selection strategy). All are implemented at the automata level, doing a sophisticated determinization procedure to filter outputs, namely, the algorithm is the same (see Section~\ref{sec:evaluation}), but the underlying automaton is different. Then for CORE we use \texttt{LAST} selection strategy, which it is the most heavy selection strategy compared with \texttt{NEXT}, \texttt{MAX}, and \texttt{ALL} (i.e., regarding determinization of the automaton) and it produces a single output per pattern occurrence, similar to all other systems.

\begin{figure}[t]
	\centering
	\includegraphics[width=8.5cm]{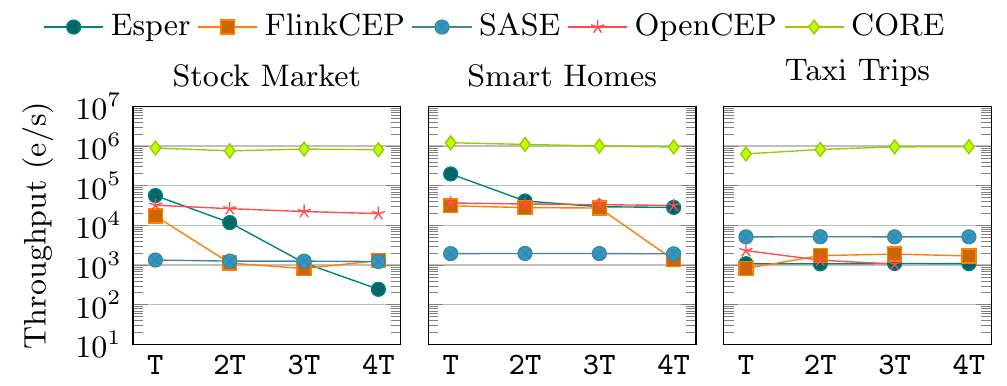}
	\vspace{-3mm}
	\caption{Throughput as a function of time-window size when using selection strategies.}
	\label{plot:seq-strategies}
\end{figure} 

In Figure~\ref{plot:seq-strategies}, we display the throughput, separated by dataset, of each system running with a selection strategy. From here, we can conclude that the use of a selection strategy improves the performance of Esper, SASE, and OpenCEP. Esper still suffers with the time windows size, decreasing its performance exponentially when the window size increases (e.g., Stock Market). SASE and OpenCEP do not incur on this cost, having a stable throughput and, moreover, increasing it up to $10^3$ and $10^4$, respectively.  Despite this improvement, CORE's throughput is at least two \oom above other systems (i.e., $10^6$ e/s) and stable when the window size increases. Therefore, we can conclude then that the advantage of CORE is in the  evaluation algorithm rather than in the use of selection strategies.